\documentclass[aps,prl,showpacs,nofootinbib]{revtex4}
\usepackage{graphicx}
\usepackage{latexsym}
\def\beq{\begin{equation}}
\def\eeq{\end{equation}}
\def\bey{\begin{eqnarray}}
\def\eey{\end{eqnarray}}

\def\lsim{\mathrel{\raise.3ex\hbox{$<$\kern-.75em\lower1ex\hbox{$\sim$}}}}
\def\gsim{\mathrel{\raise.3ex\hbox{$>$\kern-.75em\lower1ex\hbox{$\sim$}}}}

\newcommand{\be}{\begin{equation}}
\newcommand{\ee}{\end{equation}}

\begin{document}

\title{PAMELA, FGST and Sub-TeV Dark Matter}
\author{Dan Hooper$^{1,2}$ and Kathryn M. Zurek$^{1,3}$}
\address{$^1$Center for Particle Astrophysics, Fermi National Accelerator Laboratory, Batavia, IL \\ $^2$Department of Astronomy and Astrophysics, University of Chicago, Chicago, IL  \\ $^3$Department of Physics, University of Michigan, Ann Arbor, MI}

\date{\today}

\begin{abstract}

PAMELA's observation that the cosmic ray positron fraction increases rapidly with energy implies the presence of primary sources of energetic electron-positron pairs. Of particular interest is the possibility that dark matter annihilations in the halo of the Milky Way provide this anomalous flux of antimatter. The recent measurement of the cosmic ray electron spectrum by the Fermi Gamma Ray Space Telescope, however, can be used to constrain the nature of any such dark matter particle. In particular, it has been argued that in order to accommodate the observations of Fermi and provide the PAMELA positron excess, annihilating dark matter particles must be as massive as $\sim$1 TeV or heavier. In this article, we revisit Fermi's electron spectrum measurement within the context of annihilating dark matter, focusing on masses in the range of 100-1000 GeV, and considering effects such as variations in the astrophysical backgrounds from the presence of local cosmic ray accelerators, and the finite energy resolution of the Fermi Gamma Ray Space Telescope. When these factors are taken into account, we find that dark matter particles as light as $\sim$300 GeV can be capable of generating the positron fraction observed by PAMELA.

\end{abstract}
\pacs{95.35.+d; 95.85.Ry; FERMILAB-PUB-09-441-A}
\maketitle

\section{Introduction}

Last year, the publication of PAMELA's measurement of the cosmic ray positron fraction (the ratio of positrons to electrons plus positrons)~\cite{pamela} (see also Refs.~\cite{heat,ams}) created a great deal of interest among both cosmic ray astrophysicists and particle physicists. In particular, PAMELA observed the positron fraction to rise rapidly with energy, from approximately 5-6\% at 10 GeV, to 10-20\% by 100 GeV. Despite uncertainties involved in the predictions of the secondary positron flux ({\it ie.}~positrons originating from the interactions of cosmic ray protons with the interstellar medium), it appears that this result cannot be explained without the existence of primary sources of cosmic ray positrons~\cite{serpico}. One possibility that has received a great deal of attention is that these positrons are the products of dark matter annihilations~\cite{history,ann1,ann2,leptonlight,lepton,ann5,theory} or decays~\cite{decay} in the local halo of the Milky Way. Alternatively, astrophysical sources may also be capable of producing energetic electron-positron pairs. Nearby pulsars, for example, are a particularly promising candidate for the origin of PAMELA's positron excess~\cite{pulsar}.

In addition, several experiments have measured the cosmic ray electron (plus positron) spectrum over a similar range of energies. The balloon-based experiment ATIC reported an electron spectrum which rose rapidly up to approximately 600 GeV, at which point it fell suddenly~\cite{atic}. The spectral feature implied by this result was particularly exciting within the context of dark matter particles which annihilate or decay to monoenergetic electron-positron pairs~\cite{edge1,kkdmpamela}. More recently, however, the Fermi Gamma Ray Space Telescope (FGST) has reported their measurement of the cosmic ray electron spectrum, and find no evidence for such a sudden feature~\cite{fgstelectron}.  The ground-based gamma ray telescope HESS has also reported its measurement of the cosmic ray electron spectrum, again failing to confirm the presence of a edge-like feature~\cite{hessedge}. Despite the absense of a sudden edge, however, the spectrum observed by FGST does contain an overall excess between $\sim 100-1000$ GeV relative to the predictions of simple, homogeneous cosmic ray models.

The rapid rise of the positron fraction observed by PAMELA requires that the source of these particles produce an extremely hard spectrum. Within the context of dark matter, such particles must produce mostly charged leptons ($e^+ e^-$, $\mu^+ \mu^-$, and/or $\tau^+ \tau^-$) in their decays or annihilations if they are to accommodate this observation~\cite{ann1,ann2,leptonlight,lepton}. This requirement has been made even more stringent as a result of the FGST electron spectrum measurement, which found the electron spectrum to be only slightly softer than $dN_e/dE_e \propto E_e^{-3}$ over the range of the PAMELA positron fraction measurement. Furthermore, dark matter annihilations to non-leptonic final states are tightly constrained by PAMELA's measurement of the cosmic ray antiproton-to-proton ratio~\cite{pamelaantiproton,antiprotons}, which is consistent with the predictions of cosmic ray models with no exotic or primary antiproton contributions.

Following the publication of FGST's cosmic ray electron spectrum measurement, several groups performed studies to determine the range of dark matter masses, cross sections, and annihilation channels that could potentially account for these observations~\cite{fgstinter,bergstrom,meade}. In these studies, the authors of Refs.~\cite{bergstrom} and~\cite{meade} found that good fits to both the FGST electron spectrum and the PAMELA positron fraction could be found, but only in the case that the dark matter particles are very heavy ($m_{\rm DM} \sim 1-4$~TeV), and annihilate/decay to muons or possibly taus. This conclusion, however, depends critically on the astrophysical backgrounds that are present in the cosmic ray spectrum. As the number of astrophysical sources (supernova remnants, etc.) that contribute to the cosmic ray electron spectrum above $\sim 10^2$ GeV is expected to be small (of order unity), the spectrum observed at the Solar System can significantly depart from the time-averaged spectrum as a result of fluctuations in the brightness and spectra of nearby sources. This makes it very difficult to predict the shape of the relevant background over the range of energies measured by FGST (and HESS). In this article, we revisit the question of what types of annihilating dark matter could potentially account for the positron excess observed by PAMELA, while remaining consistent with the cosmic ray electron spectrum observed by FGST. We find that a wide variety of scenarios are consistent with these sets of data, even in cases in which the dark matter mass is well below 1 TeV. In particular, upon relaxing the assumptions pertaining to the cosmic ray electron background, we are able to find excellent fits to both PAMELA and FGST for dark matter particles annihilating to either $\mu^+ \mu^-$ or $\tau^+ \tau^-$ with masses as low as $\sim$300 GeV. We also find that when the energy resolution of FGST is taken into account, dark matter particles annihilating to $e^+ e^-$ can accommodate the data for masses in the range of approximately 400 to 600 GeV.

\section{Cosmic Ray Electrons and Positrons From Annihilating Dark Matter}

To calculate the cosmic ray electron/positron spectrum taking into account the effects of diffusion and energy losses, we solve the steady-state propagation equation~\cite{prop}:
\begin{eqnarray}
\frac{\partial}{\partial t}\frac{dN_{e}}{dE_{e}}(\vec{x},E_e)=0=Q(\vec{x},E_e) + \vec{\nabla}\cdot [K(E_e,\vec{x})\vec{\nabla}\frac{dN_{e}}{dE_{e}}(\vec{x},E_e)]
 + \frac{\partial}{\partial E_e}[B(E_e)\,\frac{dN_{e}}{dE_{e}}(\vec{x},E_e)],  
\end{eqnarray}
where $dN_{e}/dE_{e}$ is the number density of electrons/positrons per unit energy, $K(E_{e},\vec{x})$ is the diffusion coefficient, and the source term, $Q(E_e, \vec{x})$, reflects both the distribution of sources and the spectrum that is injected from those sources.  $B(E_e)$ is the energy loss rate of electrons due to synchrotron and inverse Compton processes. In the relativistic limit, this rate is related to the radiation field and magnetic field energy densities by
\begin{eqnarray}
B(E_e) &=& \frac{4}{3}\sigma_T \rho_{\rm rad} \bigg(\frac{E_e}{m_e}\bigg)^2    + \frac{4}{3} \sigma_T \rho_{\rm mag} \bigg(\frac{E_e}{m_e}\bigg)^2 \nonumber \\
&\approx& 1.02 \times 10^{-16}\,{\rm GeV/s}\,\bigg(\frac{\rho_{\rm rad}+\rho_{\rm mag}}{{\rm eV}/{\rm cm}^3}\bigg) \times  \bigg(\frac{E_e}{{\rm GeV}}\bigg)^2 \nonumber \\
&\equiv& \frac{1}{\tau} \times \frac{E^2_e}{(1\,{\rm GeV})},
\end{eqnarray}
where $\sigma_T$ is the Thompson cross section and $\tau$ is the representative energy loss time. Throughout our study, we adopt $\tau =10^{16}$ seconds, which corresponds to a total energy density of approximately 1 eV/cm$^3$. This density consists of contributions from the cosmic microwave background ($\rho_{\rm CMB}\approx 0.23$ eV/cm$^3$), the Galactic Magnetic Field ($\rho_{B} \approx 0.2$ eV/cm$^3\times (B/3\mu G)^2$), and starlight ($\rho_{\rm SL} \sim 1$ eV/cm$^3$).

For the diffusion coefficient and diffusion zone boundary conditions, we adopt three sets of parameters that are consistent with all measured stable and unstable secondary-to-primary ratios in the cosmic ray spectrum~\cite{simet}. We will refer to the following as Models A, B and C:
\begin{center}
\begin{itemize}
\item{Model A: $K(E_e)=5.3\cdot 10^{28}$ cm$^2$/s ($E_e/4 \, {\rm GeV})^{0.43}$, $L=$4 kpc}
\item{Model B: $K(E_e)=1.4\cdot 10^{28}$ cm$^2$/s ($E_e/4 \, {\rm GeV})^{0.43}$, $L=$1 kpc}
\item{Model C: $K(E_e)=7.3\cdot 10^{28}$ cm$^2$/s ($E_e/4 \, {\rm GeV})^{0.43}$, $L=$6 kpc},
\end{itemize}
\end{center}
where $L$ is the distance above or below the Galactic Plane at which charged particles are able to freely escape the Galactic Magnetic Field. When fitting the PAMELA and FGST data with a particular dark matter model, we use the propagation model (of the above A, B, or C) that provides the best overall fit. For the source term, $Q(E_e, \vec{x})$, we adopt an Narvarro-Frenk-White (NFW) halo profile~\cite{NFW} with a local density of 0.3 GeV/cm$^3$. 

As charged cosmic rays approach the Solar System, the solar wind can cause them to lose energy, distorting their spectrum. This effect varies with time according to the solar cycle, and has the greatest impact on the lowest energy cosmic rays. The largest effects of solar modulation are independent of charge ({\it ie.}~they effect electrons and positrons in the same way) and thus can be removed by considering ratios of different species, such as the positron fraction. Although smaller charge dependent effects of solar modulation do exist, they are significant only below about 5-10 GeV. As we only consider the positron fraction as measured above 10 GeV in this paper, we can safely neglect the effects of solar modulation on the positron fraction.

To account for the effects of charge independent solar modulation on the cosmic ray electron spectrum, we estimate the suppression of the electron/positron spectrum using the simple force-field approximation~\cite{forcefield} (see also Ref.~\cite{fisk}). In this treatment, we can relate the cosmic ray electron spectrum at Earth to that in the interstellar medium by
\begin{equation}
\frac{d\Phi_{\oplus}}{dE_{\oplus}} (E_{\oplus}) = \frac{E^2_{\oplus}}{E^2_{\rm ISM}} \frac{d\Phi_{\rm ISM}}{dE_{\rm ISM}} (E_{\rm ISM}),
\end{equation}
where the subscripts $\oplus$ and ISM denote quantities at Earth, and in the interstellar medium, respectively. The energies in this expression are related by
\begin{equation}
E_{\rm ISM} = E_{\oplus} + \phi_F,
\end{equation}
where $\phi_F$ is the associated potential, sometimes referred to as the solar modulation parameter. As FGST's measurement of the cosmic ray electron spectrum was conducted near the period of solar minima, we adopt a potential in the range of 650 to 1000 MeV, chosen in each case (within this range) to provide the best overall fit to the data. While we acknowledge that the effects of solar modulation on the cosmic ray electron spectrum may be more complicated than can be modelled by this simple and approximate treatment, this approach provides a reasonable estimate that we consider adequate for our purposes.

\section{Results For A Generic WIMP}

To begin, we consider the simple case in which dark matter particles annihilate directly to electron-positron pairs. Along with the contribution from dark matter, we include a power-law spectrum of electrons from supernova remnants or other such cosmic ray accelerators (we chose the spectral index and normalization of this background such that we provide the best overall fit in each case).   We also include a component of secondary positrons from hadronic interactions with the interstellar medium, given by $d\Phi_{e^+}/dE_{e^+} = 0.0045 \, E_{e^+}^{-3.5}$ GeV$^{-1}$ cm$^{-2}$ s$^{-1}$ sr$^{-1}$, which is estimated from the observed spectrum of cosmic ray protons and other cosmic ray measurements~\cite{secondaries}.  When comparing the electron spectrum to the measurements of FGST, we have convolved the spectrum with a gaussian of width approximately given by the energy resolution of FGST ($\Delta E_e/E_e \approx 0.15$ at 150 GeV, increasing to $\Delta E_e/E_e \approx 0.20$ above $E_e \approx 500$~GeV; see figure 1 of Ref.~\cite{fgstelectron}).

\begin{figure}[!]
\begin{center}
{\includegraphics[angle=0,width=0.42\linewidth]{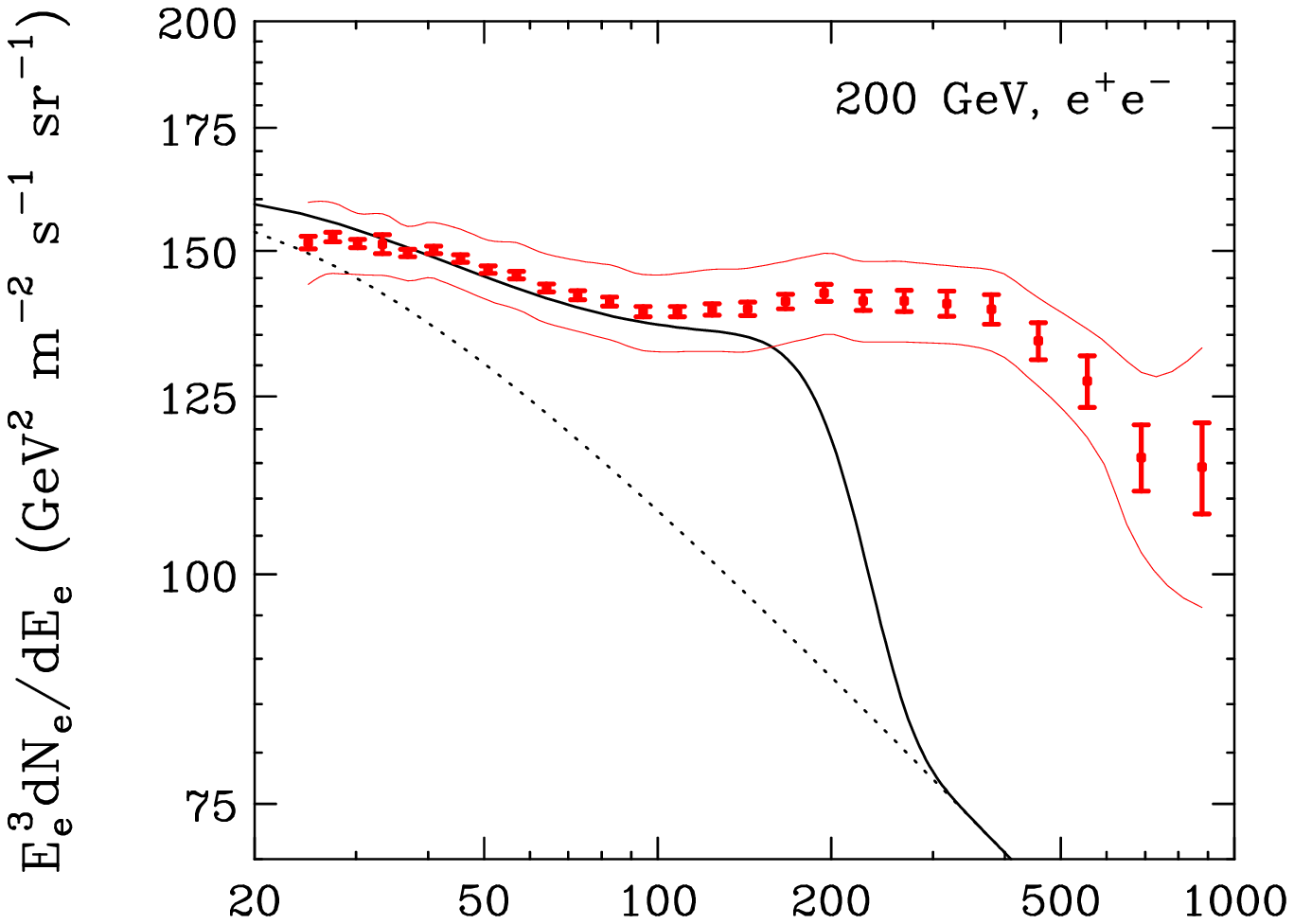}}
\hspace{0.2cm}
{\includegraphics[angle=0,width=0.42\linewidth]{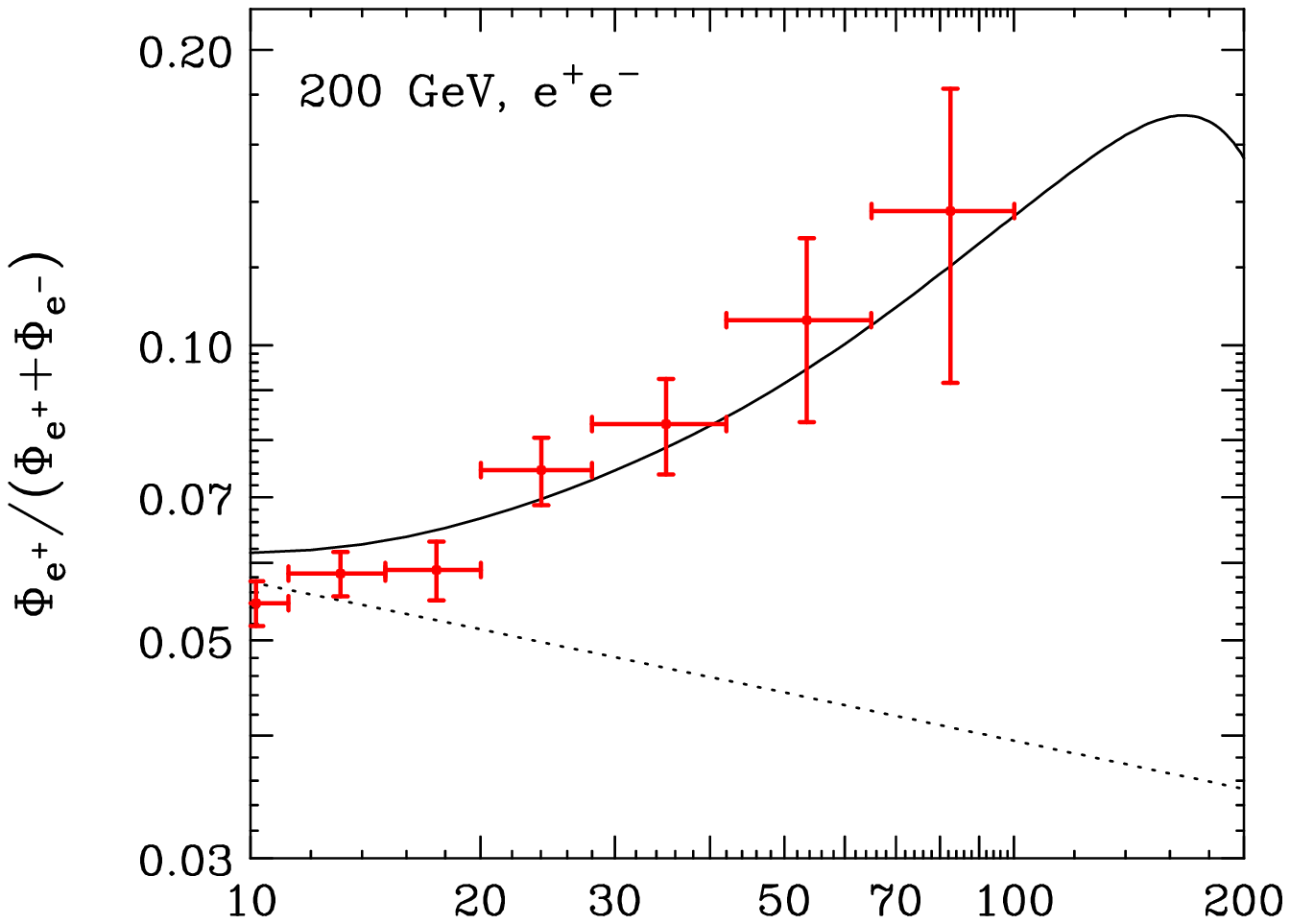}}\\
{\includegraphics[angle=0,width=0.42\linewidth]{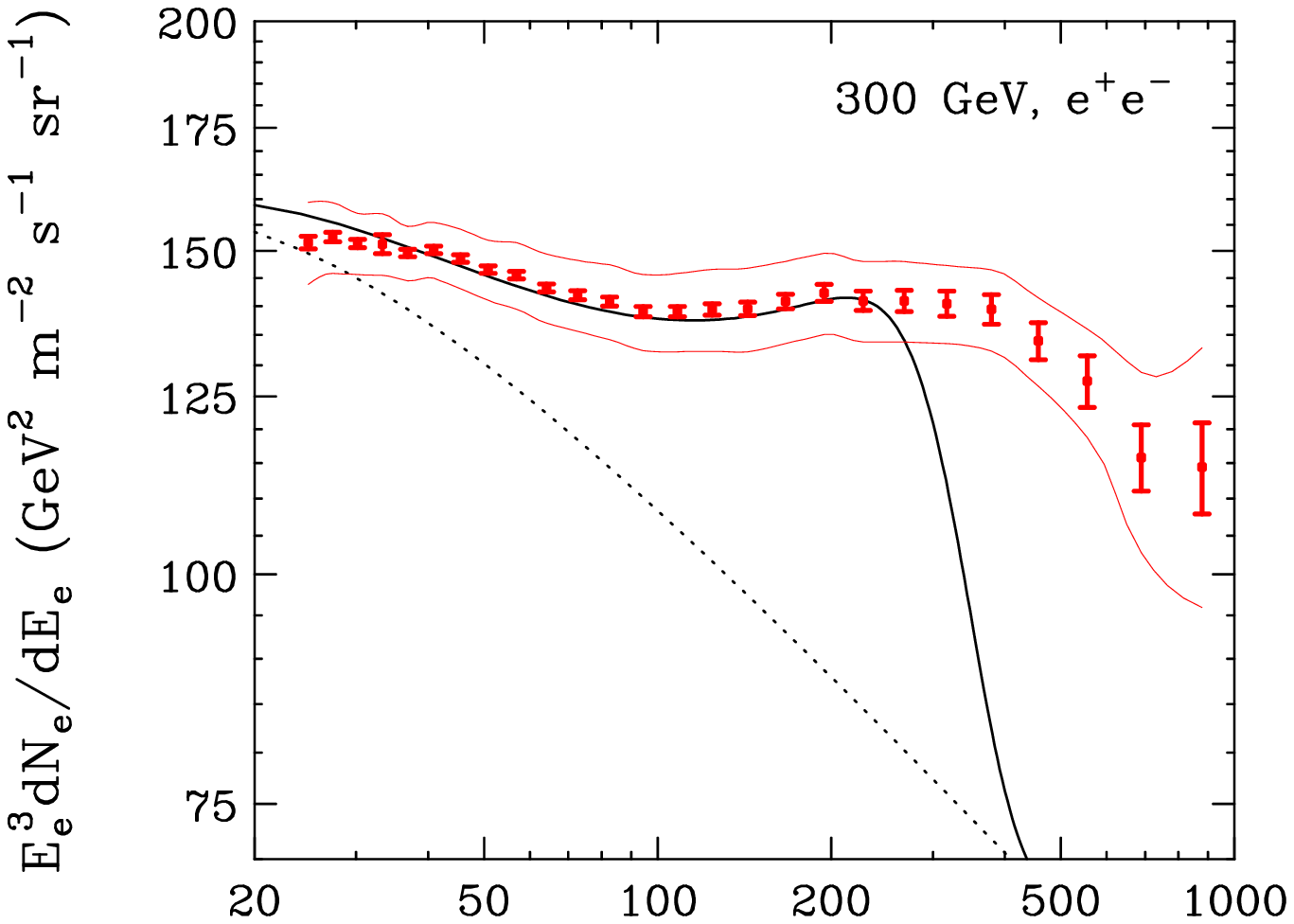}}
\hspace{0.2cm}
{\includegraphics[angle=0,width=0.42\linewidth]{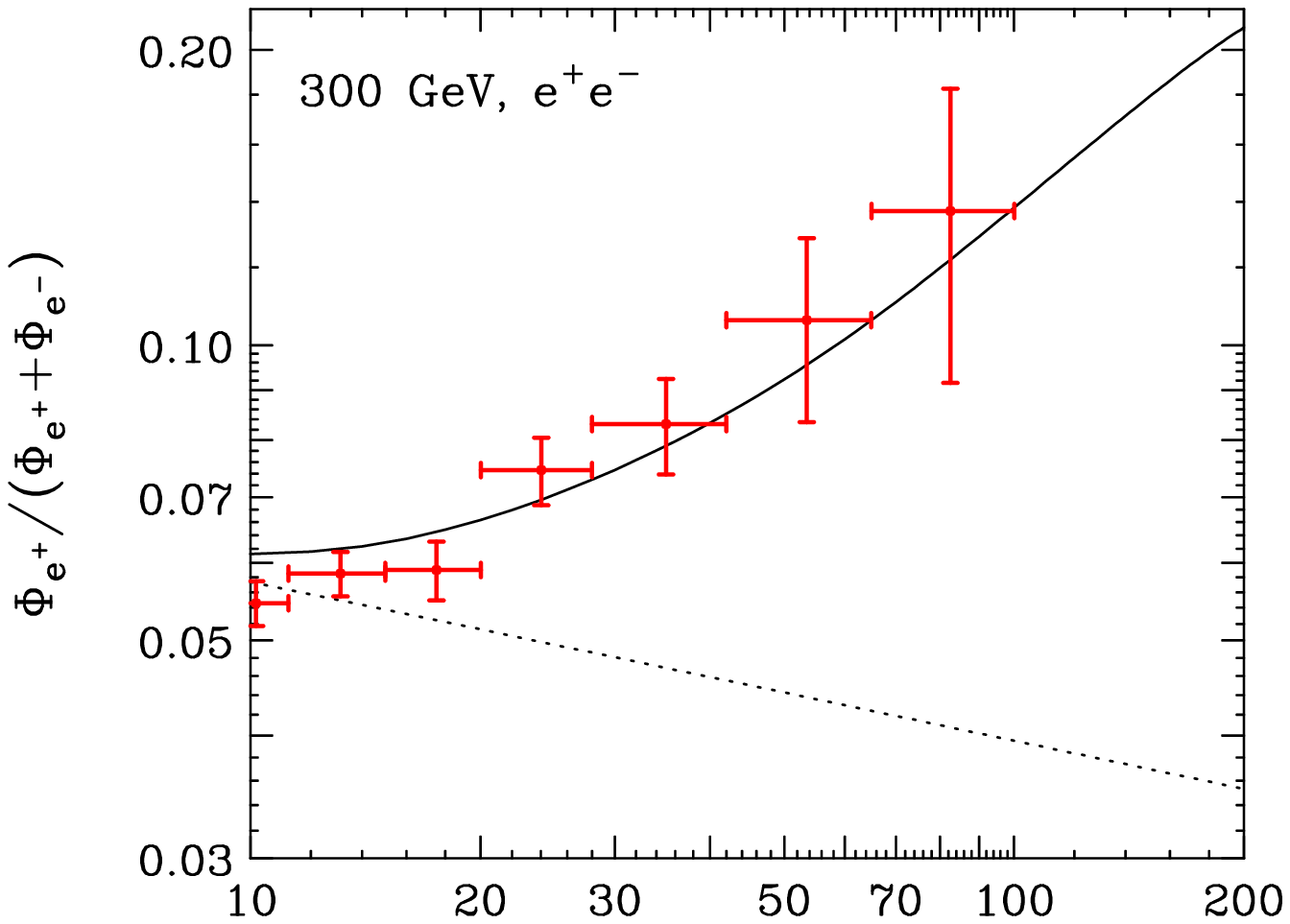}}\\
{\includegraphics[angle=0,width=0.42\linewidth]{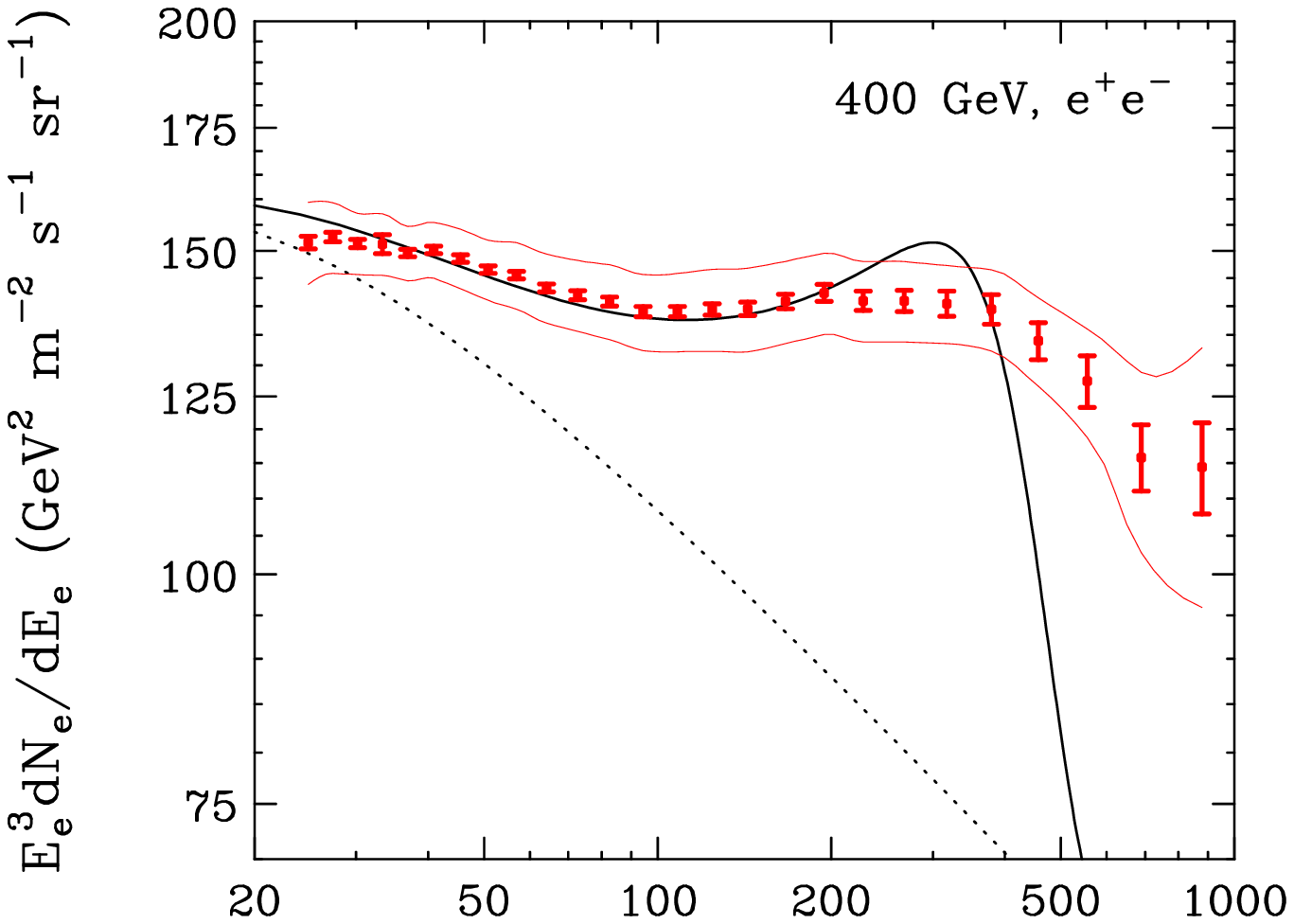}}
\hspace{0.2cm}
{\includegraphics[angle=0,width=0.42\linewidth]{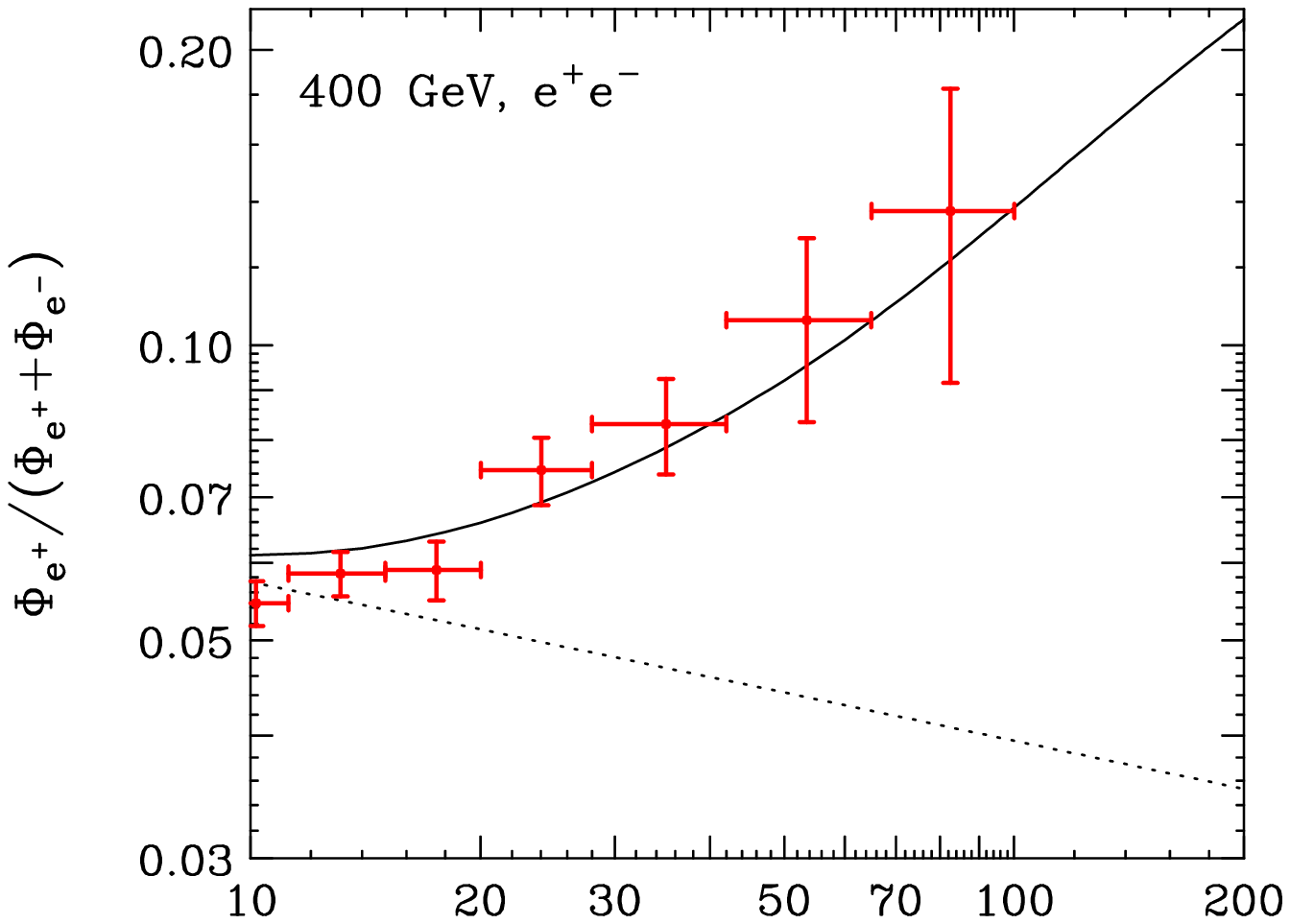}}\\
{\includegraphics[angle=0,width=0.42\linewidth]{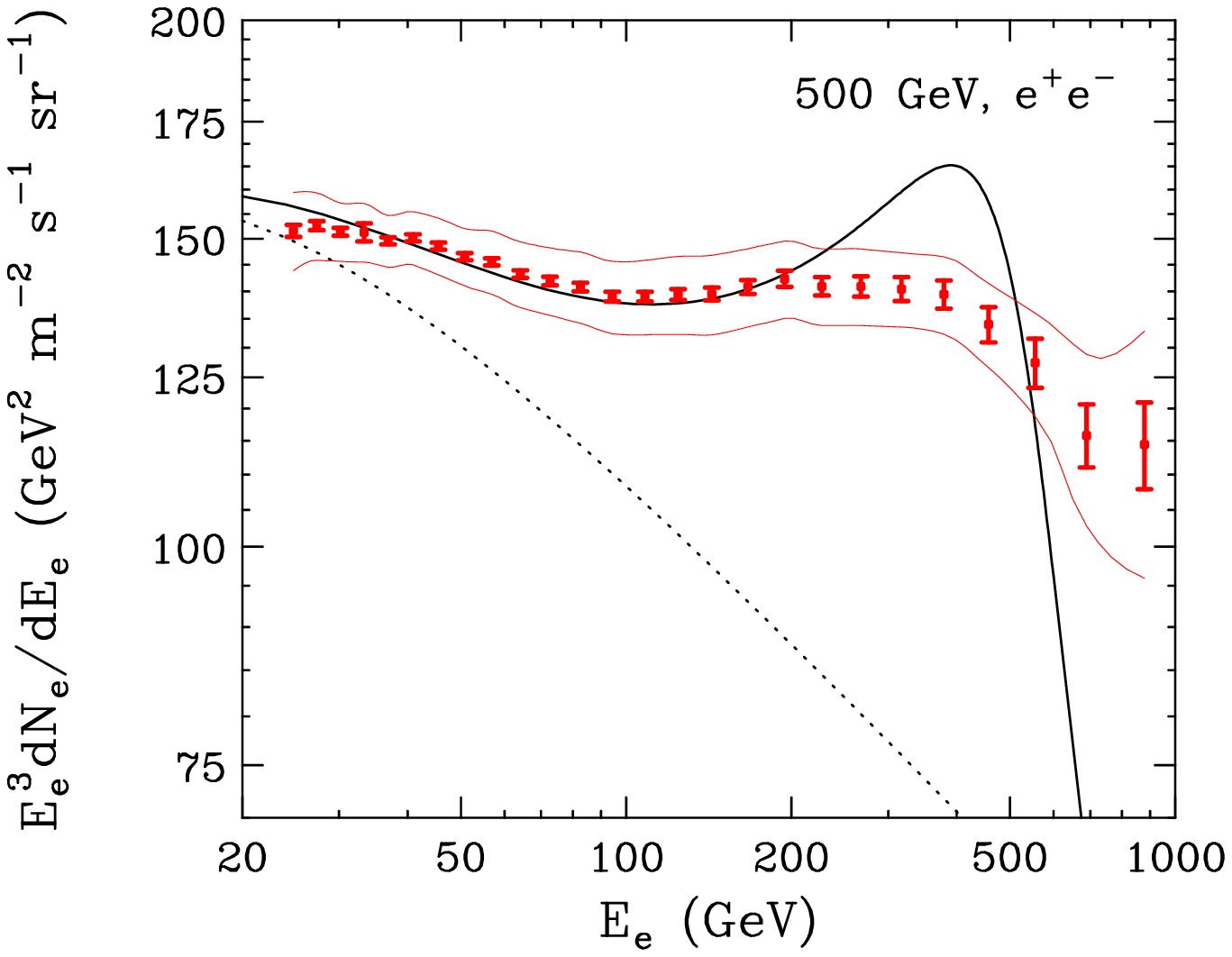}}
\hspace{0.2cm}
{\includegraphics[angle=0,width=0.4\linewidth]{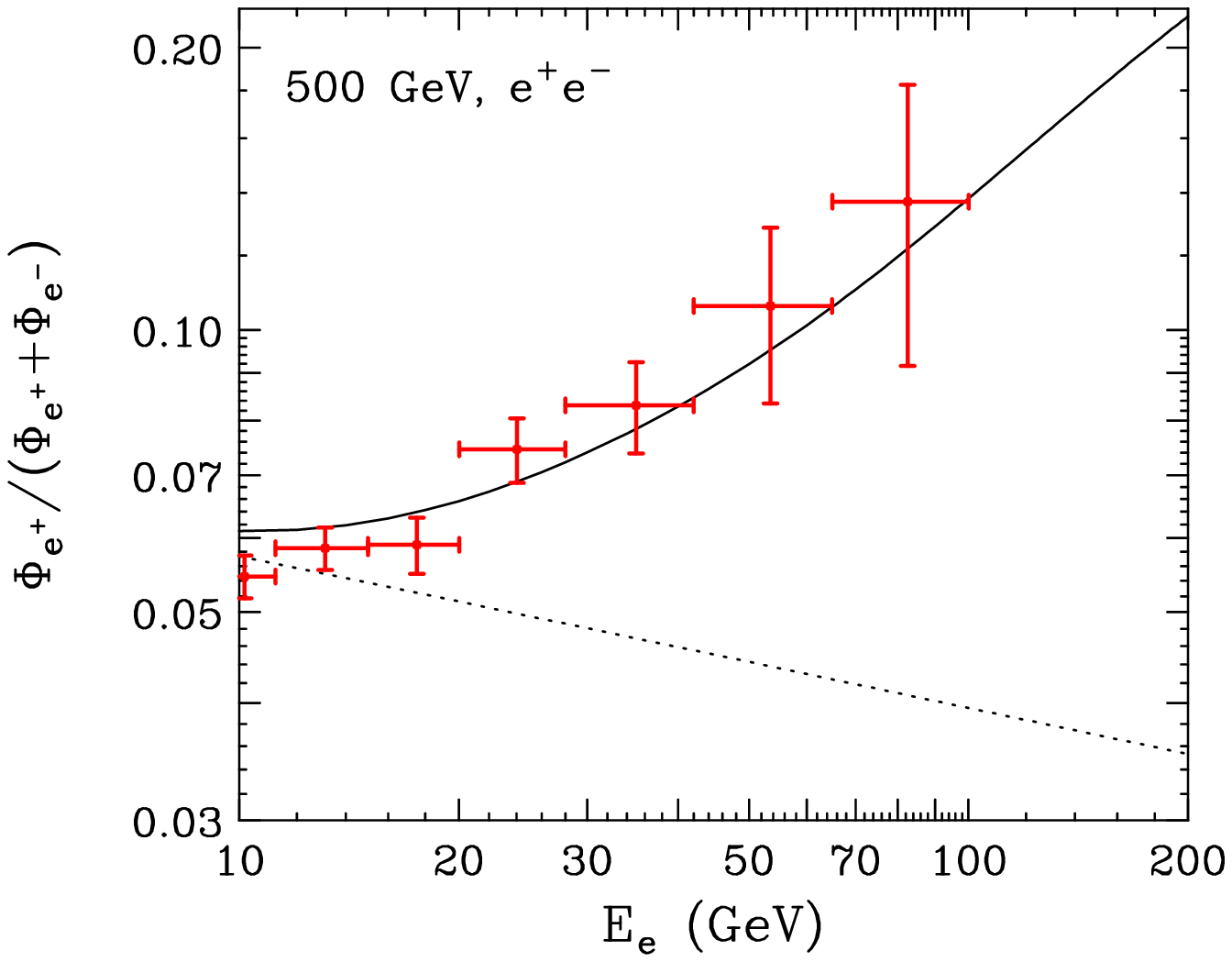}}\\
\caption{The best fits found to the PAMELA data above 10 GeV, and the FGST data {\it below 100 GeV}, for a WIMP annihilating to $e^+ e^-$ along with a power-law spectrum of cosmic ray electrons from astrophysical sources. The dotted lines denote the astrophysical background used, without the contribution from dark matter. In each case, we found a very good fit to the data ($\chi^2 \lsim 6$ distributed over 20 error bars). To normalize the dark matter annihilation rate, we have used a boost factor (relative to the rate predicted for $\sigma v = 3 \times 10^{-26}$ cm$^3$/s and $\rho_{0}=0.3$ GeV/cm$^2$) of 18.5, 41.6, 72.9, and 112 from top-to-bottom, respectively. For each case shown, we have used propagation model A, and $\phi_F=1000$ MeV. See text for more details.}
\label{lt100}
\end{center}
\end{figure}

\begin{figure}[!]
\begin{center}
{\includegraphics[angle=0,width=0.45\linewidth]{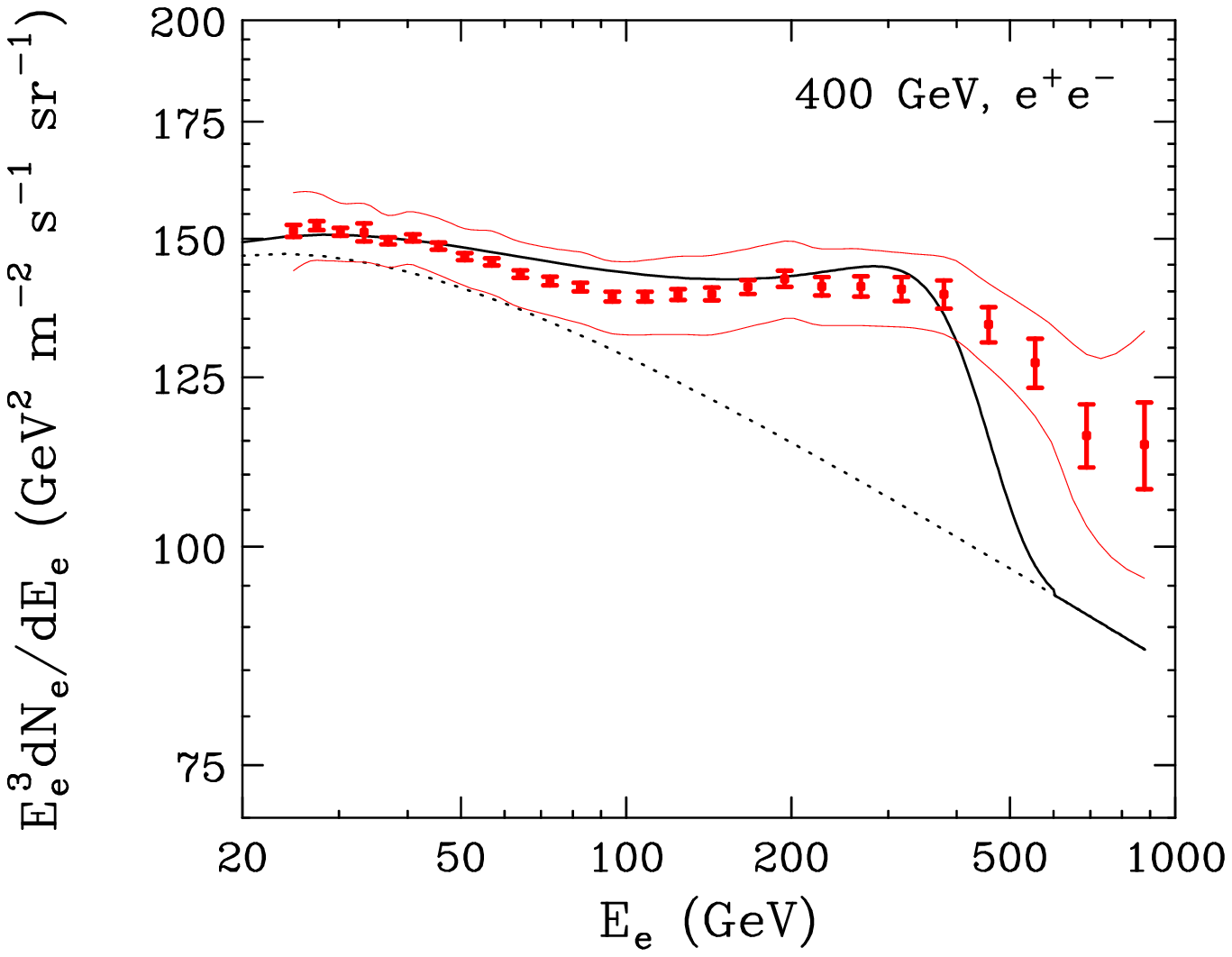}}
\hspace{0.2cm}
{\includegraphics[angle=0,width=0.45\linewidth]{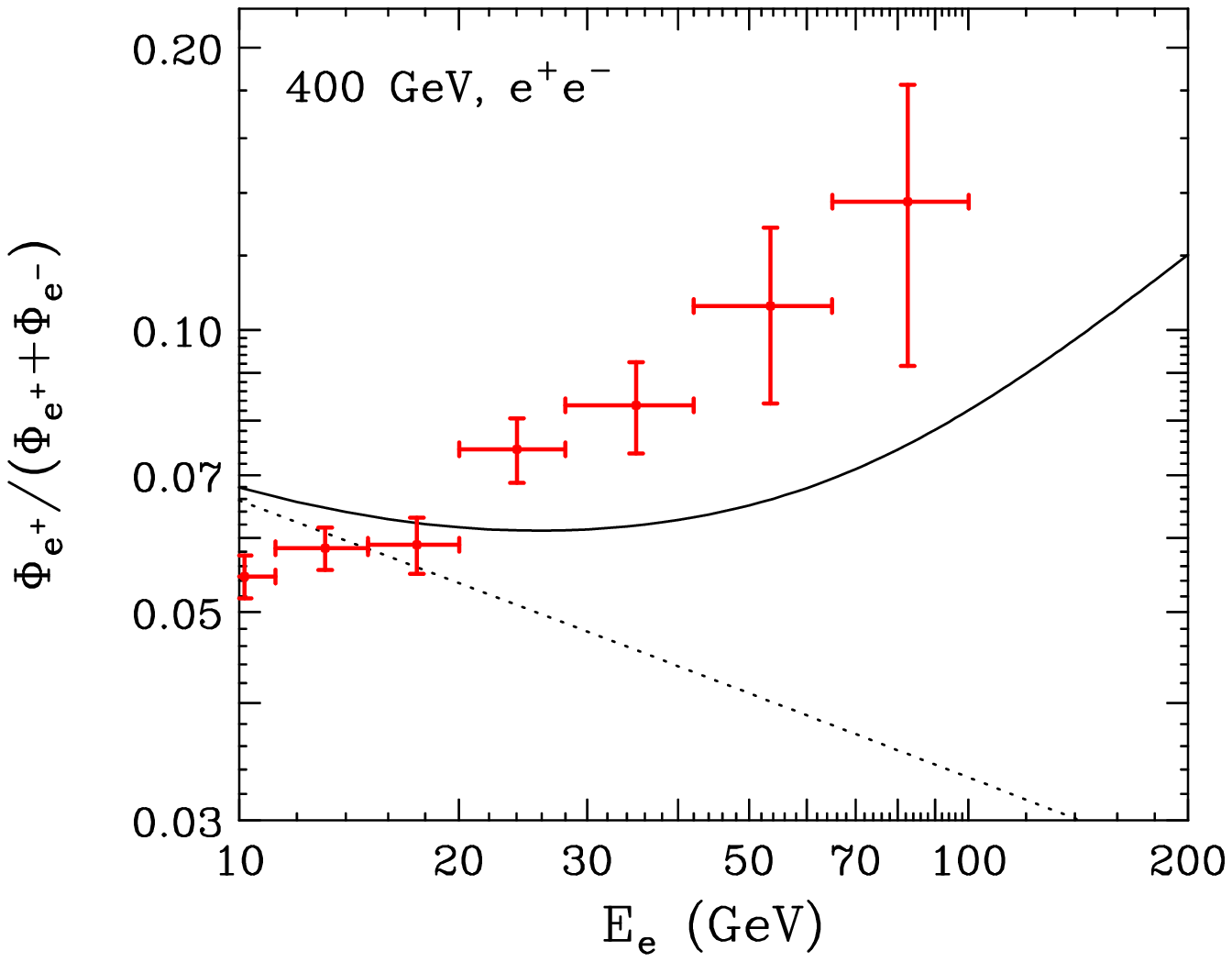}}\\
{\includegraphics[angle=0,width=0.45\linewidth]{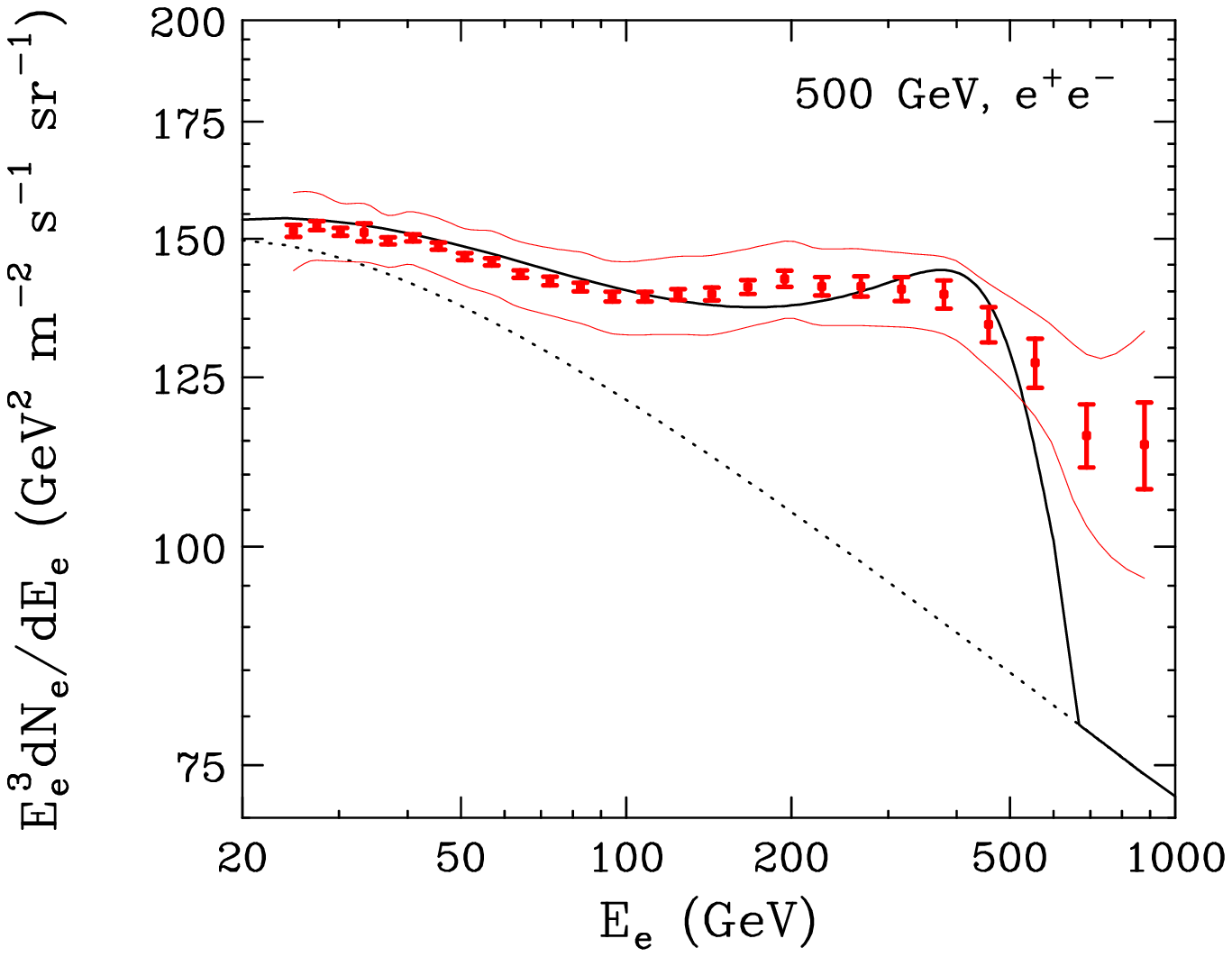}}
\hspace{0.2cm}
{\includegraphics[angle=0,width=0.45\linewidth]{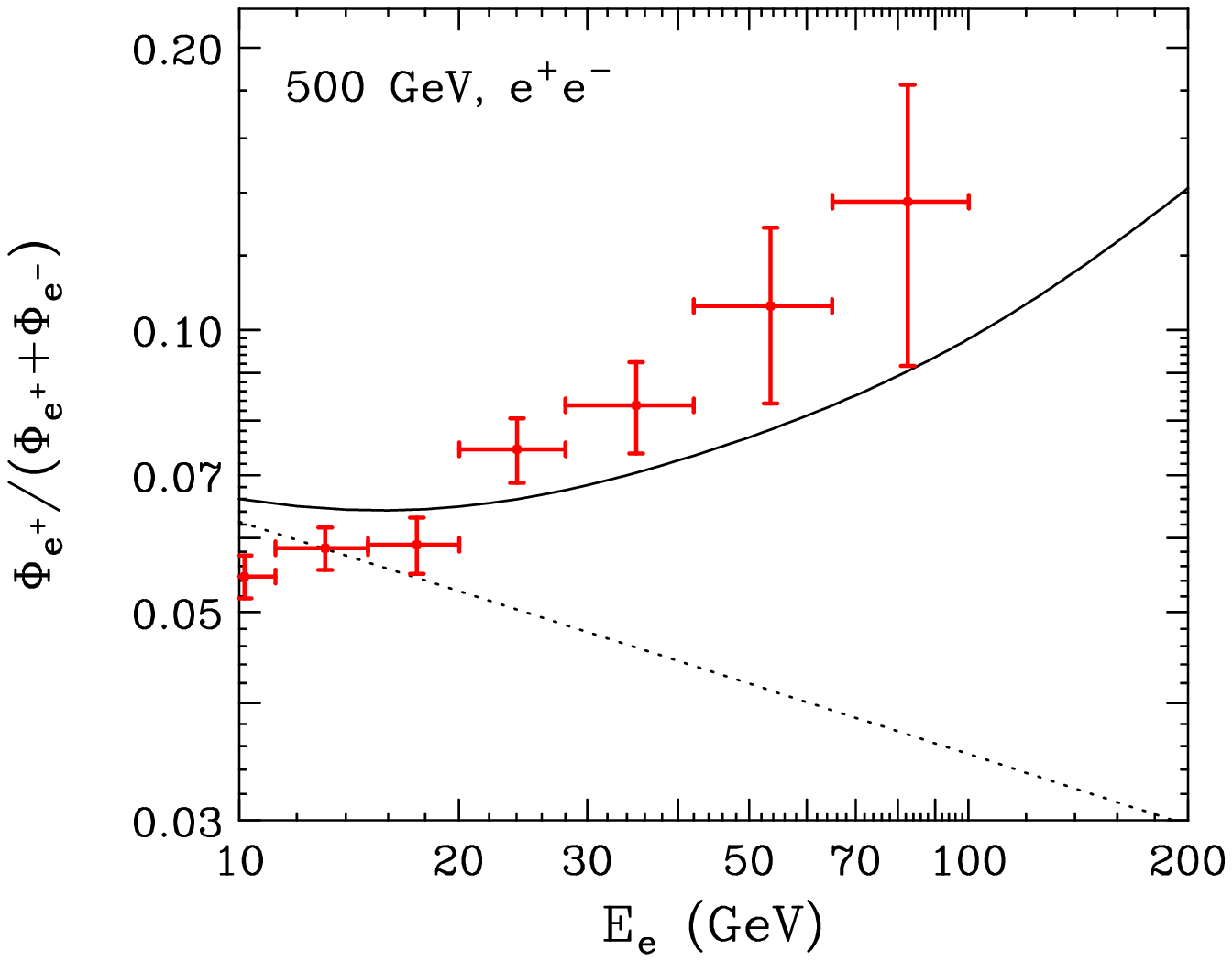}}\\
{\includegraphics[angle=0,width=0.45\linewidth]{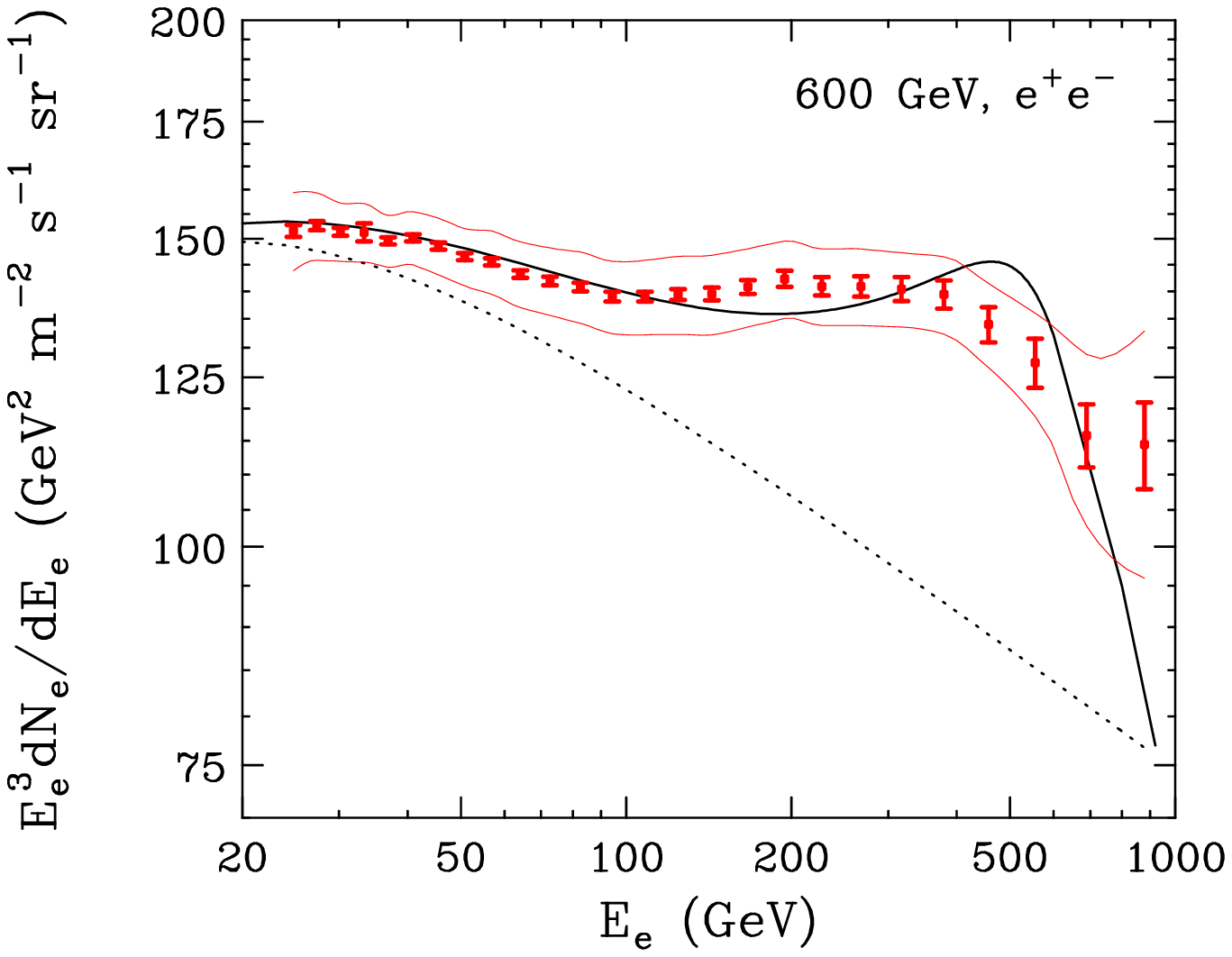}}
\hspace{0.2cm}
{\includegraphics[angle=0,width=0.45\linewidth]{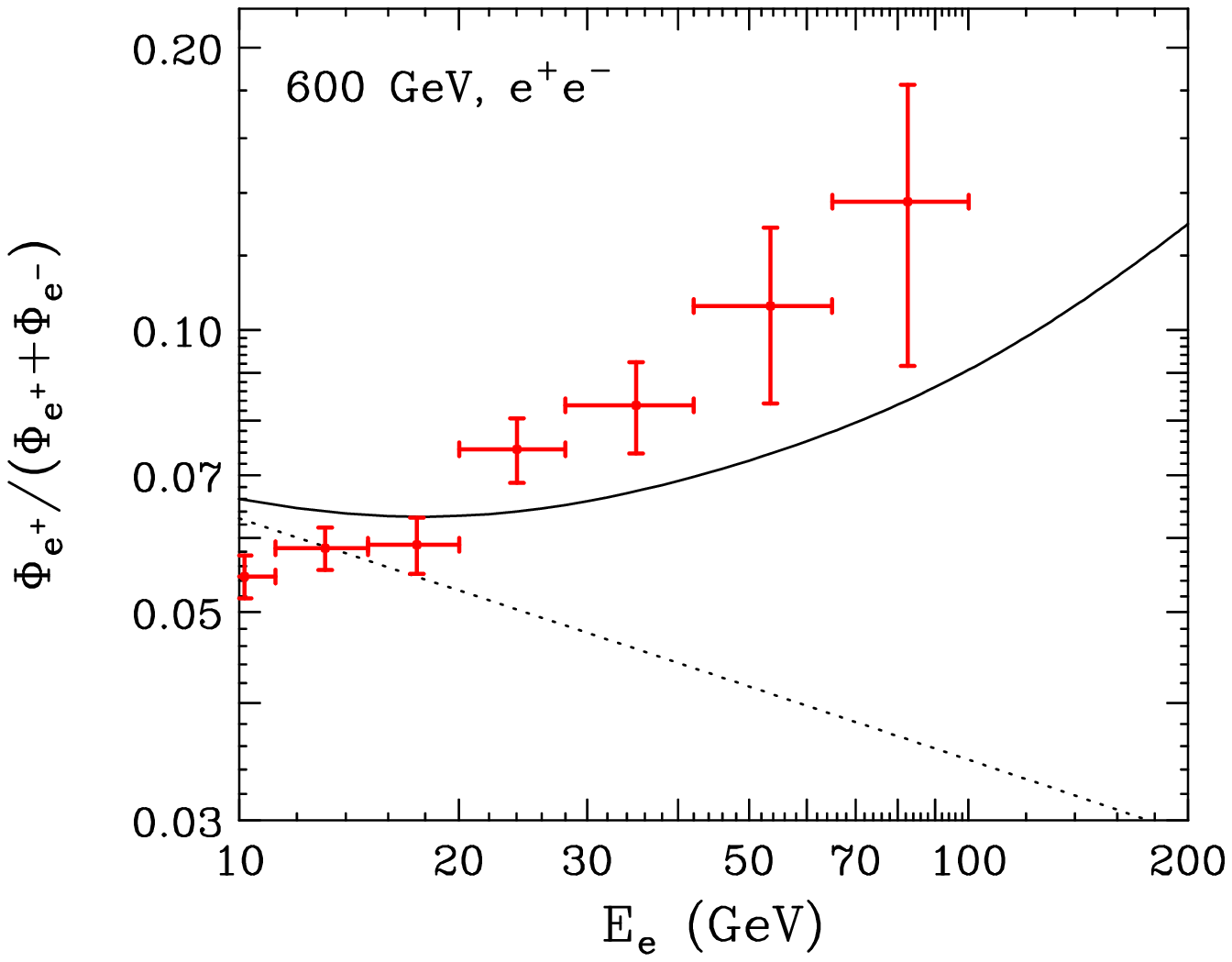}}
\caption{The best fits found to the PAMELA data above 10 GeV, and the (total) FGST data, for a WIMP annihilating to $e^+ e^-$ along with a power-law spectrum of cosmic ray electrons from astrophysical sources. The dotted lines denote the astrophysical background used, without the contribution from dark matter. For masses between 400 and 1000 GeV, we found reasonably good fits to the data ($\chi^2 \lsim 22-35$ distributed over 33 error bars). To normalize the dark matter annihilation rate, we have used a boost factor (relative to the rate predicted for $\sigma v = 3 \times 10^{-26}$ cm$^3$/s and $\rho_{0}=0.3$ GeV/cm$^2$) of 37, 72 and 91 from top-to-bottom, respectively.  From top-to-bottom, we have used propagation models A, C, and C, and in each case shown have used $\phi_F=1000$ MeV. See text for more details.}
\label{all}
\end{center}
\end{figure}

In our first figure, we focus on the range of energies that is measured by both FGST and PAMELA (approximately 25-100 GeV). Over this energy range, the combination of these two measurements provides us with determinations of both the electron and positron cosmic ray spectra (at higher energies, FGST has measured the sum of these spectra, but cannot differentiate their relative abundances). For four different dark matter masses (200, 300, 400, and 500 GeV), we show in Fig.~\ref{lt100} the best fits found to the PAMELA data above 10 GeV, and the FGST data below 100 GeV (ignoring for the moment the measurements above 100 GeV). Superb fits are found in each case (the total $\chi^2$ is less than 6 for each mass, distributed over 20 error bars). 

In each of the left frames, the error bars denote the statistical errors, whereas the upper and lower red lines denote the range of systematic errors, which may be correlated. In the right frames, we compare the positron fraction to the measurements of PAMELA. In each frame, the dotted line denotes the result without any contribution from dark matter annihilations. We have selected the free parameters (the spectral index and normalization of the electron background, the dark matter annihilation rate, the propagation model, and the solar modulation potential) to provide the best fit to the PAMELA data above 10 GeV and the FGST data below 100 GeV. 

It is clear from Fig.~\ref{lt100} that while such WIMPs can very nicely accommodate the PAMELA and FGST data between 10 and 100 GeV, very light WIMPs ($m_{\rm DM}\lsim 400$ GeV) are not capable, on their own, of accommodating the spectrum observed by FGST at higher energies. Furthermore, heavier WIMPs ($m_{\rm DM} \gsim 500$ GeV) annihilating to $e^+ e^-$ can exceed the spectrum measured by FGST at the highest energies. In Fig.~\ref{all}, we show our best fits to all of the PAMELA and FGST data (above 10 GeV) for 400, 500, and 600 GeV WIMPs annihilating to $e^+ e^-$. In each case, we find a good fit to the combined data ($\chi^2 \approx 22-35$ over 33 error bars), although the measurements of the positron fraction tend to somewhat exceed the predictions of the models. Essentially, to avoid exceeding the FGST spectrum at high energies, the annihilation rate is reduced in these cases, resulting in a reduction of the positron fraction. Nevertheless, for dark matter annihilating to $e^+ e^-$ with masses within the range of approximately 400 to 600 GeV, we find good overall fits to the PAMELA positron fraction and the entire FGST spectrum. Particularly attractive is the ability of these models to accommodate the drop in the electron spectrum observed around 400 to 600 GeV by FGST.

For lighter dark matter particles, the cosmic ray electron background from astrophysical sources must increase at high energies if the spectrum observed by FGST is to be accommodated. Such behavior in the cosmic ray electron spectrum, however, is not at all implausible from the perspective of cosmic ray propagation~\cite{sto} (see also Ref.~\cite{fgstinter}). In particular, a cosmic ray electron will lose most of its energy over a time scale of $\sim \tau/E_e \approx 3 \times 10^{13}$~seconds $\times (300~\rm{GeV}/E_e)$. Over this length of time, an electron will travel a distance on the order of $D \sim \sqrt{K(E_e) \tau/E_e} \sim 1~$kpc $\times (300~{\rm GeV}/E_e)^{0.3}$. As this distance is comparable to the typical spacing between active supernova remnants (those currently injecting cosmic ray electrons), it is likely that significant departures from the average cosmic ray electron spectrum will be observed at and above this approximate energy scale. It is even plausible that only one or two relatively young and nearby supernova remnants provide the dominant contribution to the highest energy portion of the cosmic ray electron spectrum measured by FGST.

\begin{figure}[!]
\begin{center}
{\includegraphics[angle=0,width=0.45\linewidth]{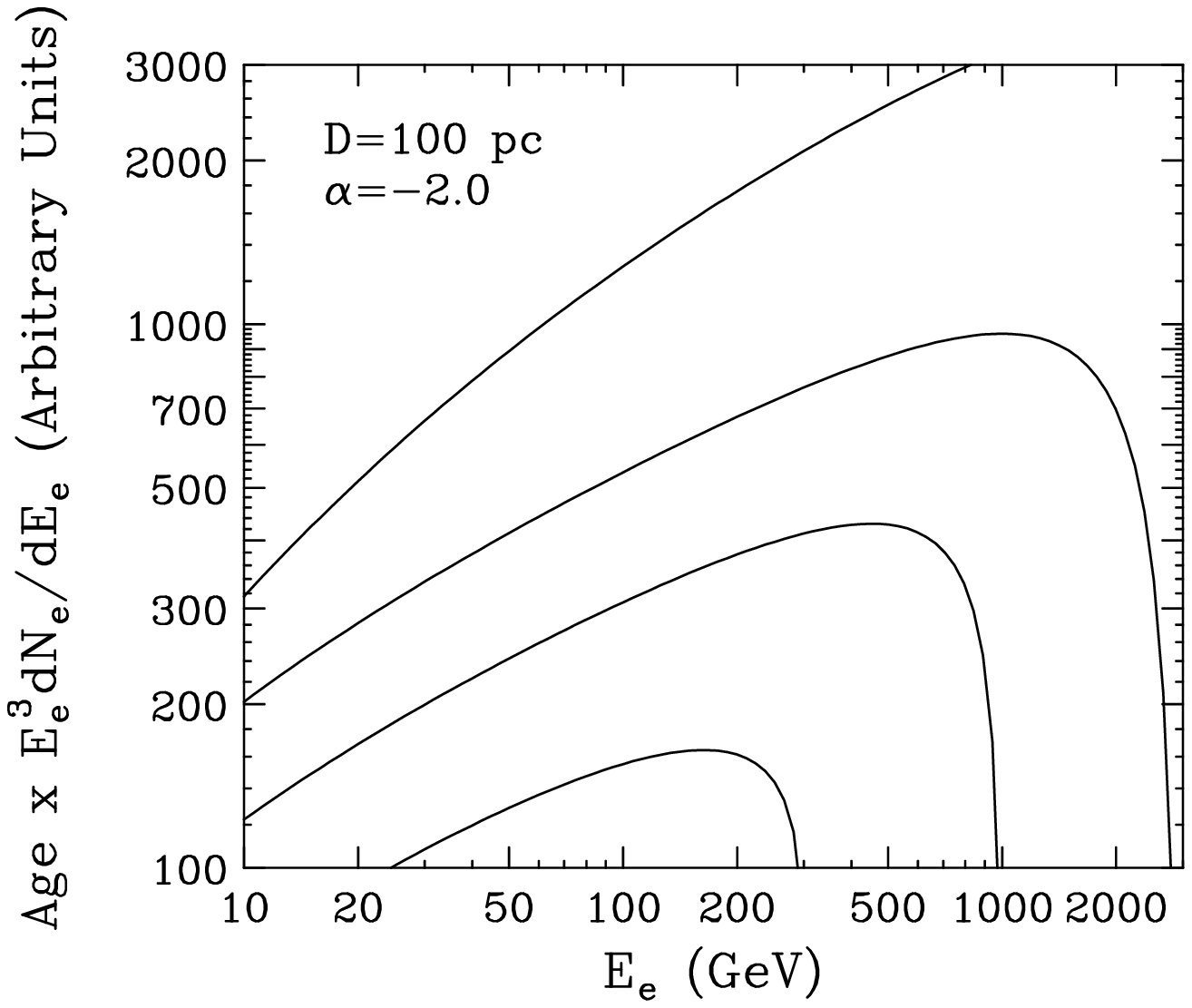}}
{\includegraphics[angle=0,width=0.45\linewidth]{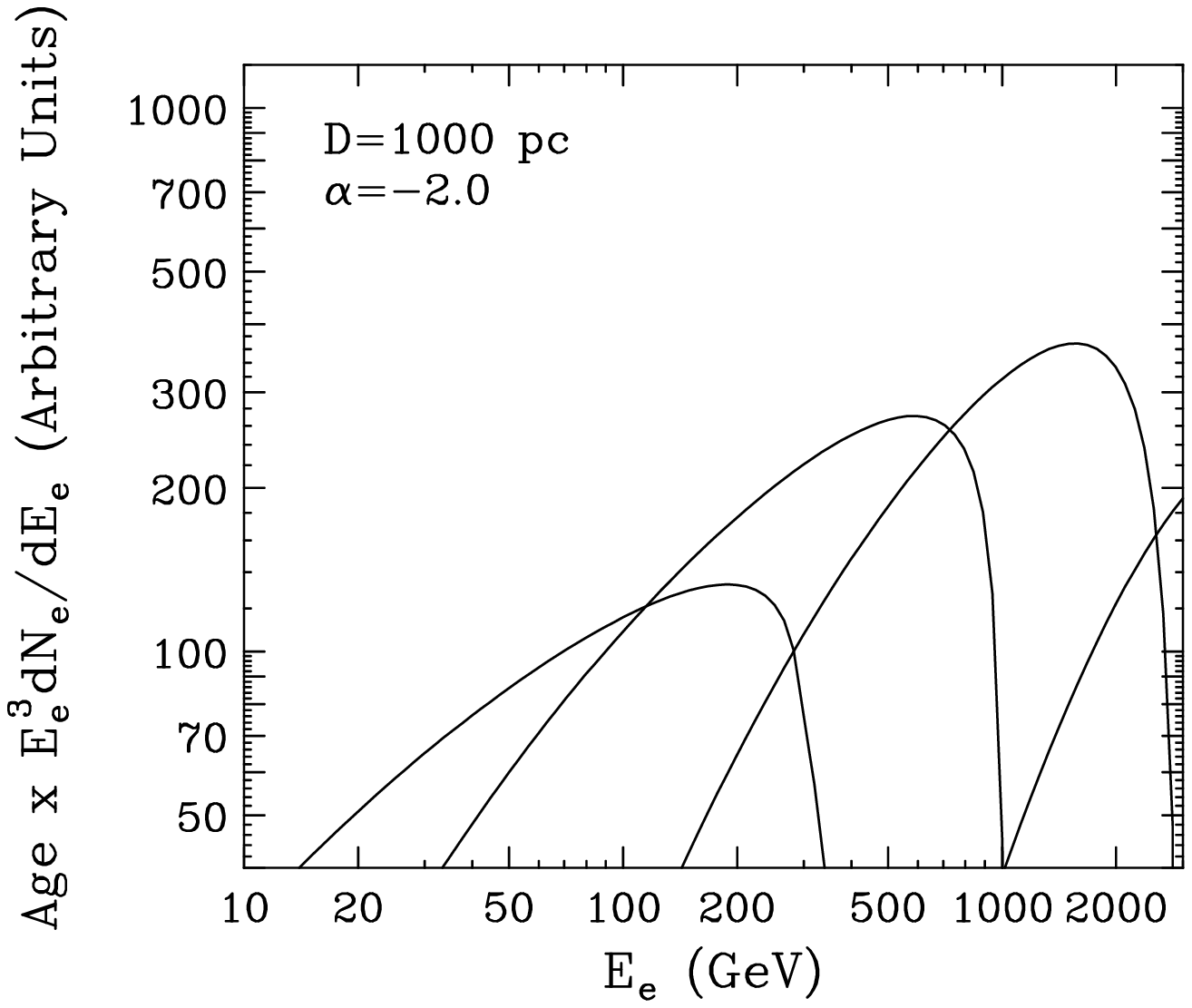}}
\caption{The spectral shape of cosmic ray electrons from a nearby and young cosmic ray accelerator. In the left and right frames, the source is 100 parsecs and 1000 parsecs away, respectively. In each case, the source is assumed to have injected a spectrum of electrons given by $dN_e/dE_e \propto E_e^{-2}$. Each line represent a different age of the source: 10,000, 100,000, 300,000 and 1,000,000 years, from top-to-bottom in the left frame, or 30,000, 100,000, 300,000 and 1,000,000 years, from right-to-left in the right frame). Here, we have used propagation model A.}
\label{100pc}
\end{center}
\end{figure}

In Fig.~\ref{100pc}, we show the electron spectral shape from a nearby (100 parsecs in left frame, 1000 parsecs in the right) and young (10,000, 100,000, 300,000 and 1,000,000 years, from top-to-bottom in the left frame, 30,000, 100,000, 300,000 and 1,000,000 years, from right-to-left in the right frame) cosmic ray electron accelerator, with an injection spectrum of $dN_e/dE_e \propto E_e^{-2}$. A spectrum similar to that injected from the source is found up to an energy of $E_e (GeV) \sim \tau/\Delta t$, where $\Delta t$ is the age of the source. Above this approximate energy, the flux begins to drop suddenly. More distant sources have a harder spectrum because of the fact that the least energetic particles have not yet diffused far enough to reach the observer.

\begin{figure}[!]
\begin{center}
{\includegraphics[angle=0,width=0.45\linewidth]{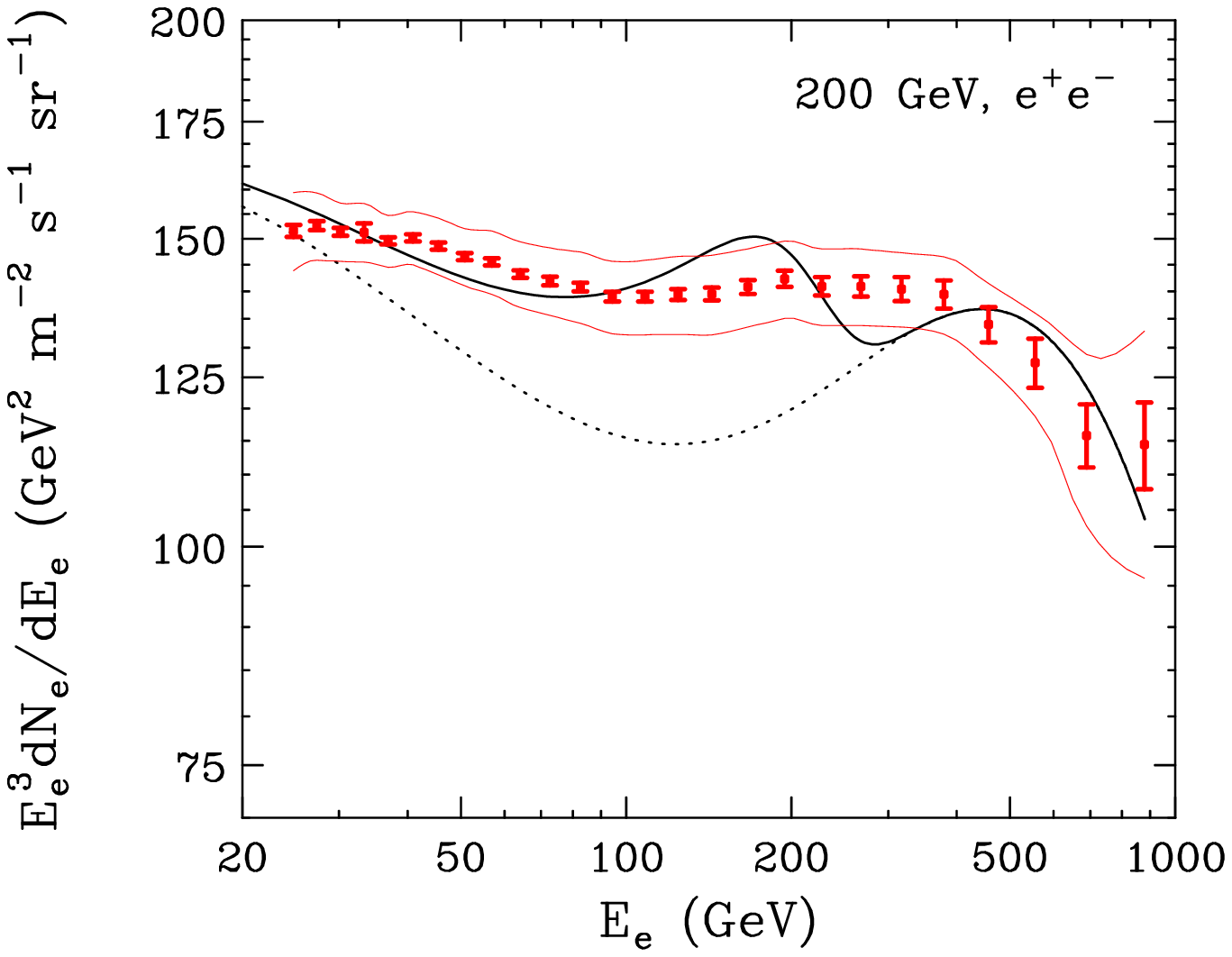}}
\hspace{0.2cm}
{\includegraphics[angle=0,width=0.45\linewidth]{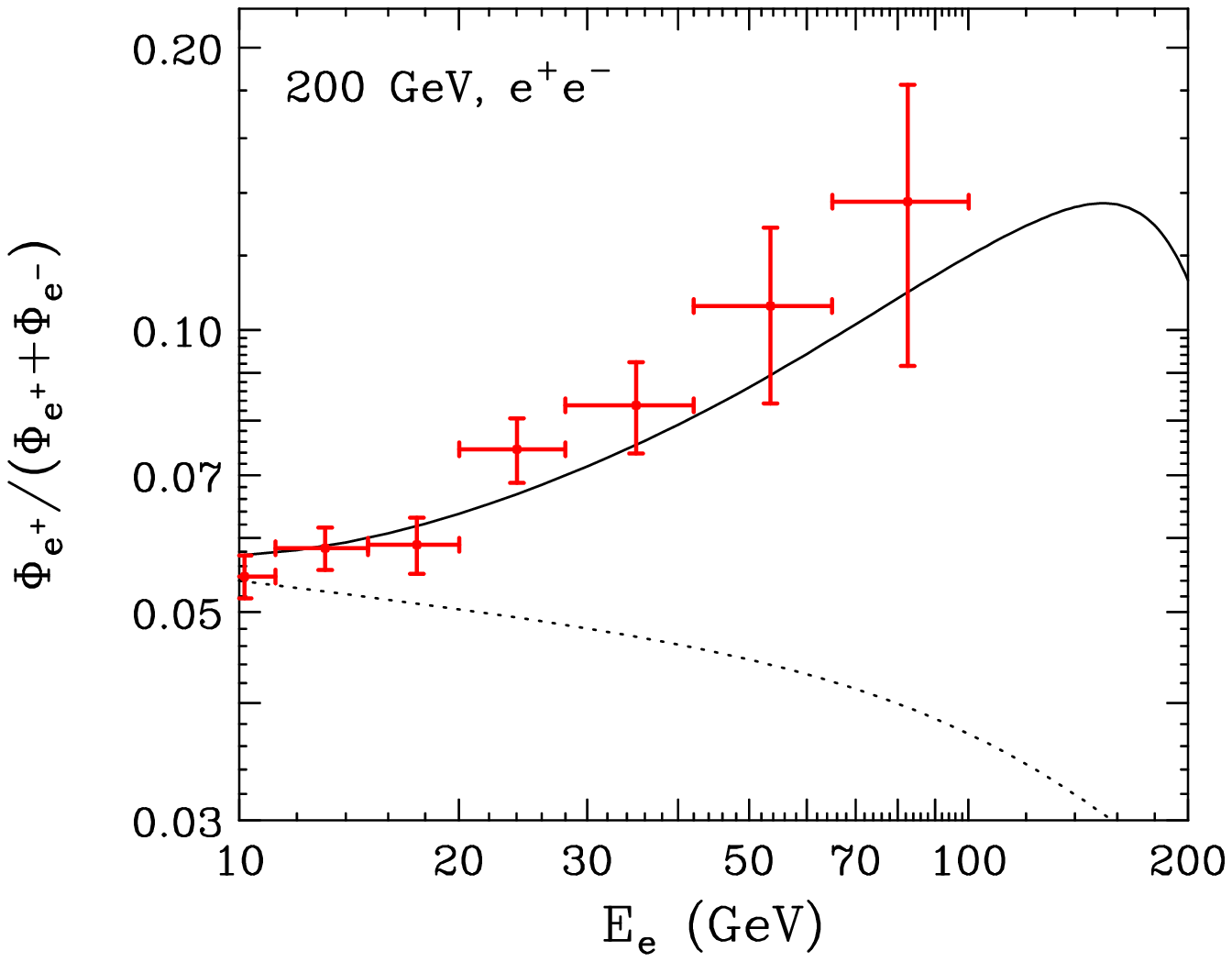}}\\
{\includegraphics[angle=0,width=0.45\linewidth]{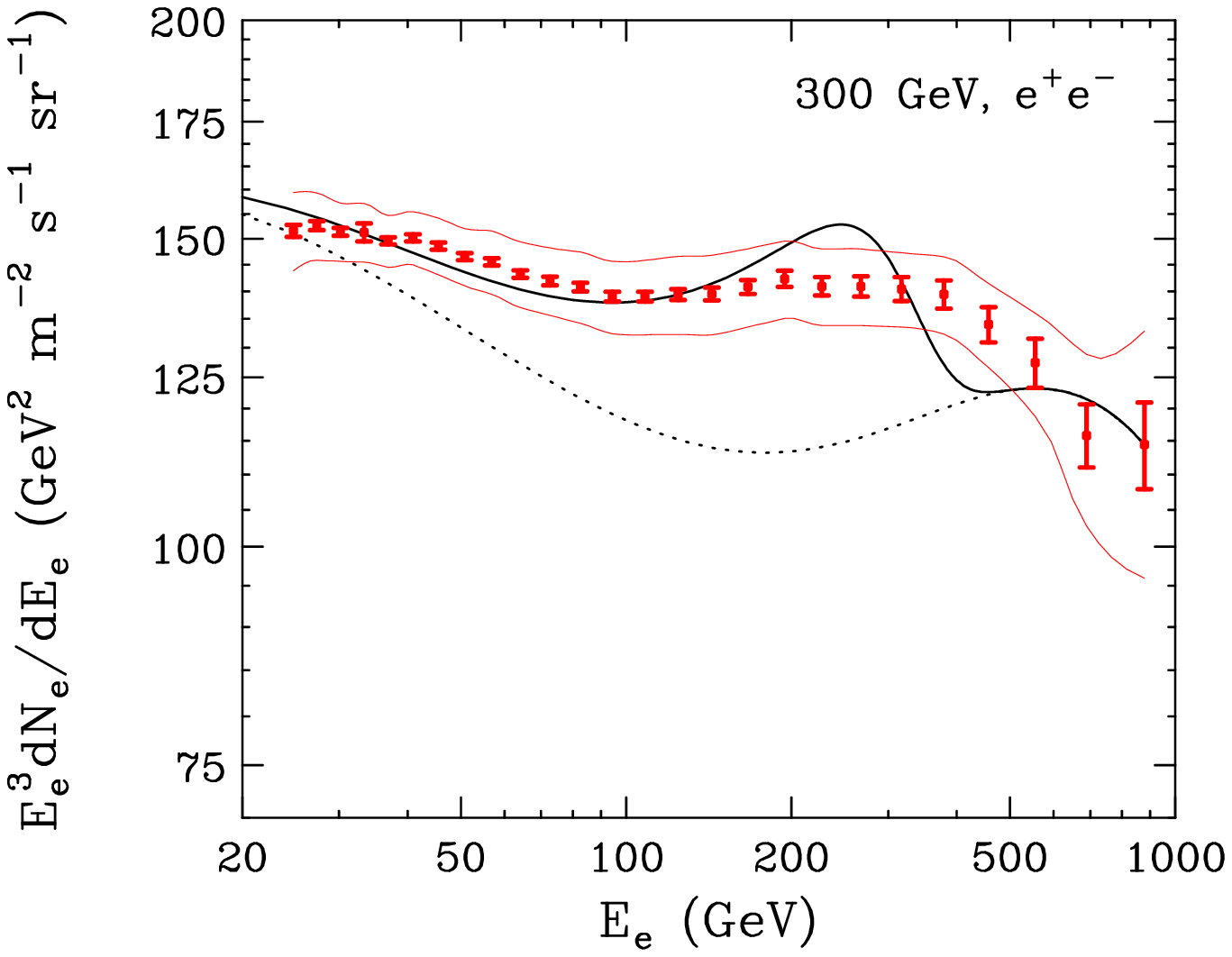}}
\hspace{0.2cm}
{\includegraphics[angle=0,width=0.45\linewidth]{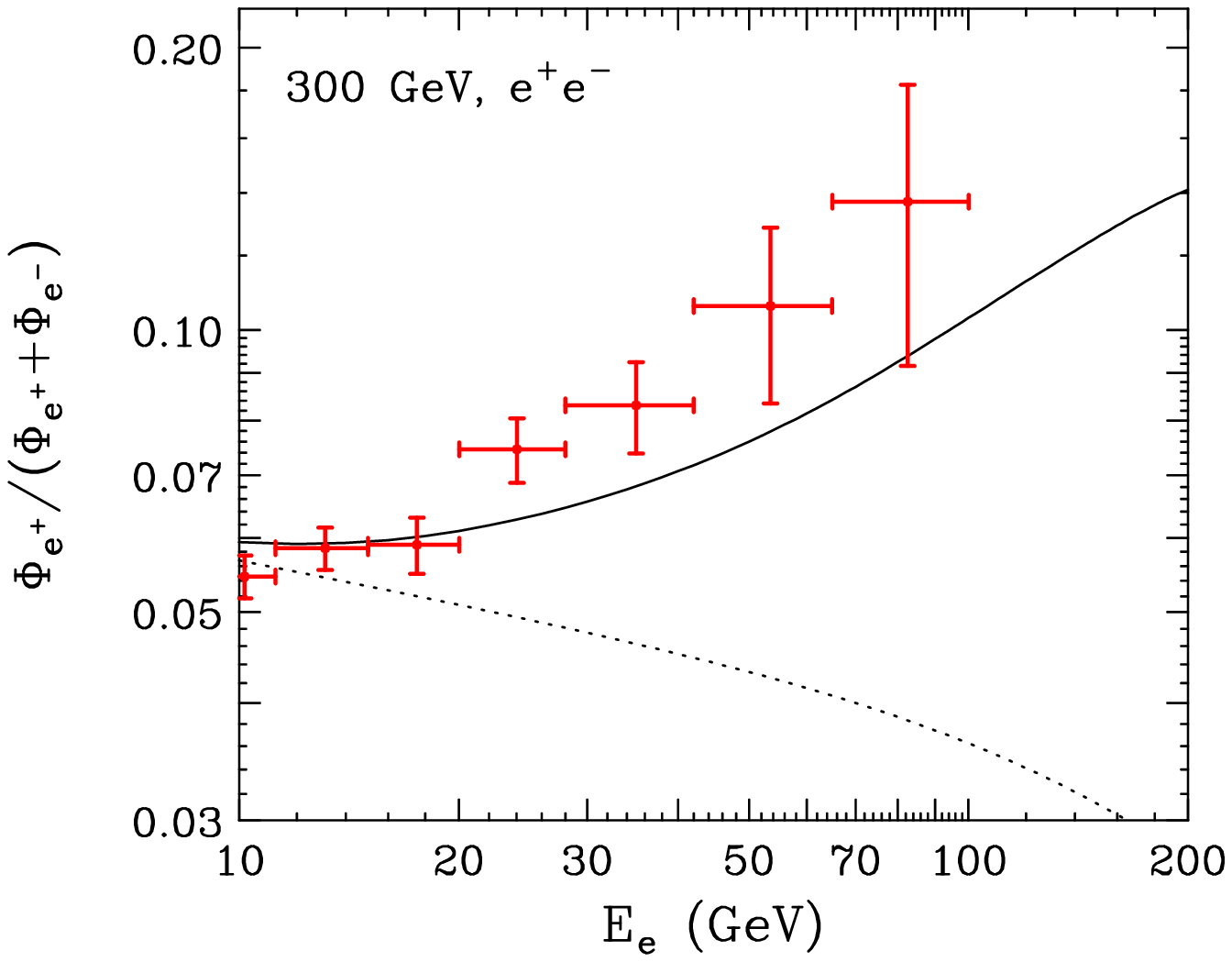}}\\
{\includegraphics[angle=0,width=0.45\linewidth]{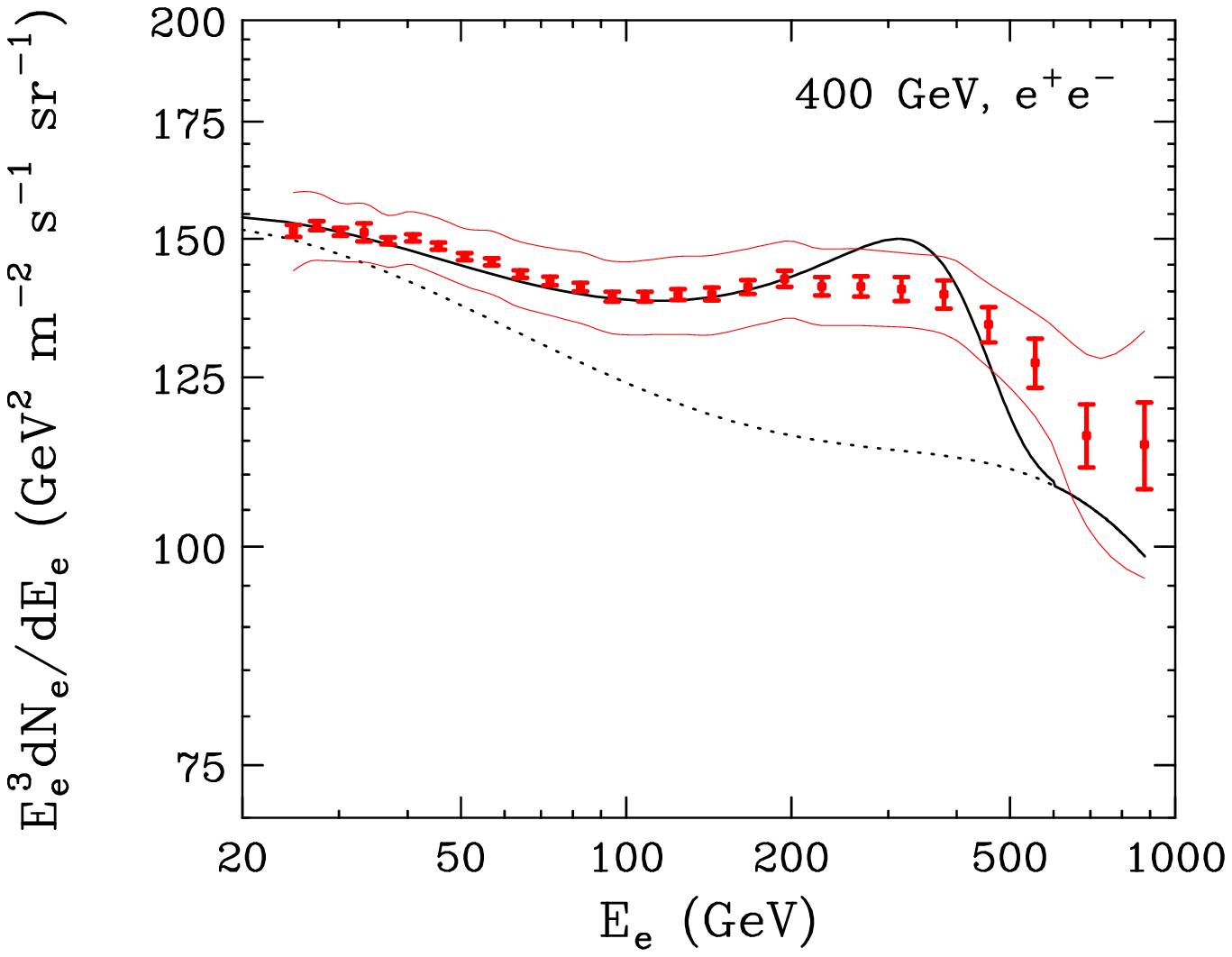}}
\hspace{0.2cm}
{\includegraphics[angle=0,width=0.45\linewidth]{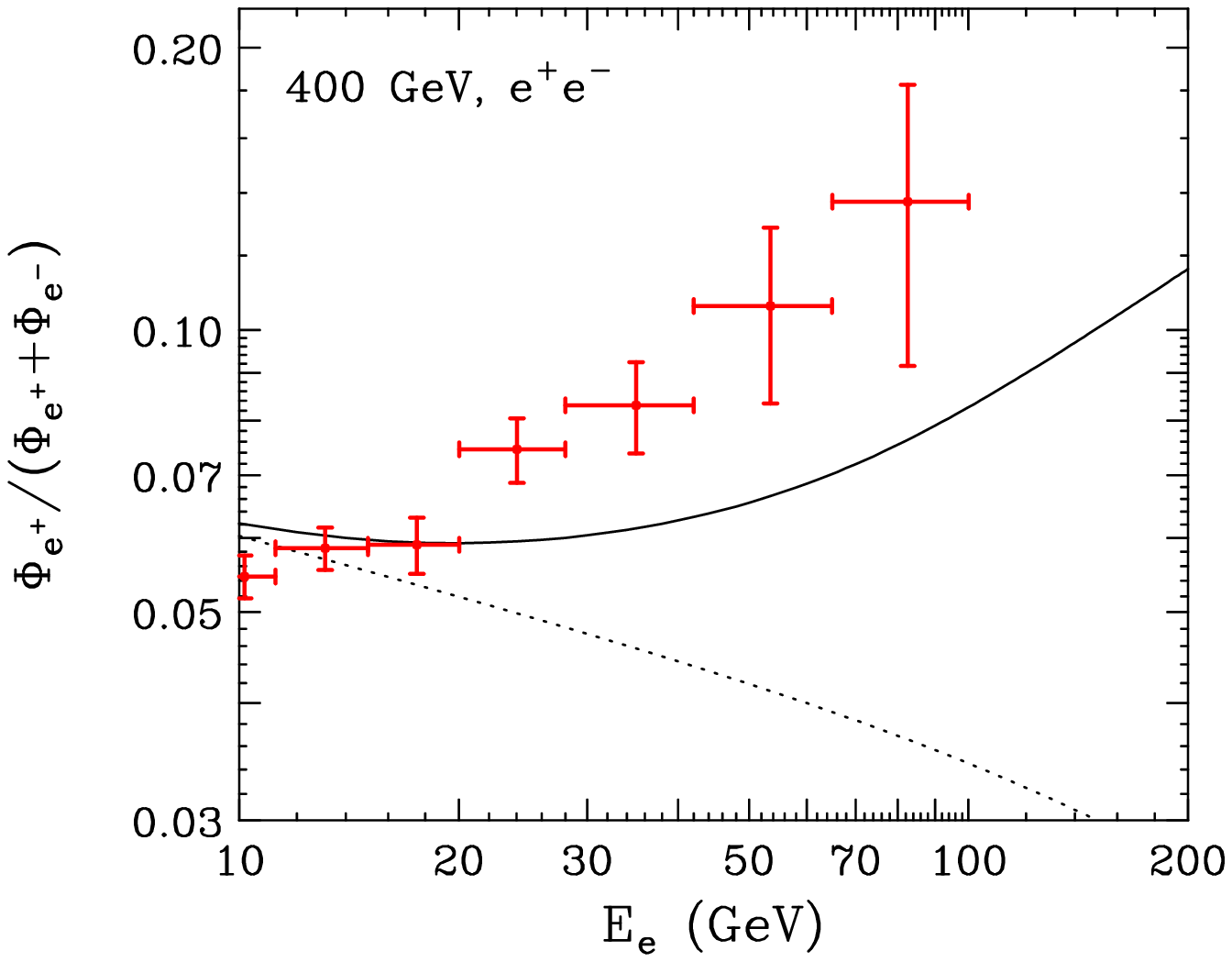}}\\
\caption{The best fits found to the PAMELA data above 10 GeV, and the (total) FGST data, for a WIMP annihilating to $e^+ e^-$ along with a power-law spectrum of cosmic ray electrons from astrophysical sources, and an additional spectrum of cosmic ray electrons from nearby, young cosmic ray accelerators. The dotted lines denote the astrophysical background used, without the contribution from dark matter. For each case shown, we found a very good fit to the data ($\chi^2 \sim 13-23$ distributed over 33 error bars). To normalize the dark matter annihilation rate, we have used a boost factor (relative to the rate predicted for $\sigma v = 3 \times 10^{-26}$ cm$^3$/s and $\rho_{0}=0.3$ GeV/cm$^2$) of 16.3, 28.1 and 35.4 from top-to-bottom, respectively. From top-to-bottom, we have used propagation models A, C, and A, and in each case shown have used $\phi_F=1000$ MeV. See text for more details.}
\label{astro}
\end{center}
\end{figure}

With the possibility in mind that a small number of local, young sources of this type might dominate the cosmic ray electron spectrum at the highest energies measured by FGST, we have allowed the astrophysical background to include an extra component (beyond the simple power-law used in Figs.~\ref{lt100} and~\ref{all}) of the form $dN_e/dE_e \propto E_e^{\alpha^{\prime}} \exp(-E_e/E_{\rm cut})$. In Fig.~\ref{astro}, we show the results including this contribution, again in the case of a WIMP annihilating to $e^+ e^-$. We find that good fits($\chi^2 \approx 13-23$ over 33 error bars) can be found for masses within the approximate range of 200 to 400 GeV. Recall (see Fig.~\ref{all}) that WIMPs with masses between approximately 400 and 600 GeV provide a good fit without an additional local contribution. Very heavy WIMPs ($m_{\rm dm}\gsim 600$~GeV) may also potentially provide a good fit to the combined data if the astrophysical background were to fall more rapidly at very high energies, corresponding to a downward fluctuation in the local injection rate of cosmic ray electrons.\footnote{Note that we are including here only stochastic effects from known sources of cosmic ray electrons, and not a new source or sources of positrons, such as has been postulated elsewhere to explain the PAMELA data~\cite{pulsar}.}

A word of caution is in order at this point. The $\chi^2$'s we have calculated here do not take into account any correlations within the FGST systematic errors. While this does not appear likely to change our conclusions significantly in the case of a moderately heavy WIMP ($\sim 400$~GeV), there is reason to expect that the fits found for lighter WIMPs ($\sim 200$~GeV) might overestimate the degree of agreement with the data. Without knowing the details of these correlations ({\it ie.} the corresponding covariance matrix), we cannot quantitatively assess this issue. That being said, it is possible that the results shown in the top frame of Fig.~\ref{astro} do not constitute a good fit the FGST spectrum.

\begin{figure}[!]
\begin{center}
{\includegraphics[angle=0,width=0.45\linewidth]{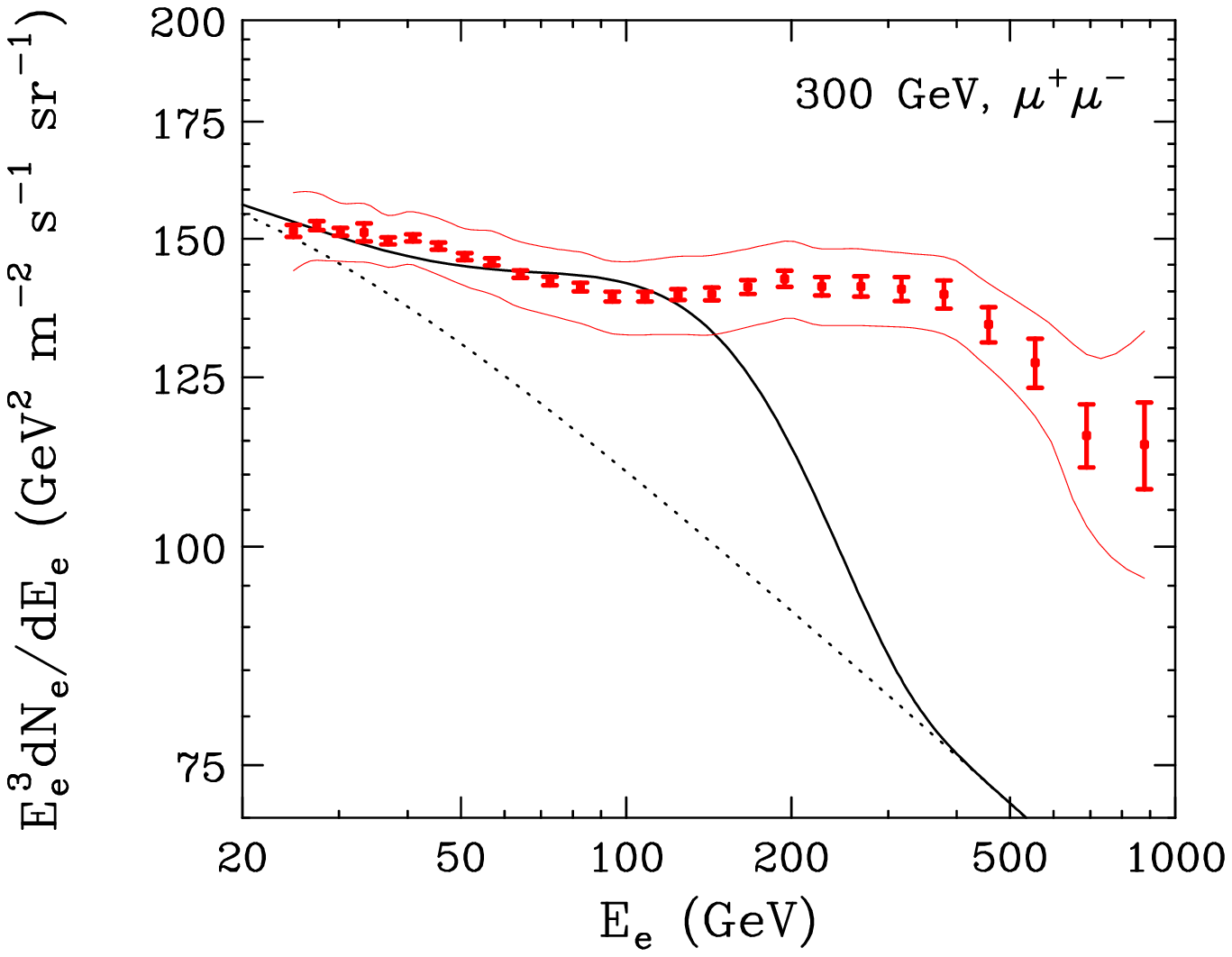}}
\hspace{0.2cm}
{\includegraphics[angle=0,width=0.45\linewidth]{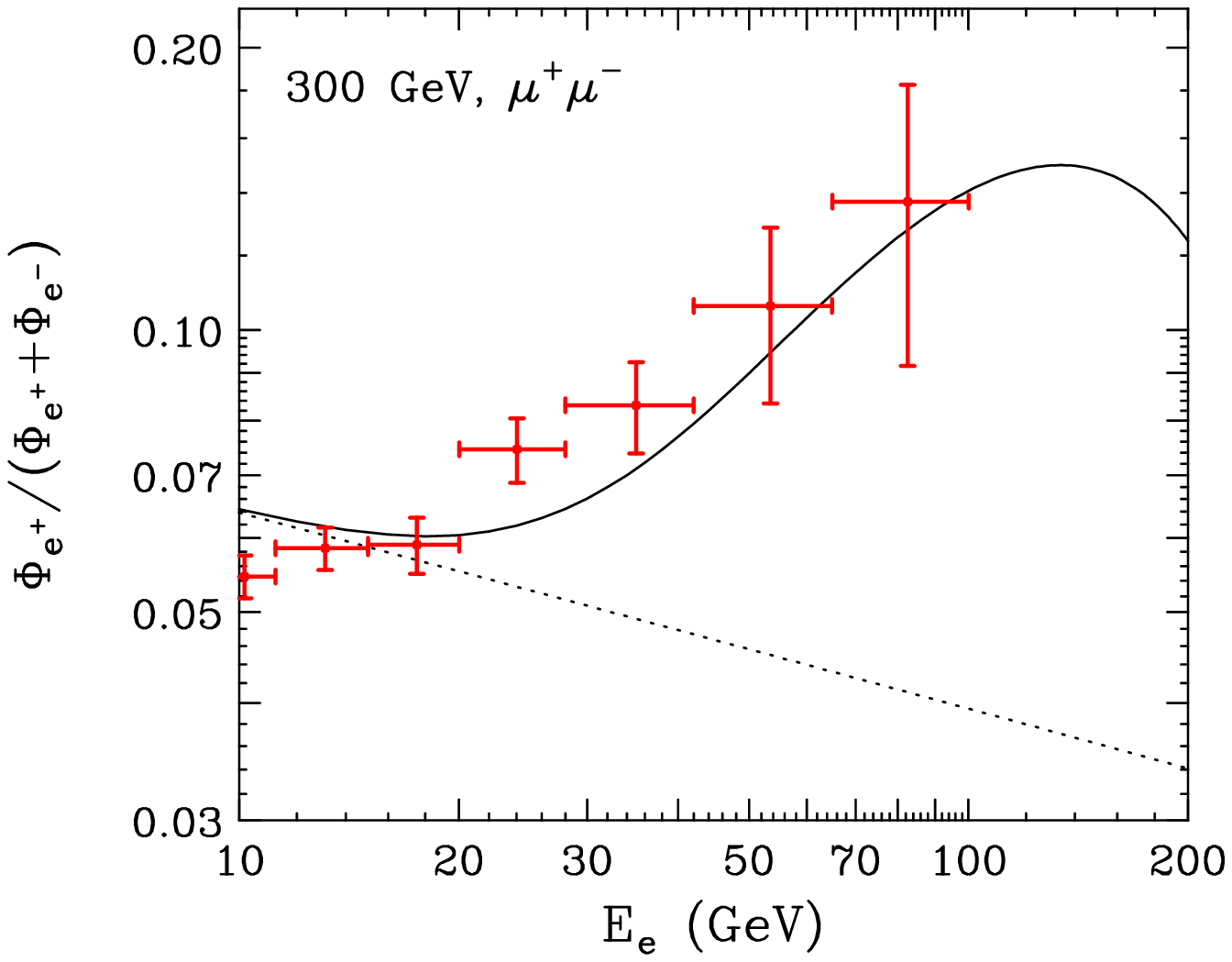}}\\
{\includegraphics[angle=0,width=0.45\linewidth]{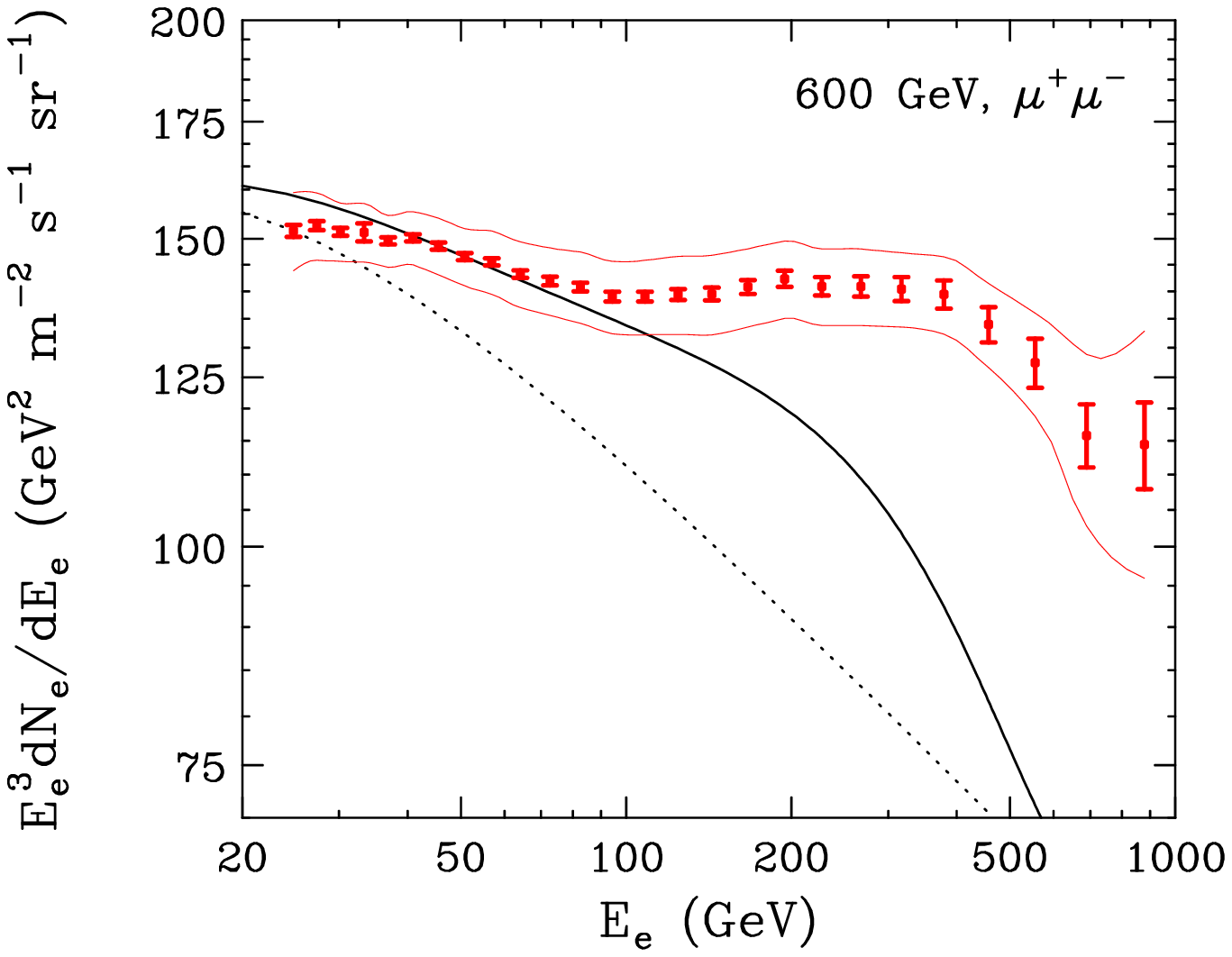}}
\hspace{0.2cm}
{\includegraphics[angle=0,width=0.45\linewidth]{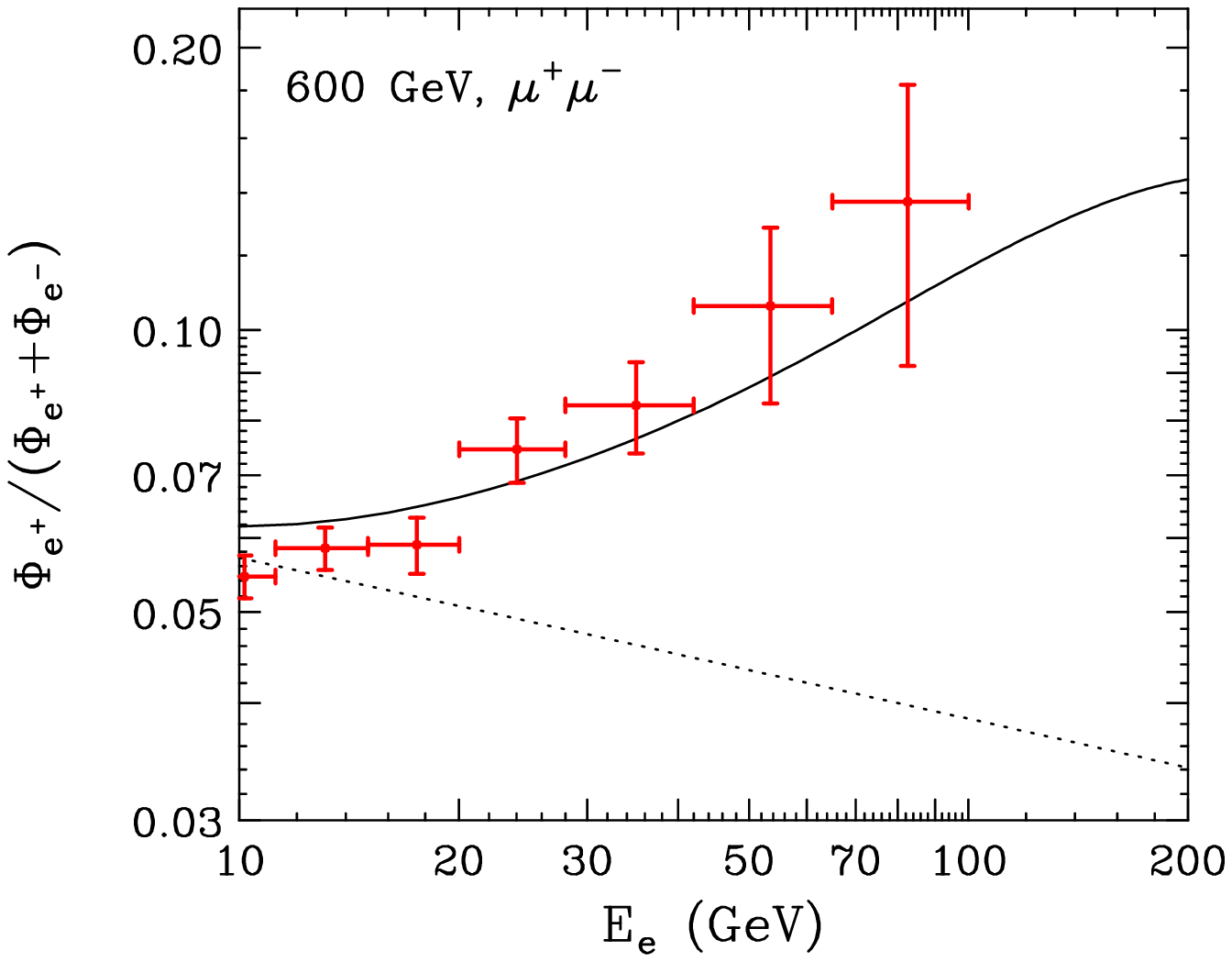}}\\
{\includegraphics[angle=0,width=0.45\linewidth]{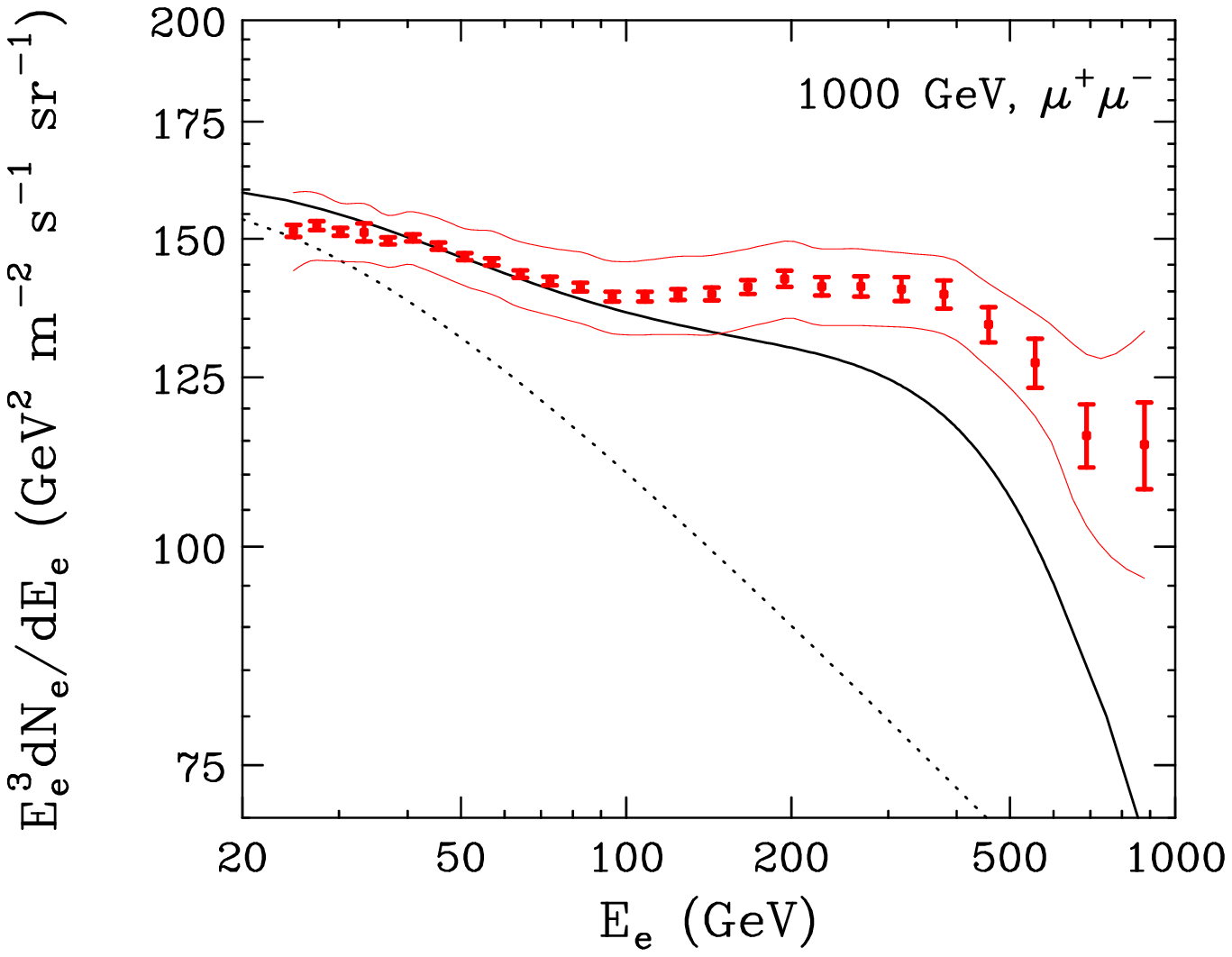}}
\hspace{0.2cm}
{\includegraphics[angle=0,width=0.45\linewidth]{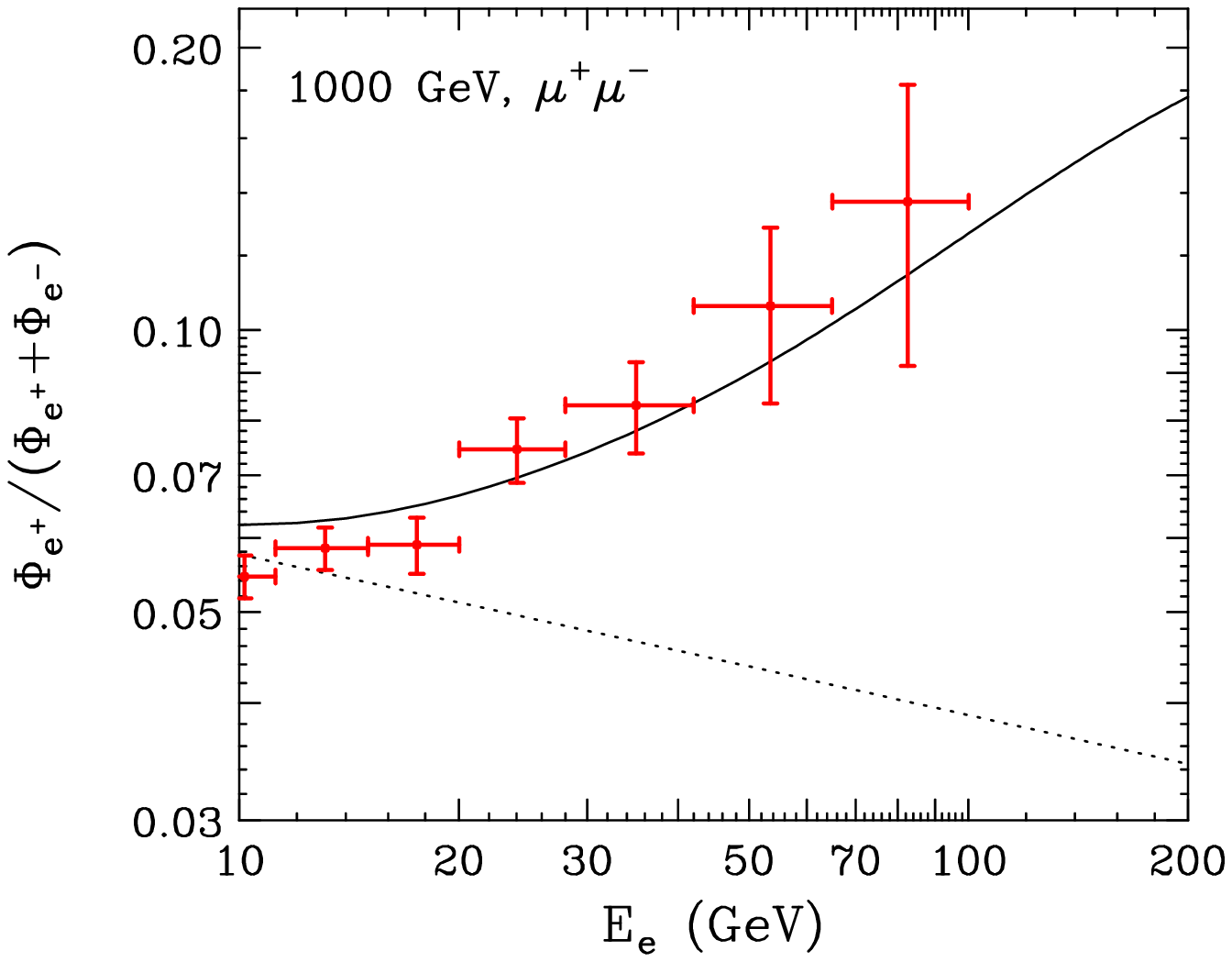}}\\
\caption{The best fits found to the PAMELA data above 10 GeV, and the FGST data {\it below 100 GeV}, for a WIMP annihilating to $\mu^+ \mu^-$ along with a power-law spectrum of cosmic ray electrons from astrophysical sources. The dotted lines denote the astrophysical background used, without the contribution from dark matter. In each case, we found a very good fit to the data ($\chi^2 \sim 7-10$ distributed over 20 error bars). To normalize the dark matter annihilation rate, we have used a boost factor (relative to the rate predicted for $\sigma v = 3 \times 10^{-26}$ cm$^3$/s and $\rho_{0}=0.3$ GeV/cm$^2$) of 119, 184 and 496 from top-to-bottom, respectively.  From top-to-bottom, we have used propagation models B, A, and A, and $\phi_F=650, 1000$, and $1000$ MeV. See text for more details.}
\label{muonslt100}
\end{center}
\end{figure}

\begin{figure}[!]
\begin{center}
{\includegraphics[angle=0,width=0.45\linewidth]{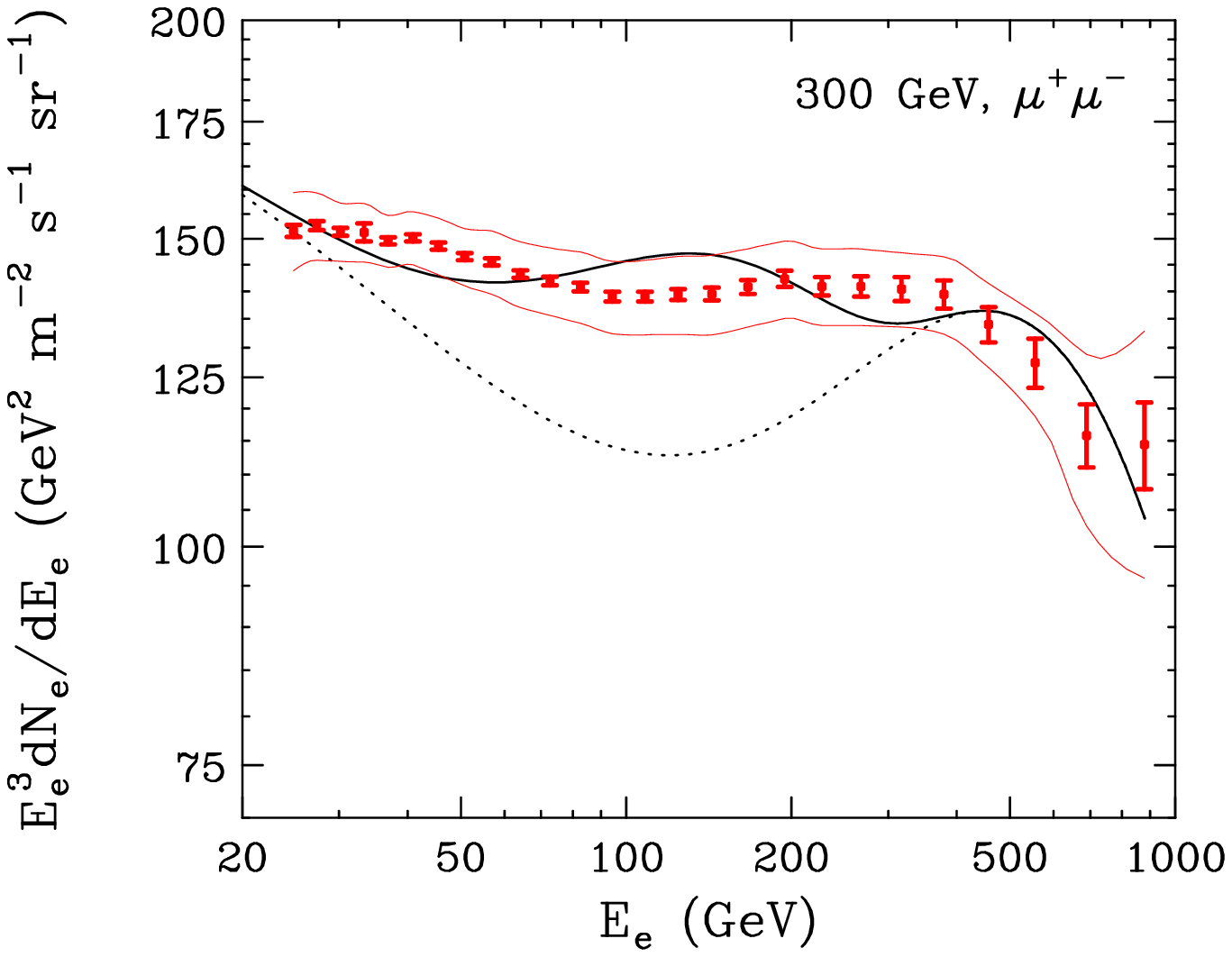}}
\hspace{0.2cm}
{\includegraphics[angle=0,width=0.45\linewidth]{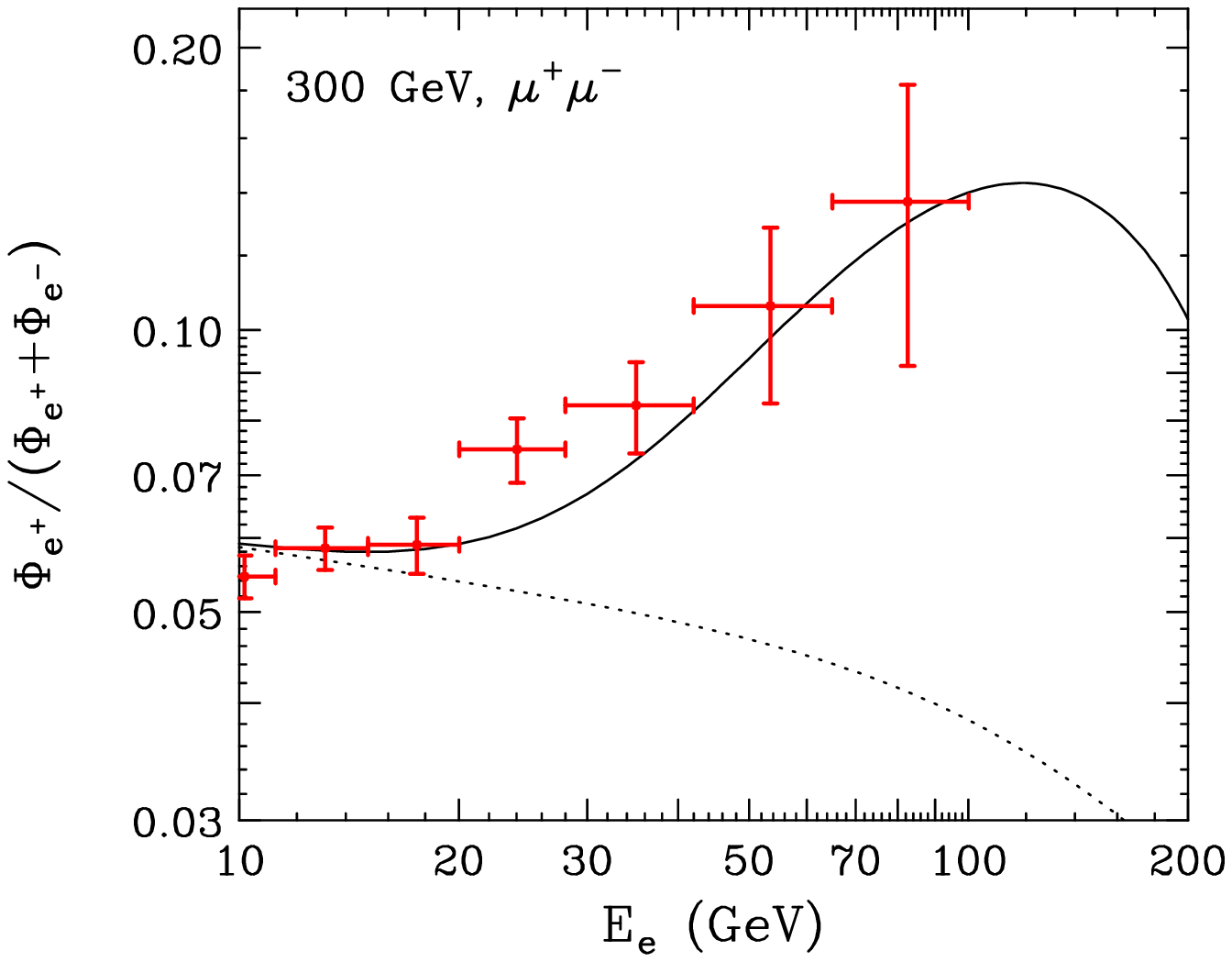}}\\
{\includegraphics[angle=0,width=0.45\linewidth]{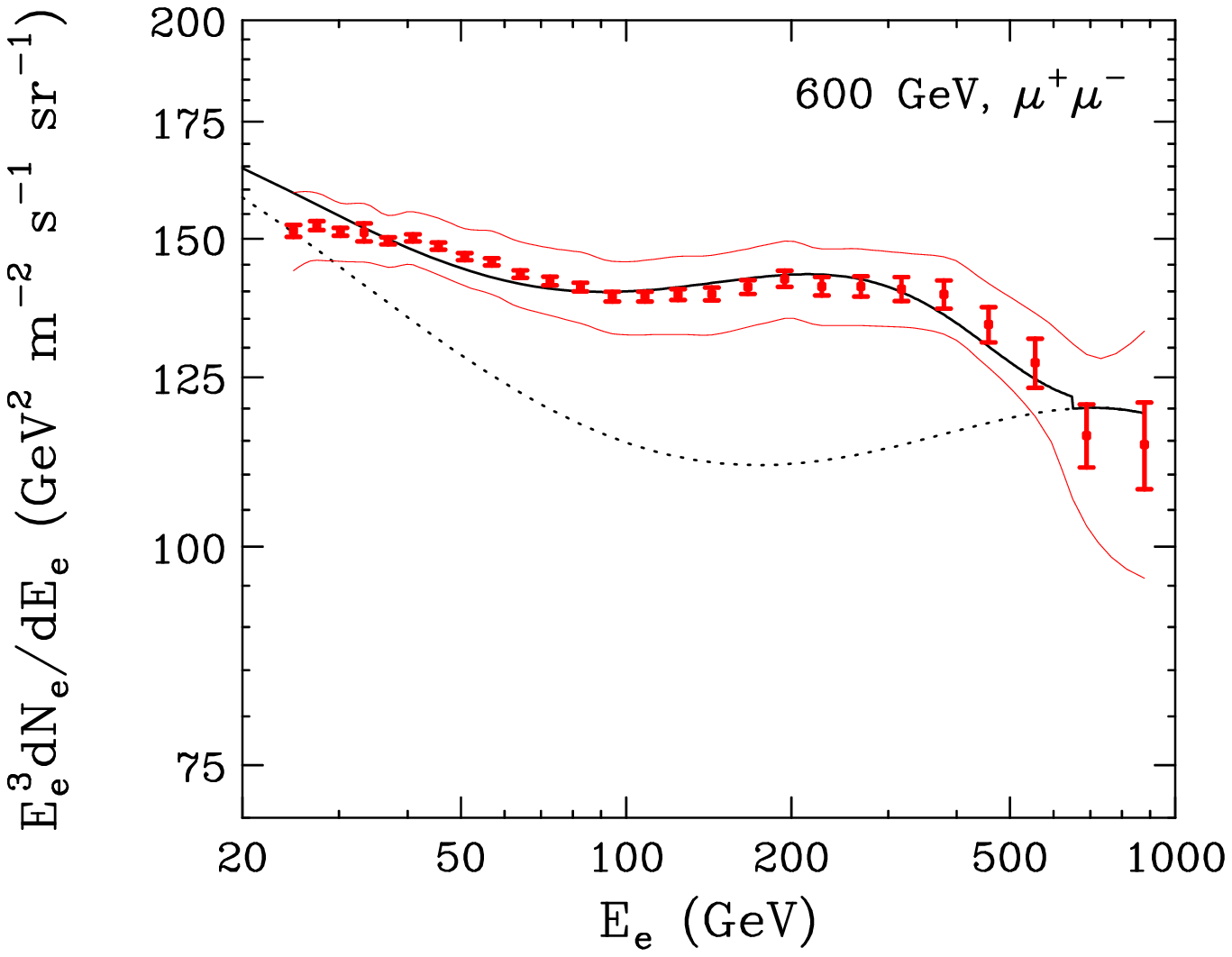}}
\hspace{0.2cm}
{\includegraphics[angle=0,width=0.45\linewidth]{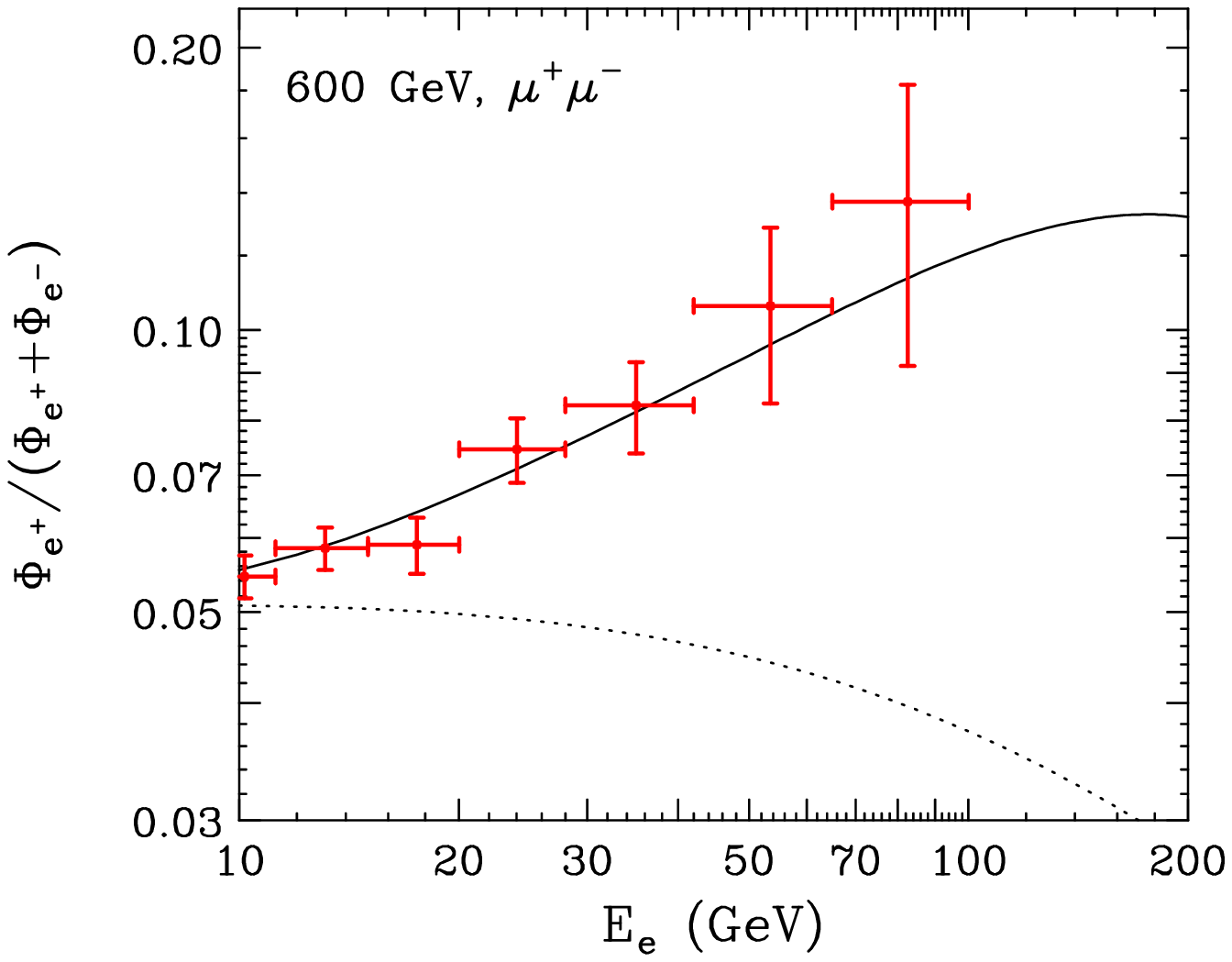}}\\
{\includegraphics[angle=0,width=0.45\linewidth]{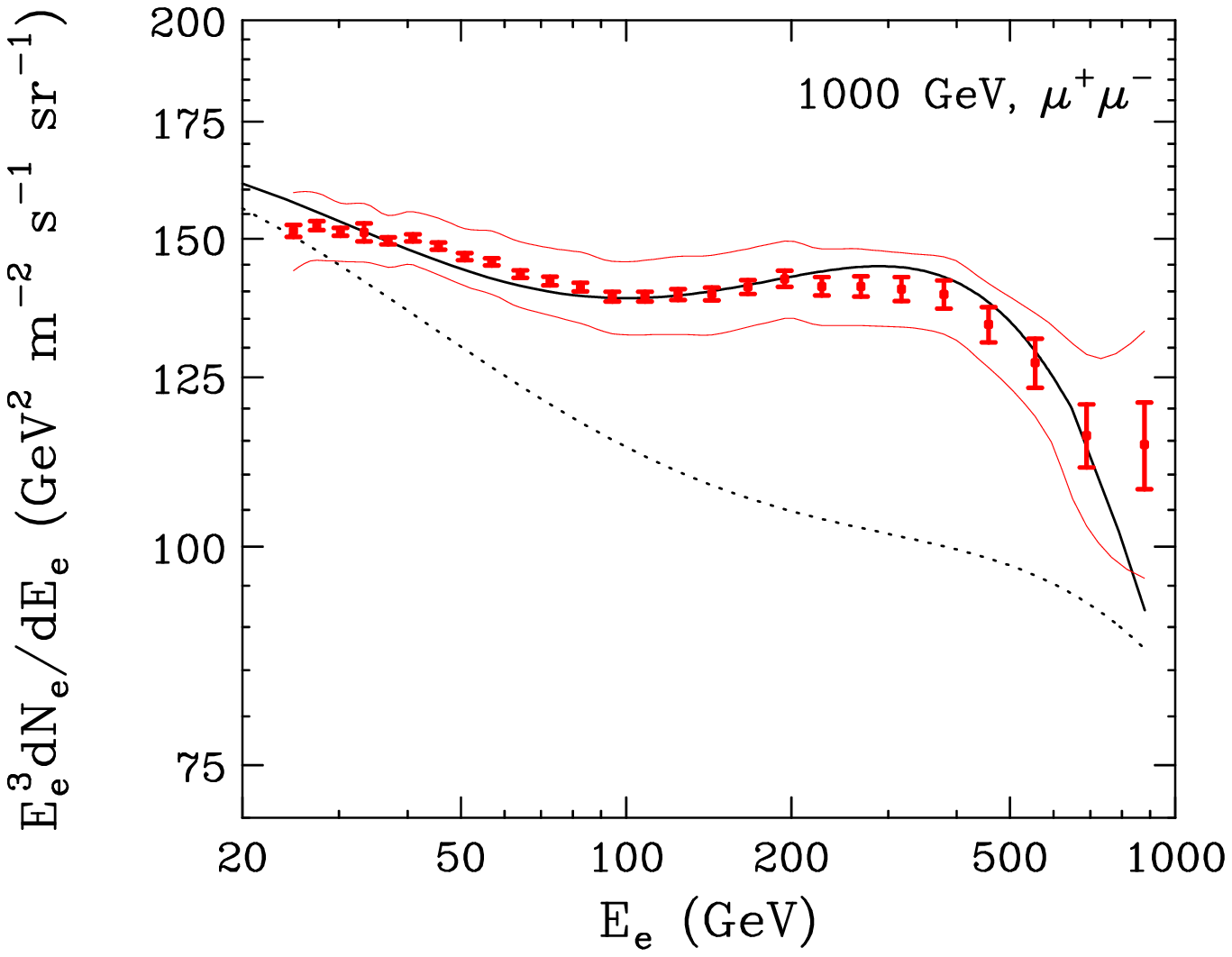}}
\hspace{0.2cm}
{\includegraphics[angle=0,width=0.45\linewidth]{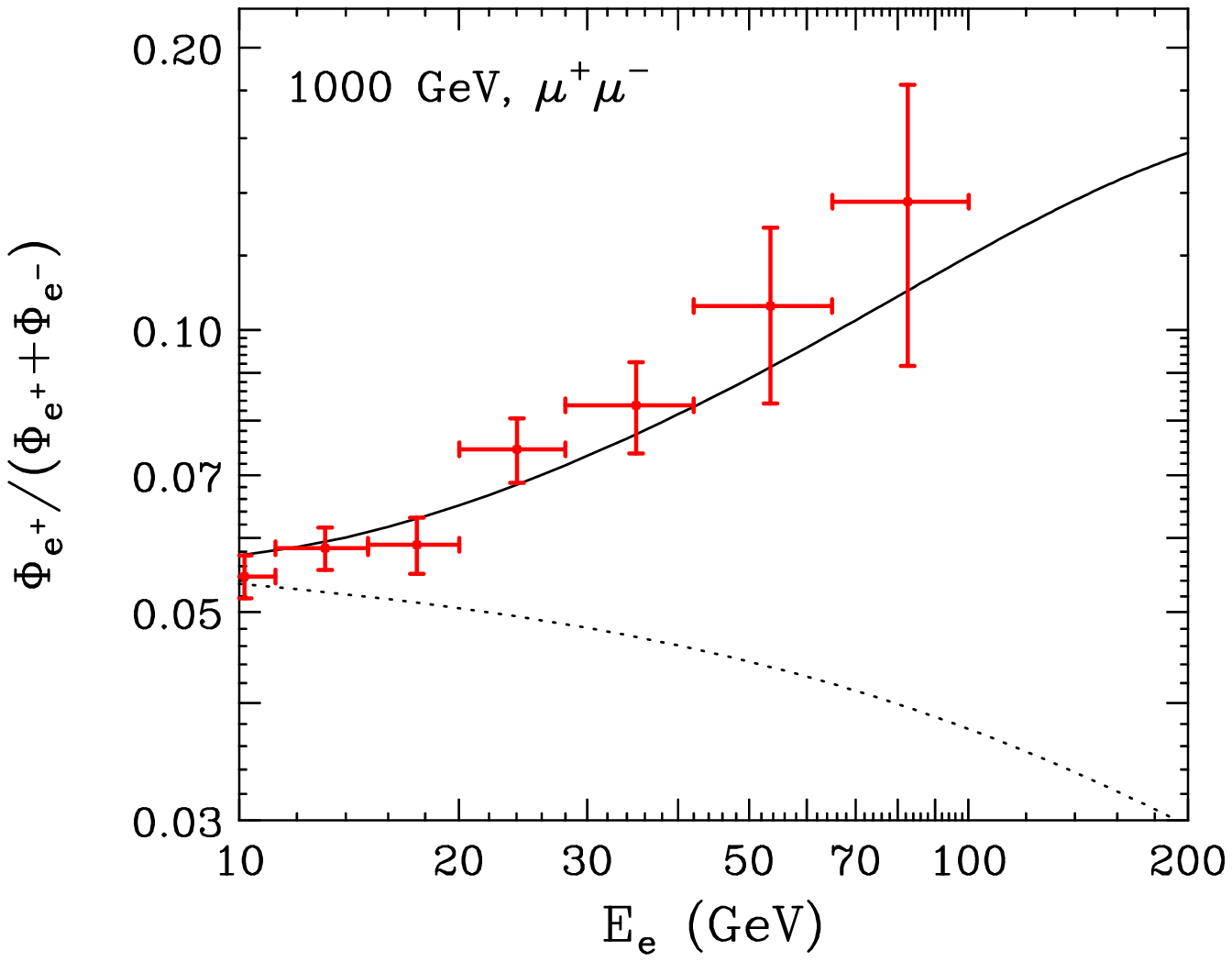}}\\
\caption{The best fits found to the PAMELA data above 10 GeV, and the (total) FGST data, for a WIMP annihilating to $\mu^+ \mu^-$ along with a power-law spectrum of cosmic ray electrons from astrophysical sources, and an additional spectrum of cosmic ray electrons from nearby, young cosmic ray accelerators. The dotted lines denote the astrophysical background used, without the contribution from dark matter. For each case shown, we found a very good fit to the data ($\chi^2 \sim 5-16$ distributed over 33 error bars). To normalize the dark matter annihilation rate, we have used a boost factor (relative to the rate predicted for $\sigma v = 3 \times 10^{-26}$ cm$^3$/s and $\rho_{0}=0.3$ GeV/cm$^2$) of 123, 205 and 472 from top-to-bottom, respectively.  From top-to-bottom, we have used propagation models B, A, and A, and $\phi_F=850, 1000$, and $1000$ MeV. See text for more details.}
\label{muonsastro}
\end{center}
\end{figure}

In Figs.~\ref{muonslt100} and \ref{muonsastro}, we show the results for WIMPs annihilating to $\mu^+ \mu^-$. In this case, we omit the set of figures fitting to the whole range of FGST data with no nearby sources of cosmic-ray electrons, as this always results in a poor fit to the combined PAMELA and FGST data. From Fig.~\ref{muonslt100} we see that the rising positron fraction can be accommodated, but additional astrophysical contributions to the cosmic ray electron spectrum at high energies are required unless the WIMP is as heavy as a TeV or more. In Fig.~\ref{muonsastro}, we include an additional local astrophysical contribution of the form described earlier and in each case find excellent fits to the FGST and PAMELA data ($\chi^2 \approx 5-16$ over 33 error bars). Indeed, the presence of the additional astrophysical flux naturally fits the hard FGST spectrum in the several hundred GeV range, without introducing any spurious features in the spectrum.

\begin{figure}[!]
\begin{center}
{\includegraphics[angle=0,width=0.45\linewidth]{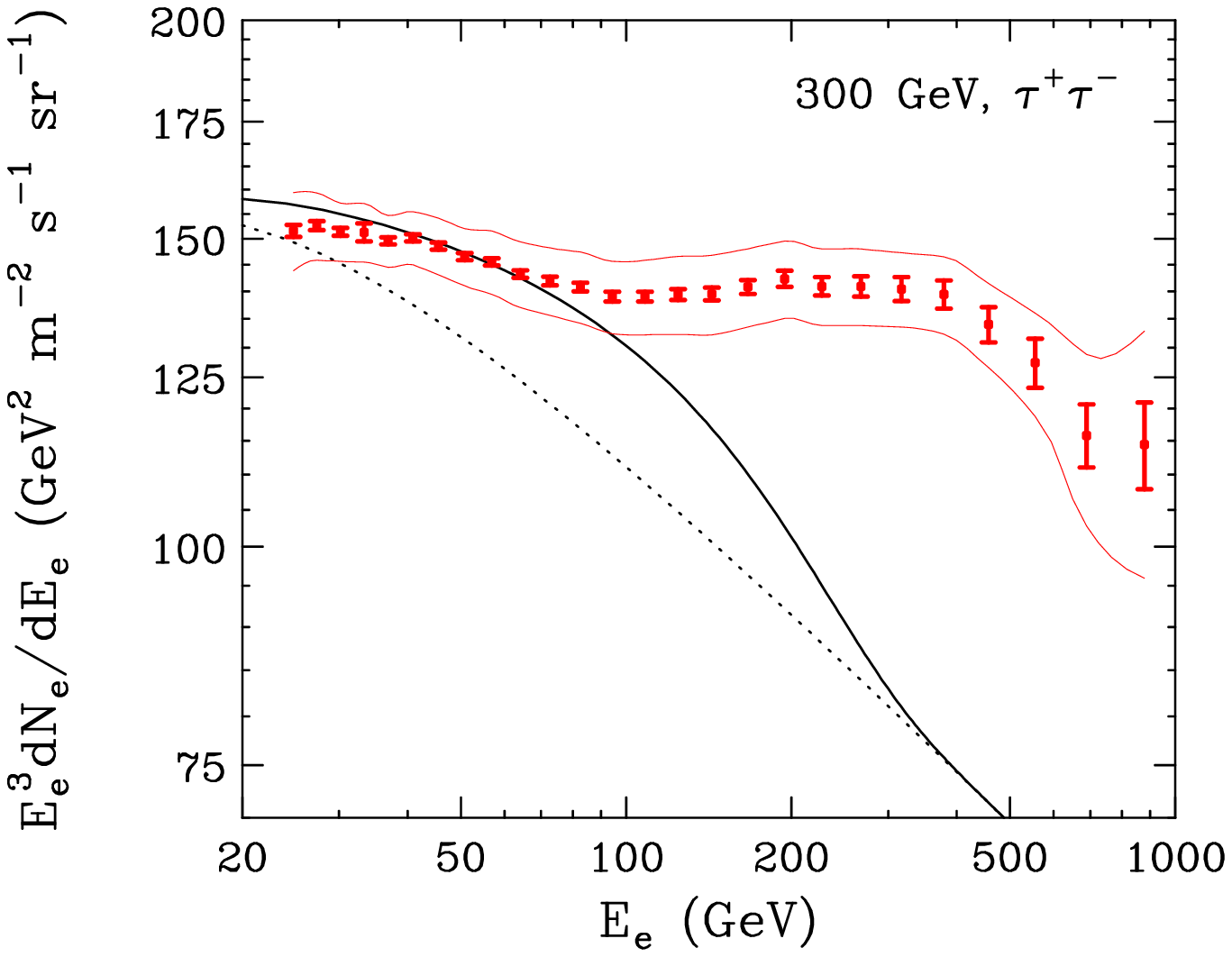}}
\hspace{0.2cm}
{\includegraphics[angle=0,width=0.45\linewidth]{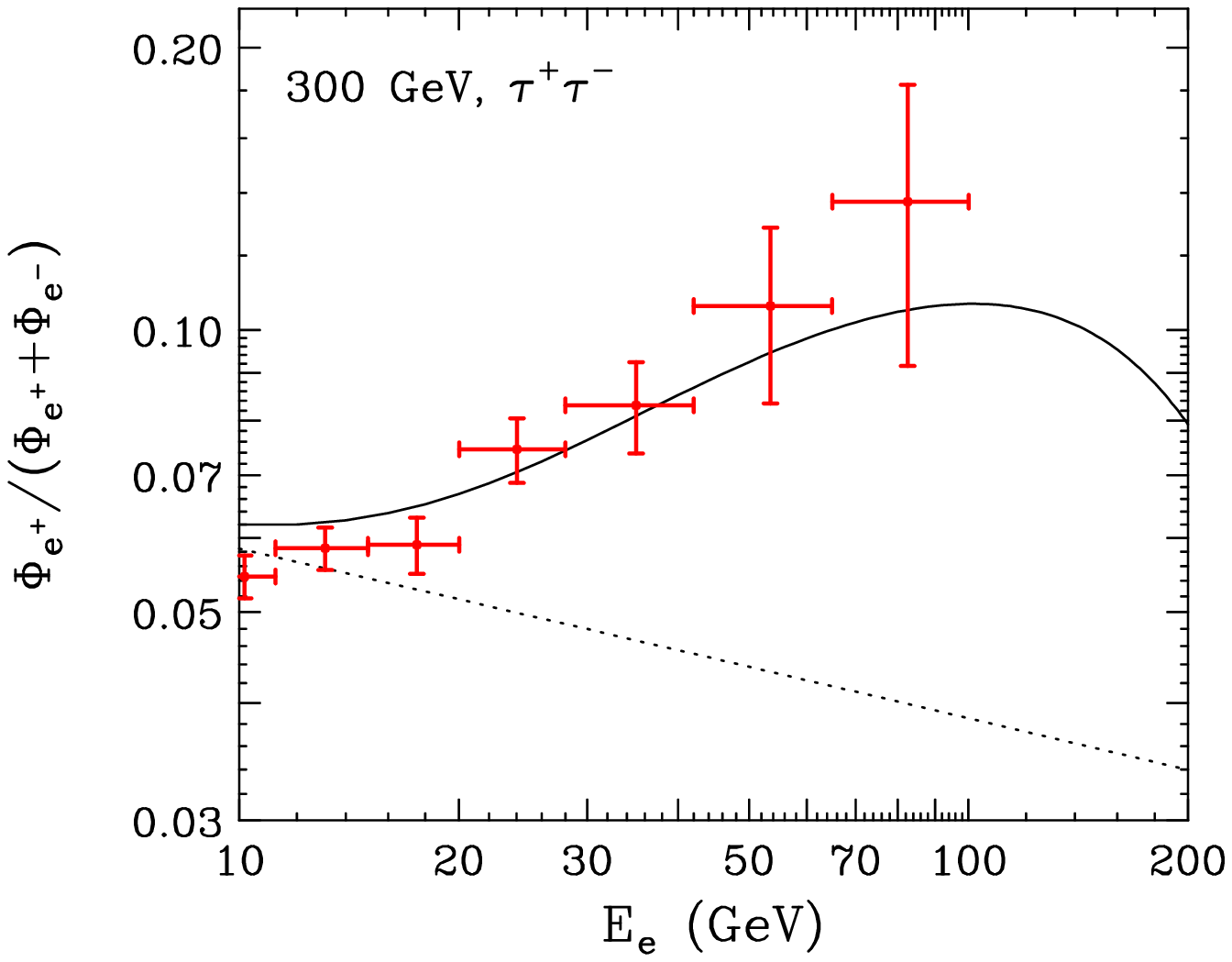}}\\
{\includegraphics[angle=0,width=0.45\linewidth]{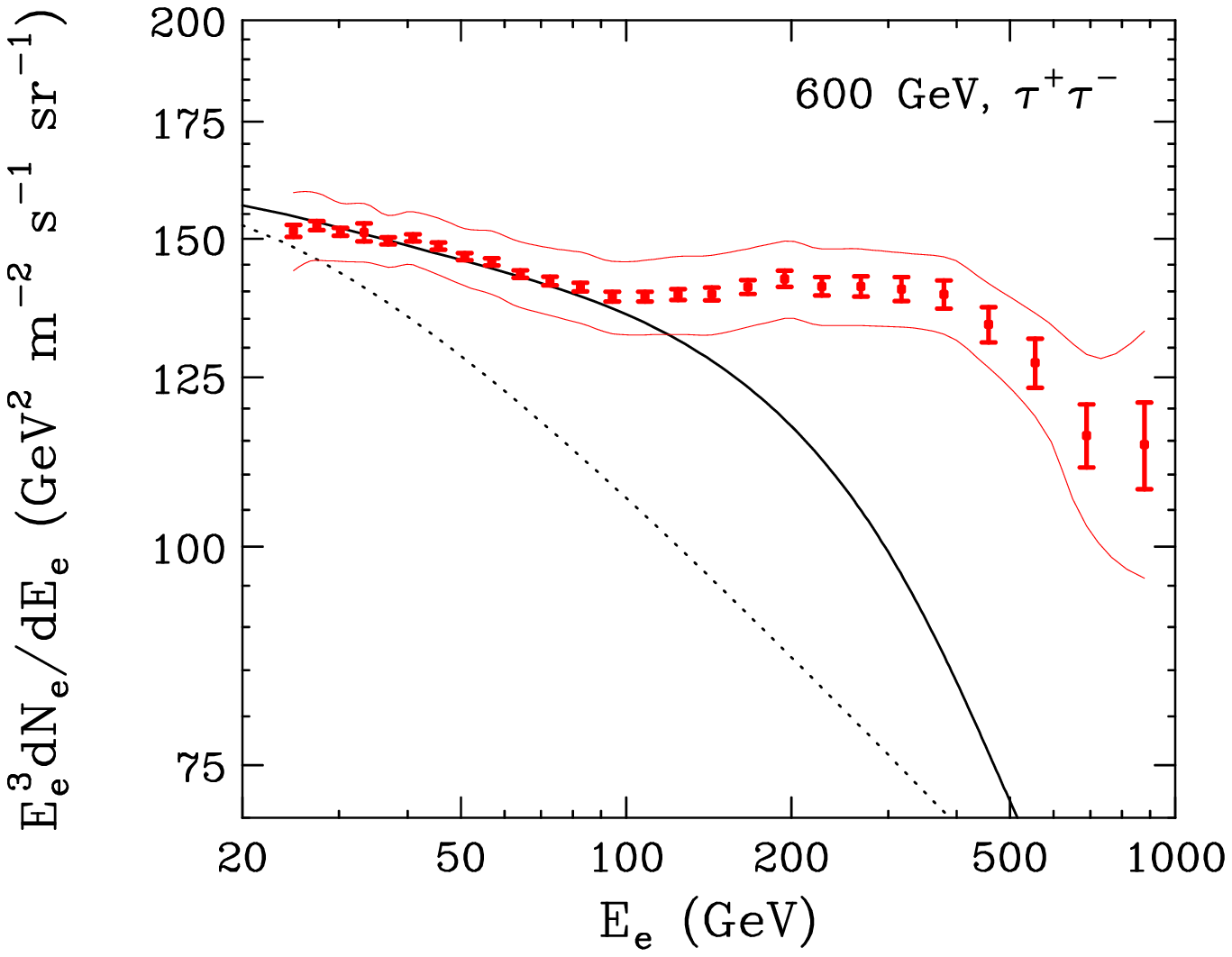}}
\hspace{0.2cm}
{\includegraphics[angle=0,width=0.45\linewidth]{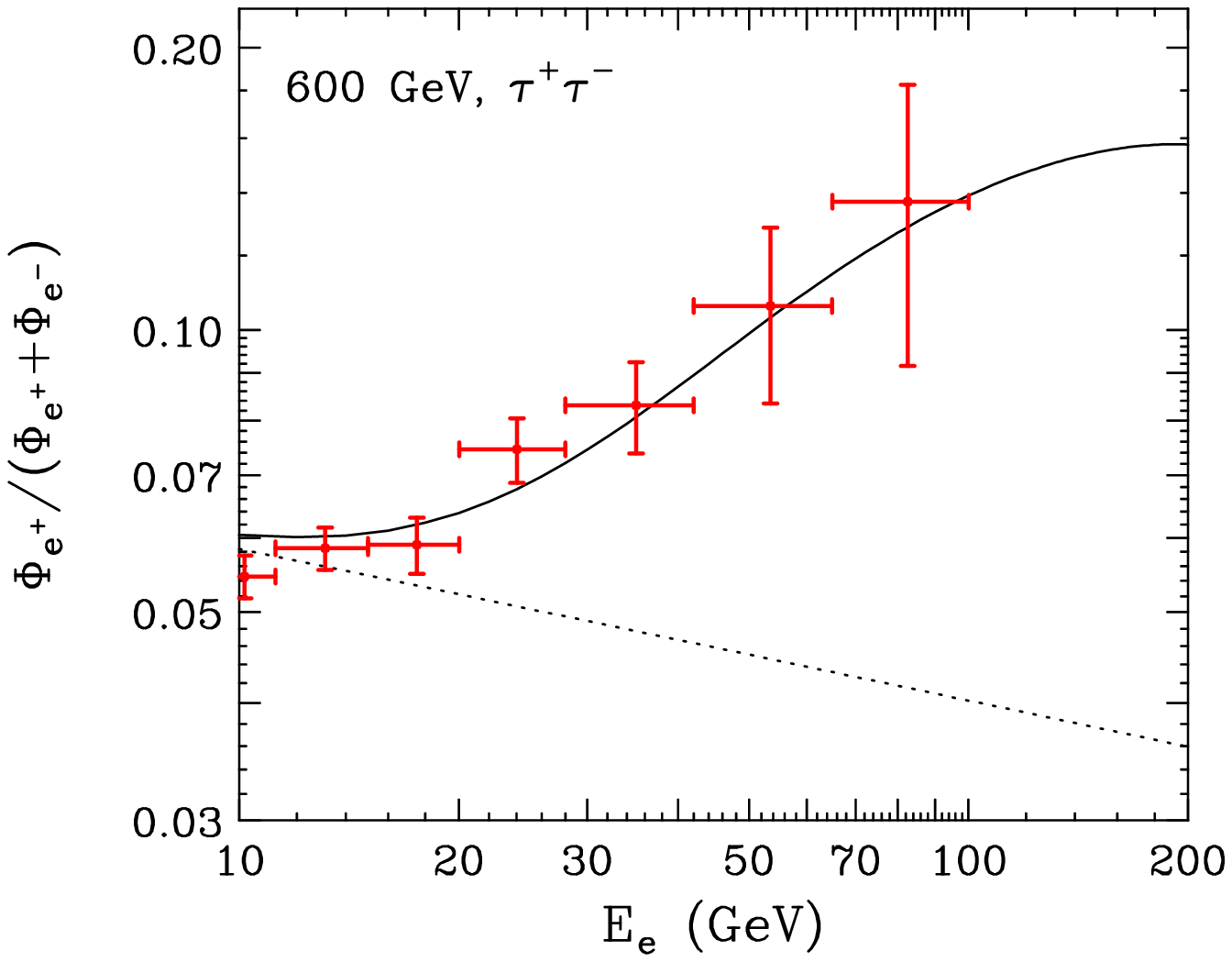}}\\
{\includegraphics[angle=0,width=0.45\linewidth]{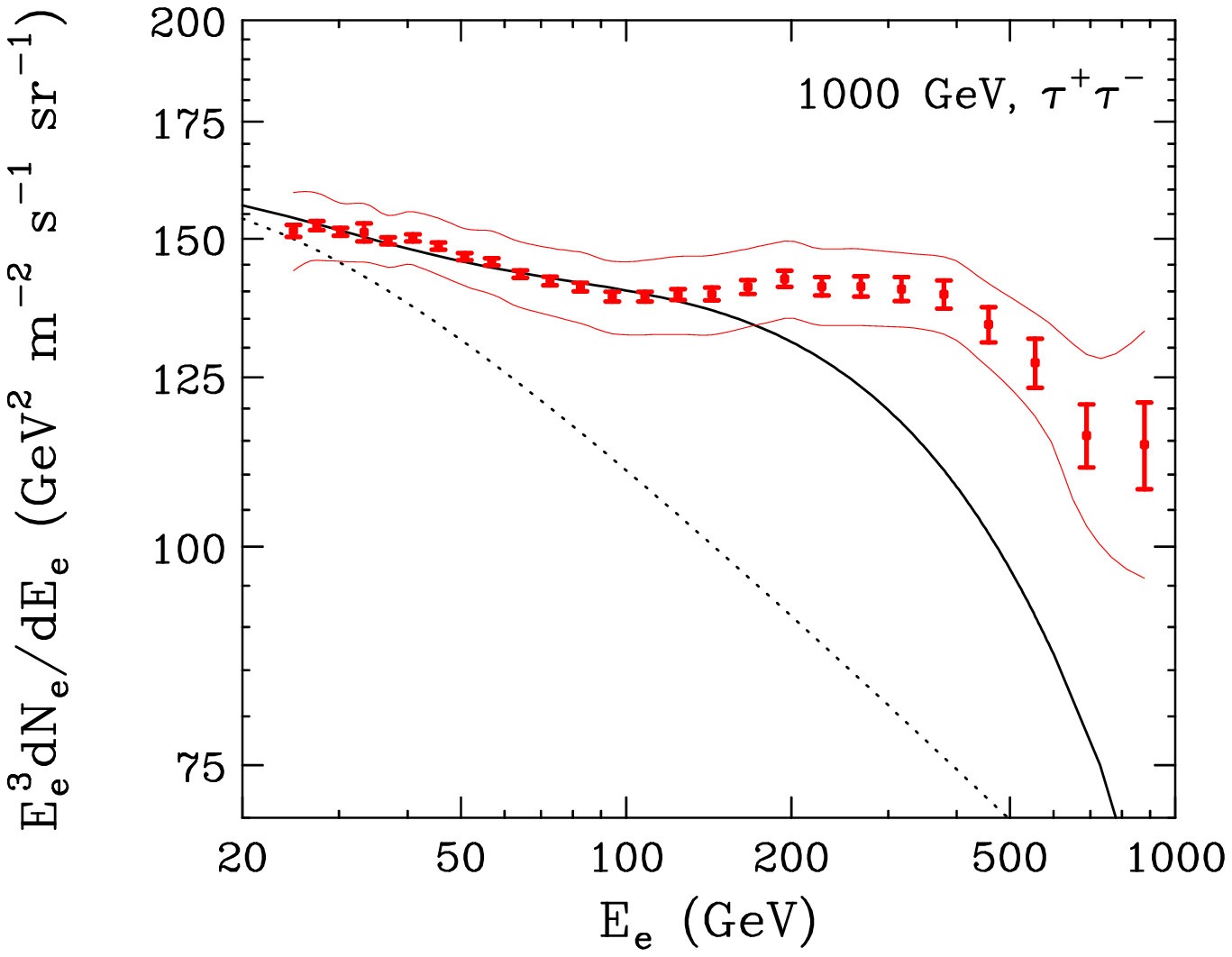}}
\hspace{0.2cm}
{\includegraphics[angle=0,width=0.45\linewidth]{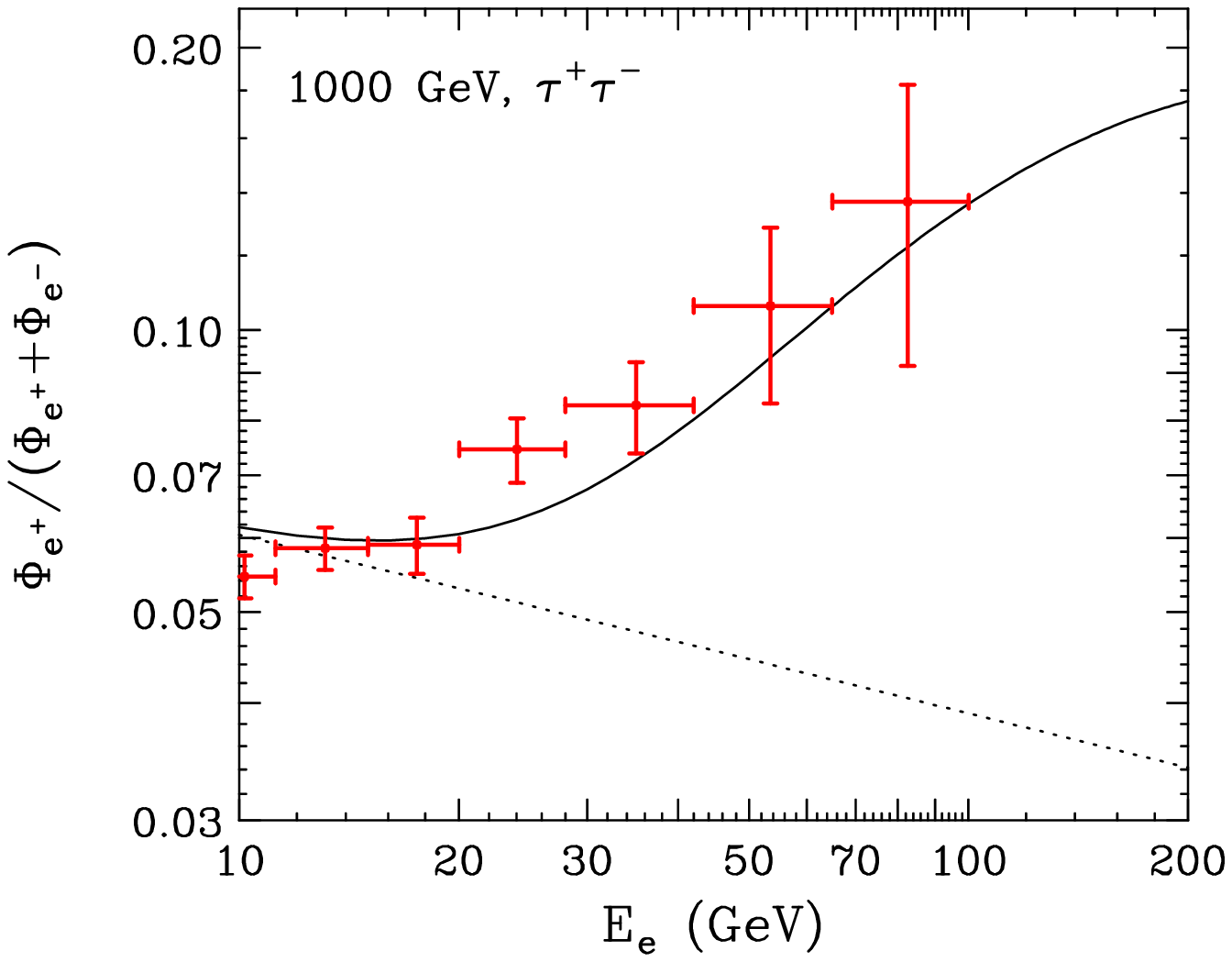}}\\
\caption{The best fits found to the PAMELA data above 10 GeV, and the FGST data {\it below 100 GeV}, for a WIMP annihilating to $\tau^+ \tau^-$ along with a power-law spectrum of cosmic ray electrons from astrophysical sources. The dotted lines denote the astrophysical background used, without the contribution from dark matter. In each case, we found a very good fit to the data ($\chi^2 \sim 3-8$ distributed over 20 error bars). To normalize the dark matter annihilation rate, we have used a boost factor (relative to the rate predicted for $\sigma v = 3 \times 10^{-26}$ cm$^3$/s and $\rho_{0}=0.3$ GeV/cm$^2$) of 261, 745 and 1407 from top-to-bottom, respectively. In each case shown, we have used propagation model B, and from top-to-bottom, have used $\phi_F=1000, 950$, and $800$ MeV. See text for more details.}
\label{tauslt100}
\end{center}
\end{figure}

\begin{figure}[!]
\begin{center}
{\includegraphics[angle=0,width=0.45\linewidth]{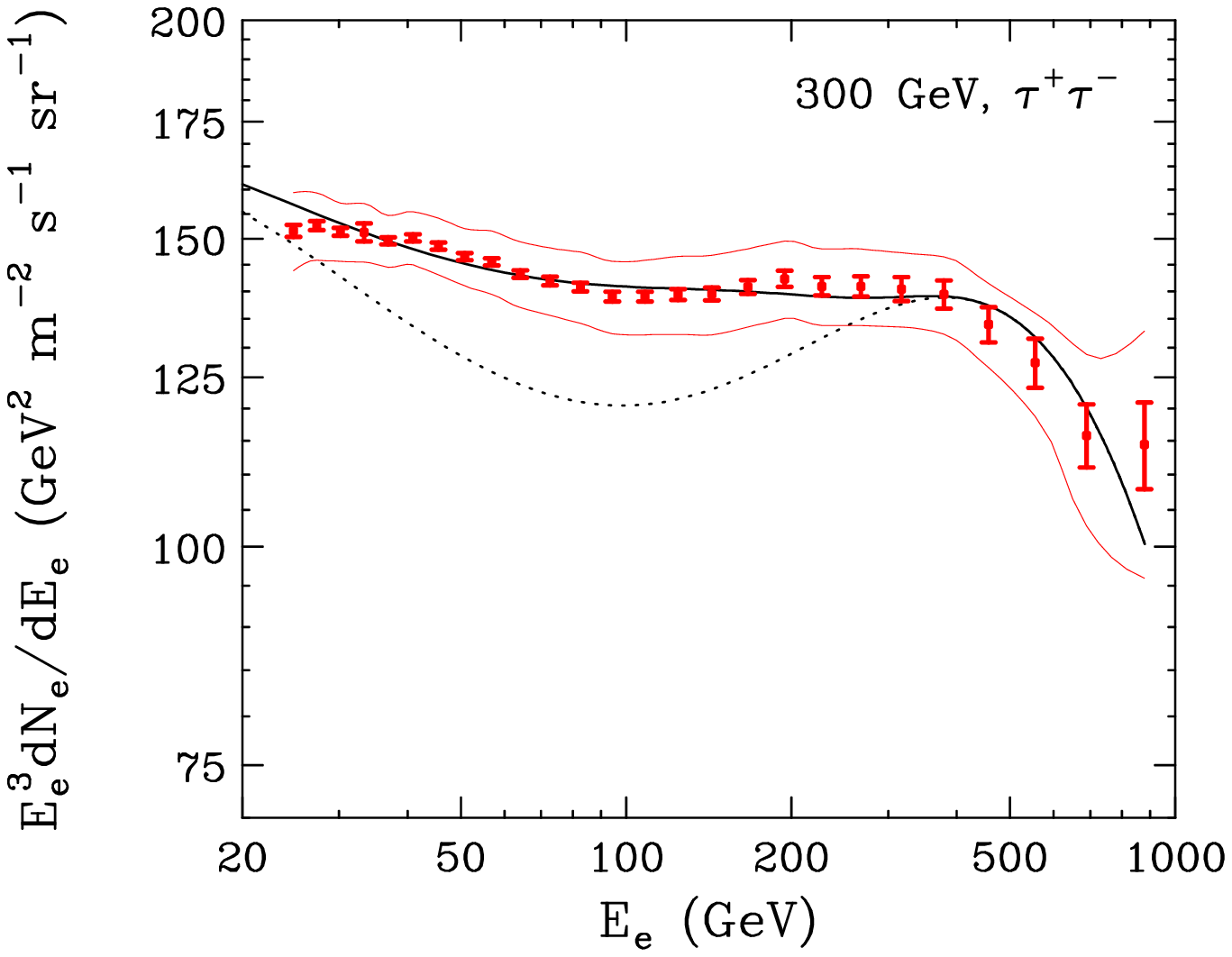}}
\hspace{0.2cm}
{\includegraphics[angle=0,width=0.45\linewidth]{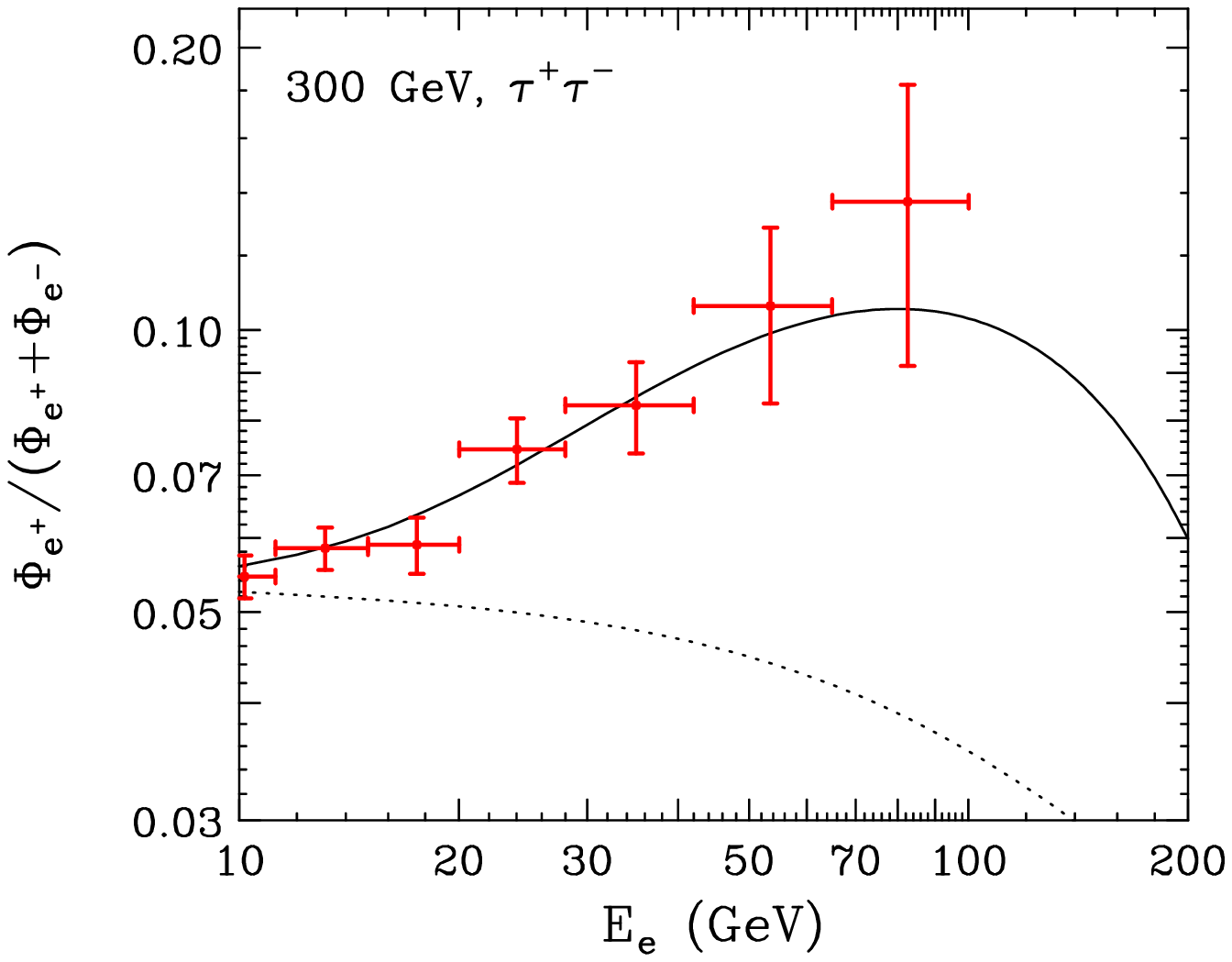}}\\
{\includegraphics[angle=0,width=0.45\linewidth]{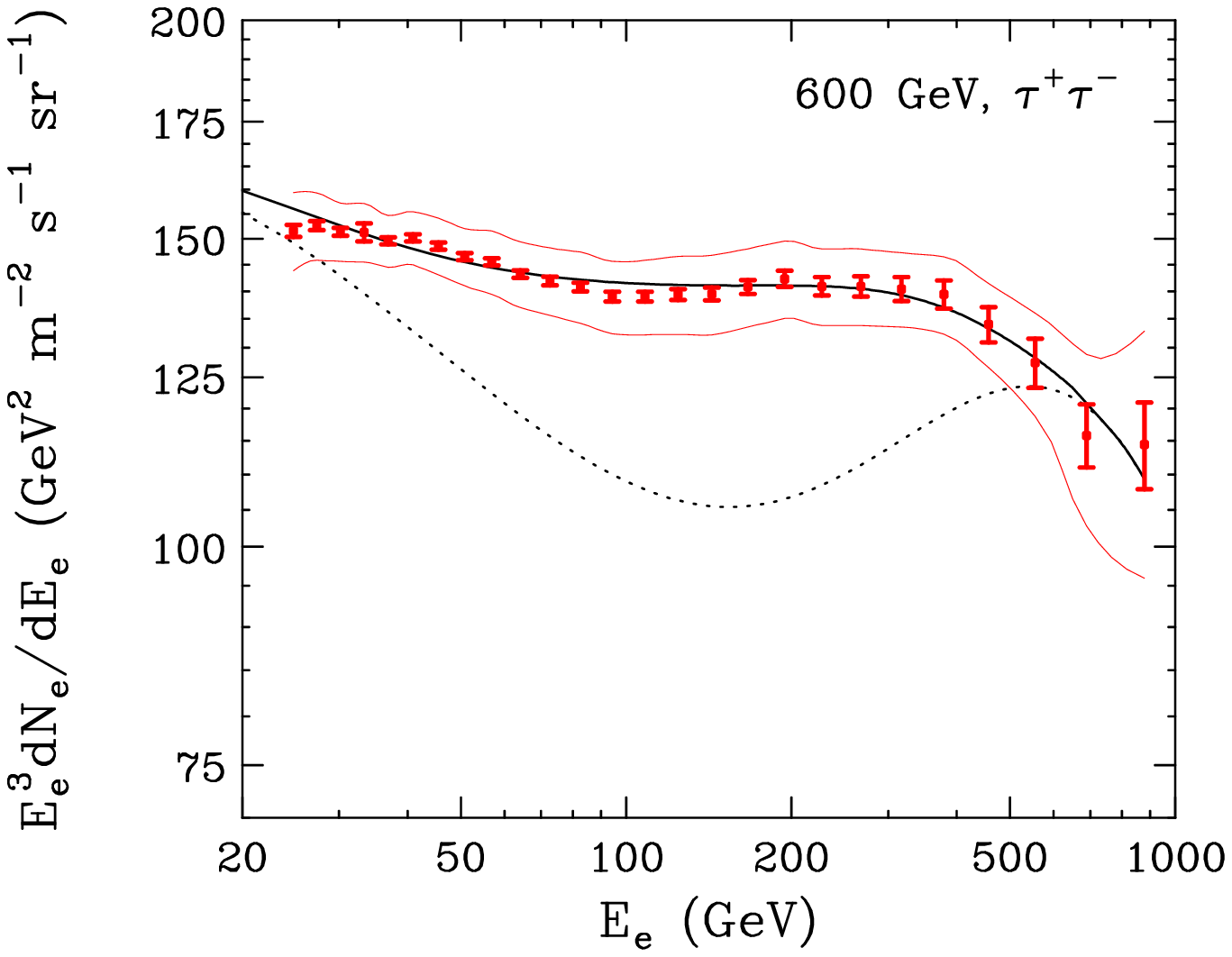}}
\hspace{0.2cm}
{\includegraphics[angle=0,width=0.45\linewidth]{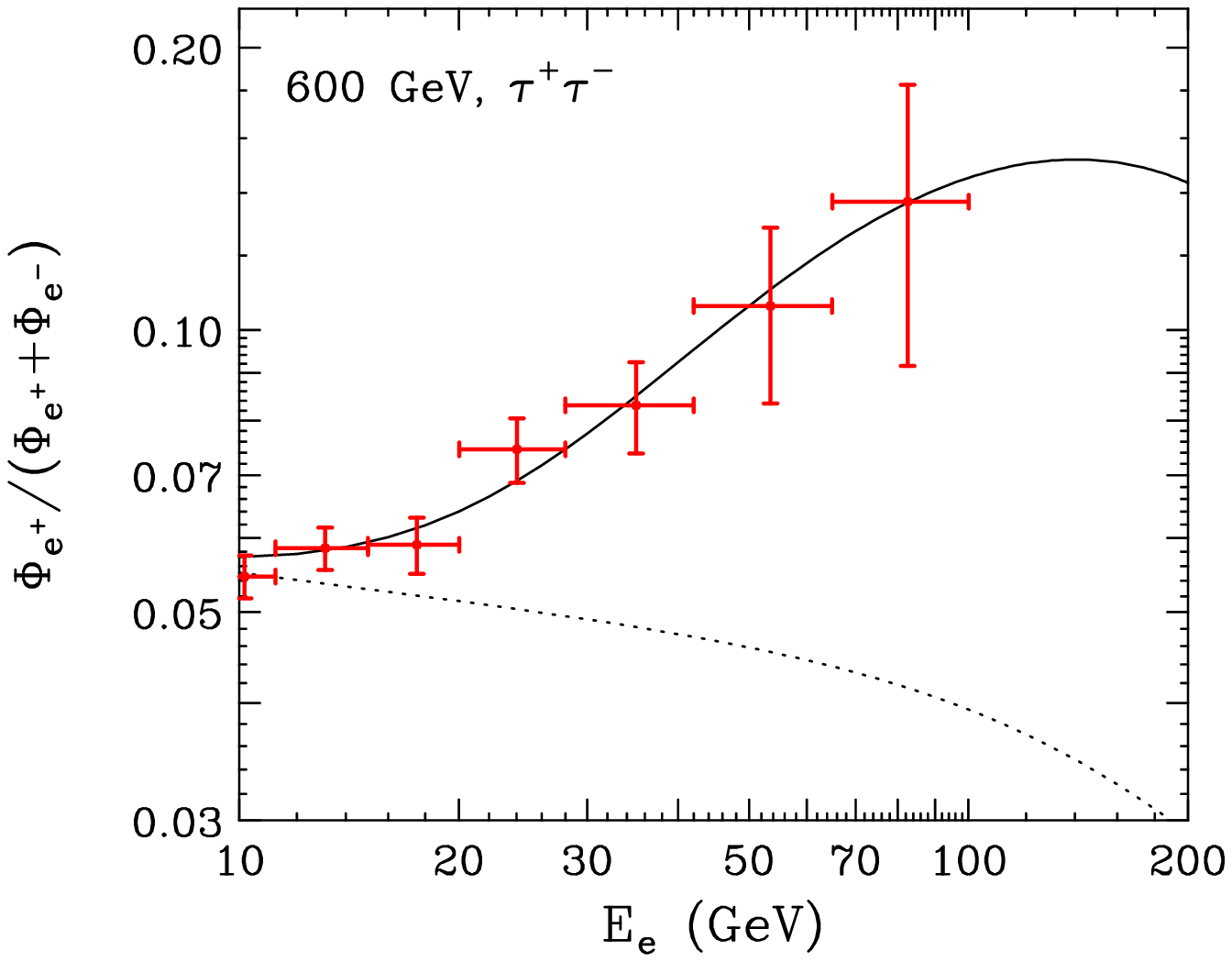}}\\
{\includegraphics[angle=0,width=0.45\linewidth]{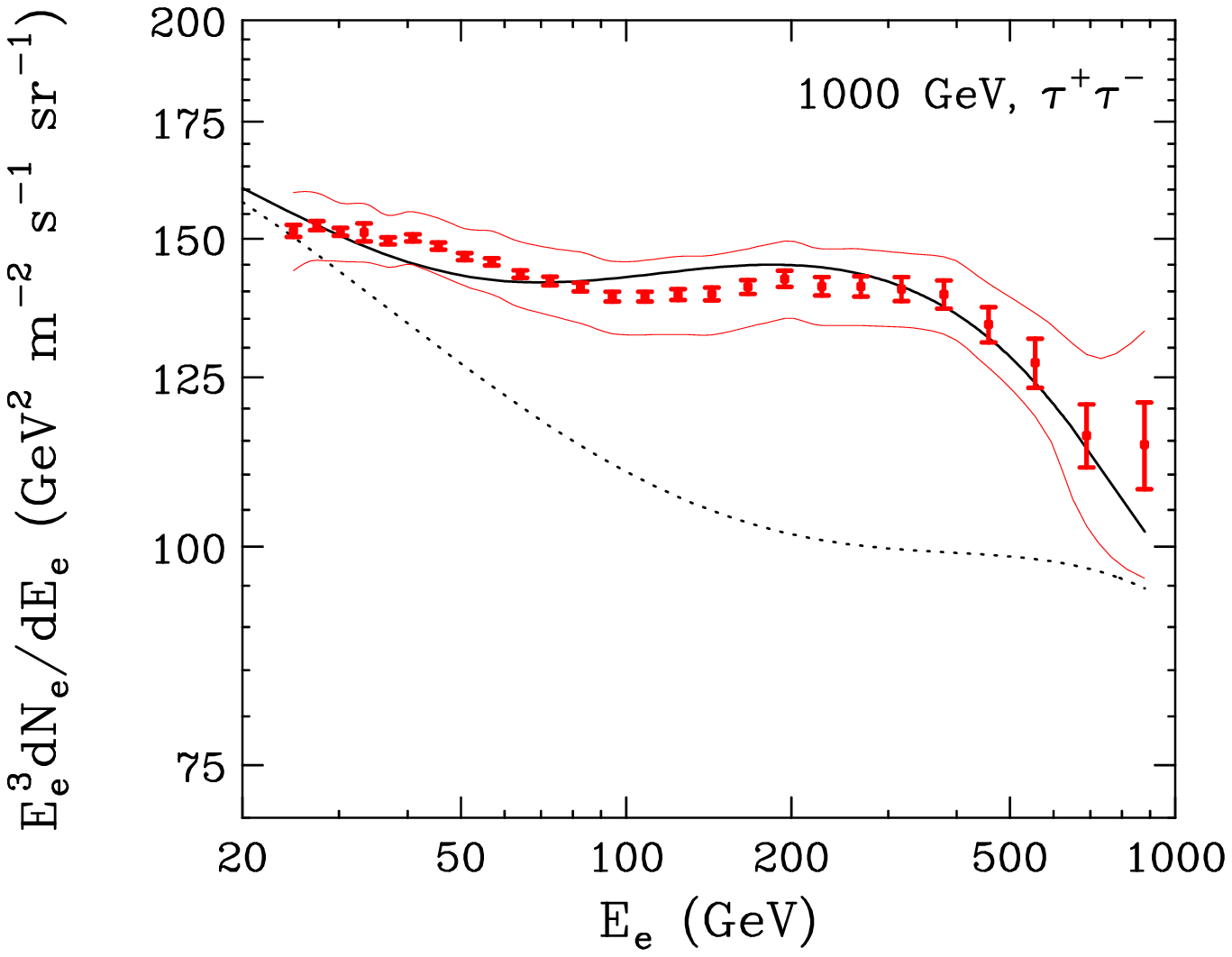}}
\hspace{0.2cm}
{\includegraphics[angle=0,width=0.45\linewidth]{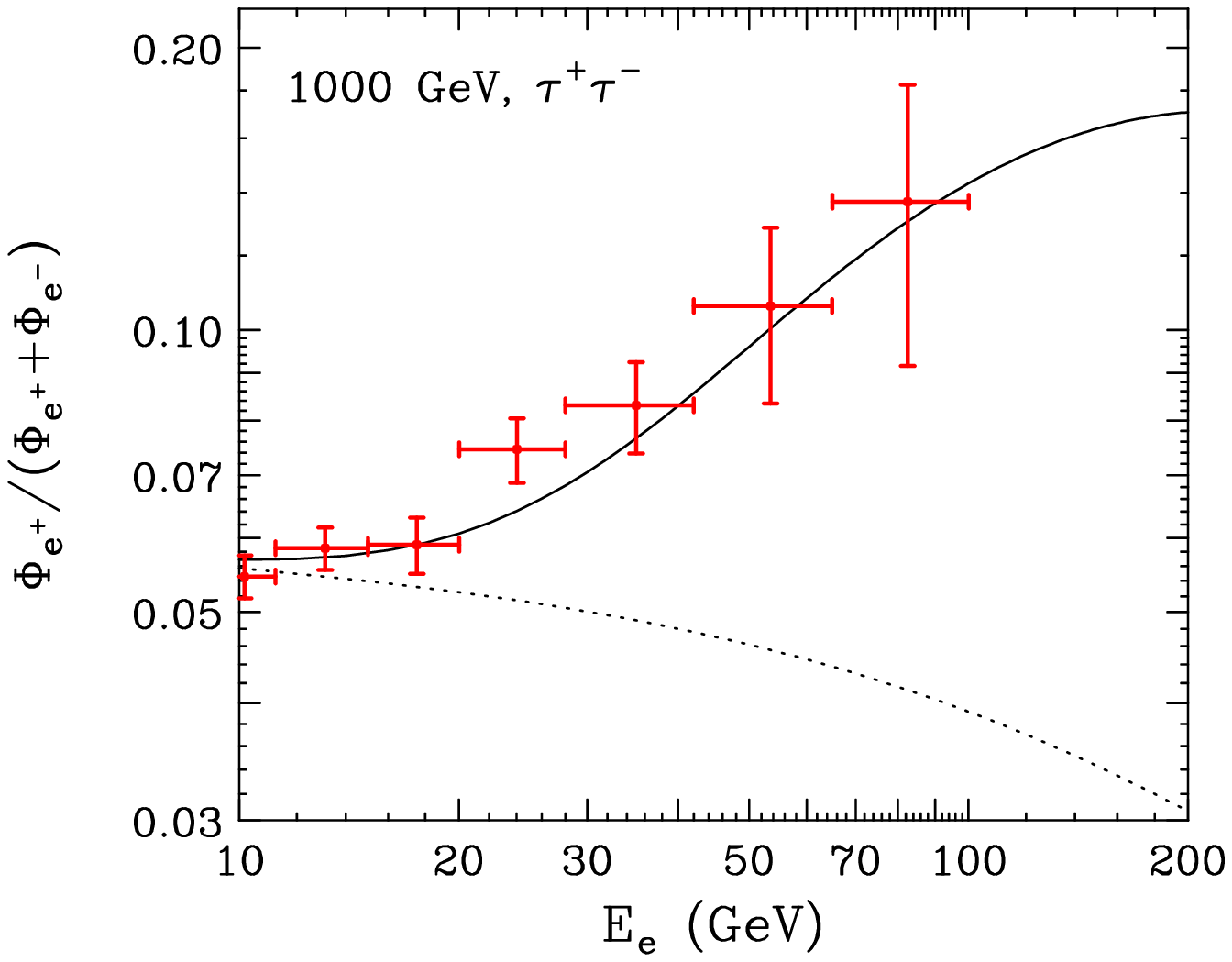}}\\
\caption{The best fits found to the PAMELA data above 10 GeV, and the (total) FGST data, for a WIMP annihilating to $\tau^+ \tau^-$ along with a power-law spectrum of cosmic ray electrons from astrophysical sources, and an additional spectrum of cosmic ray electrons from nearby, young cosmic ray accelerators. The dotted lines denote the astrophysical background used, without the contribution from dark matter. For each case shown, we found a very good fit to the data ($\chi^2 \sim 2.4-10$ distributed over 33 error bars). To normalize the dark matter annihilation rate, we have used a boost factor (relative to the rate predicted for $\sigma v = 3 \times 10^{-26}$ cm$^3$/s and $\rho_{0}=0.3$ GeV/cm$^2$) of 278, 830 and 1551 from top-to-bottom, respectively.  In each case shown, we have used propagation model B, and from top-to-bottom, have used $\phi_F=1000, 950$, and $800$ MeV. See text for more details.}
\label{tausastro}
\end{center}
\end{figure}

Finally, in Figs.~\ref{tauslt100} and \ref{tausastro}, we show our results for the case of WIMPs annihilating to $\tau^+ \tau^-$. Again, we find that the rising positron fraction can be accommodated in this channel, but that nearby astrophysical sources must contribute to the FGST spectrum at high energies to provide a good fit. To calculate the spectrum of electrons and positrons from taus, muons, or other final state particles, we have used PYTHIA~\cite{pythia} as implemented in DarkSUSY~\cite{darksusy}.

\section{Dark Matter Candidates From Other Particle Physics Scenarios}

In this section, we consider two particle physics scenarios beyond the simple annihilation to a single species of charged lepton pair. In particular, we study the lightest Kaluza-Klein state in a model with a single universal extra dimension, and the class of models in which dark matter annihilates to light states which subsequently decay into pairs of charged leptons.

\subsection{Kaluza-Klein Dark Matter}

One particle physics model that has previously been shown to provide a good fit to the PAMELA data~\cite{kkdmpamela,kkdmpos} is the first Kaluza-Klein excitation of the hypercharge gauge boson, $B^{(1)}$, which appears in models with one universal extra dimension~\cite{kkdm}. We will simply refer to this dark matter candidate by the name ``Kaluza-Klein dark matter''. 

\begin{figure}[!]
\begin{center}
{\includegraphics[angle=0,width=0.45\linewidth]{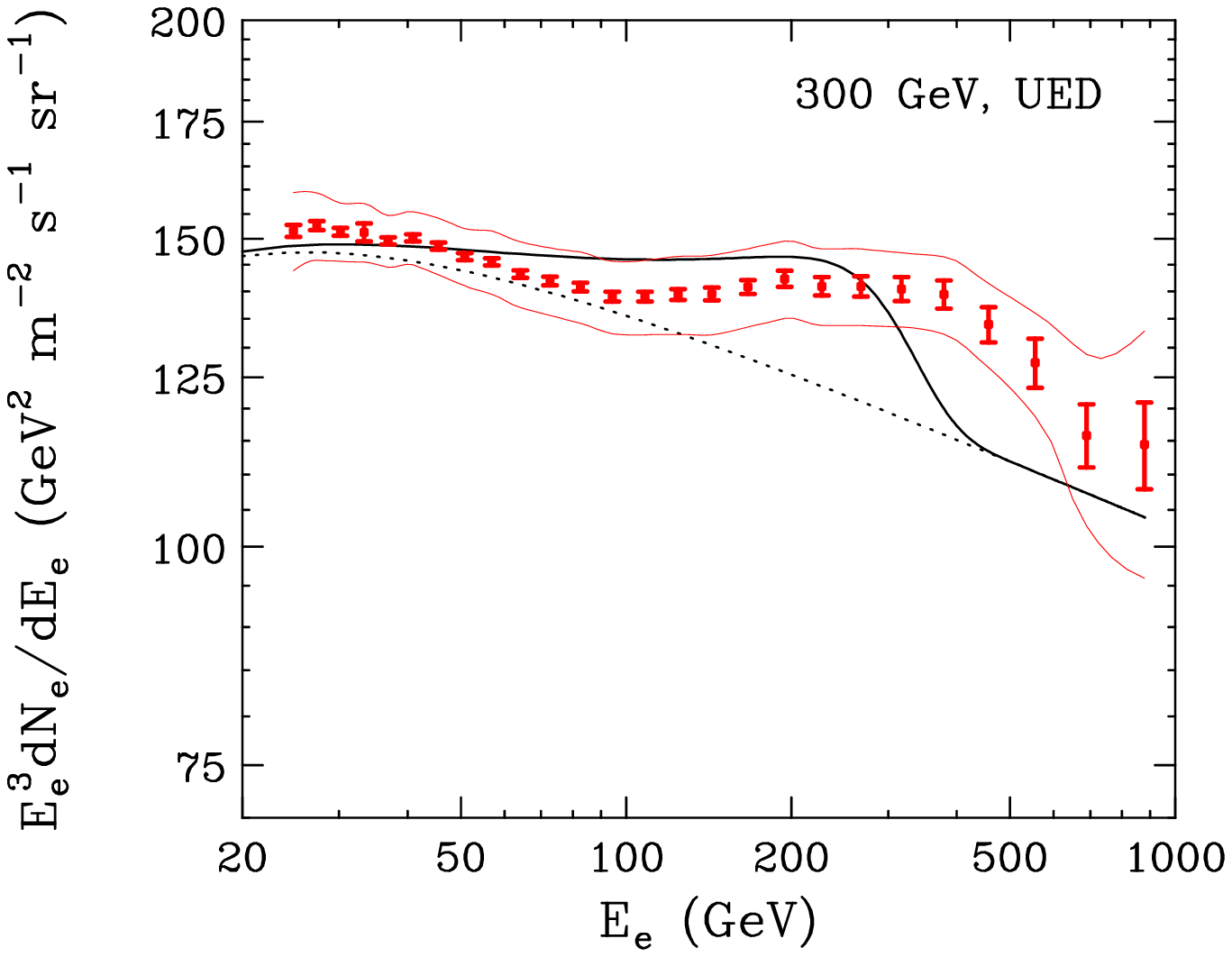}}
\hspace{0.2cm}
{\includegraphics[angle=0,width=0.45\linewidth]{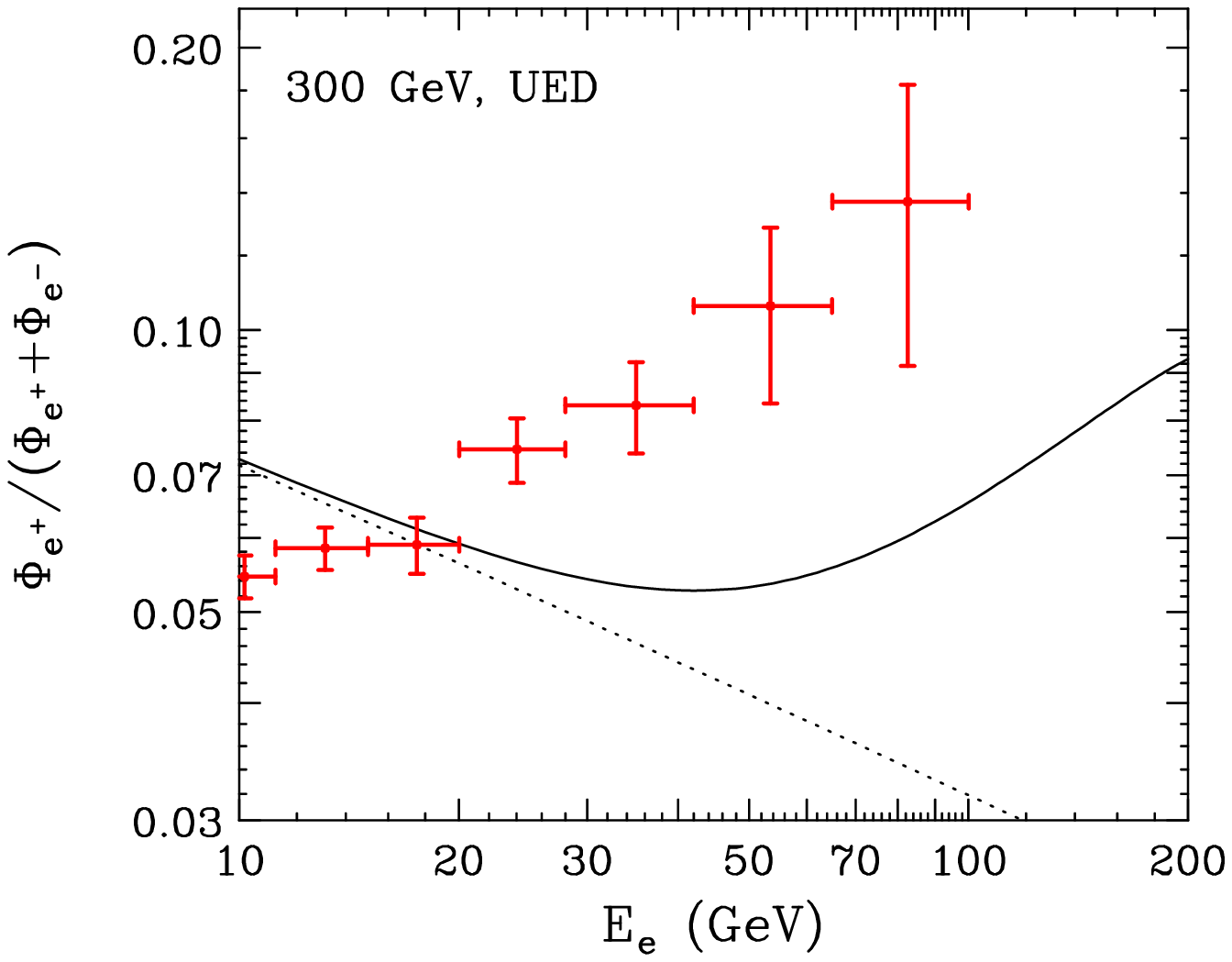}}\\
{\includegraphics[angle=0,width=0.45\linewidth]{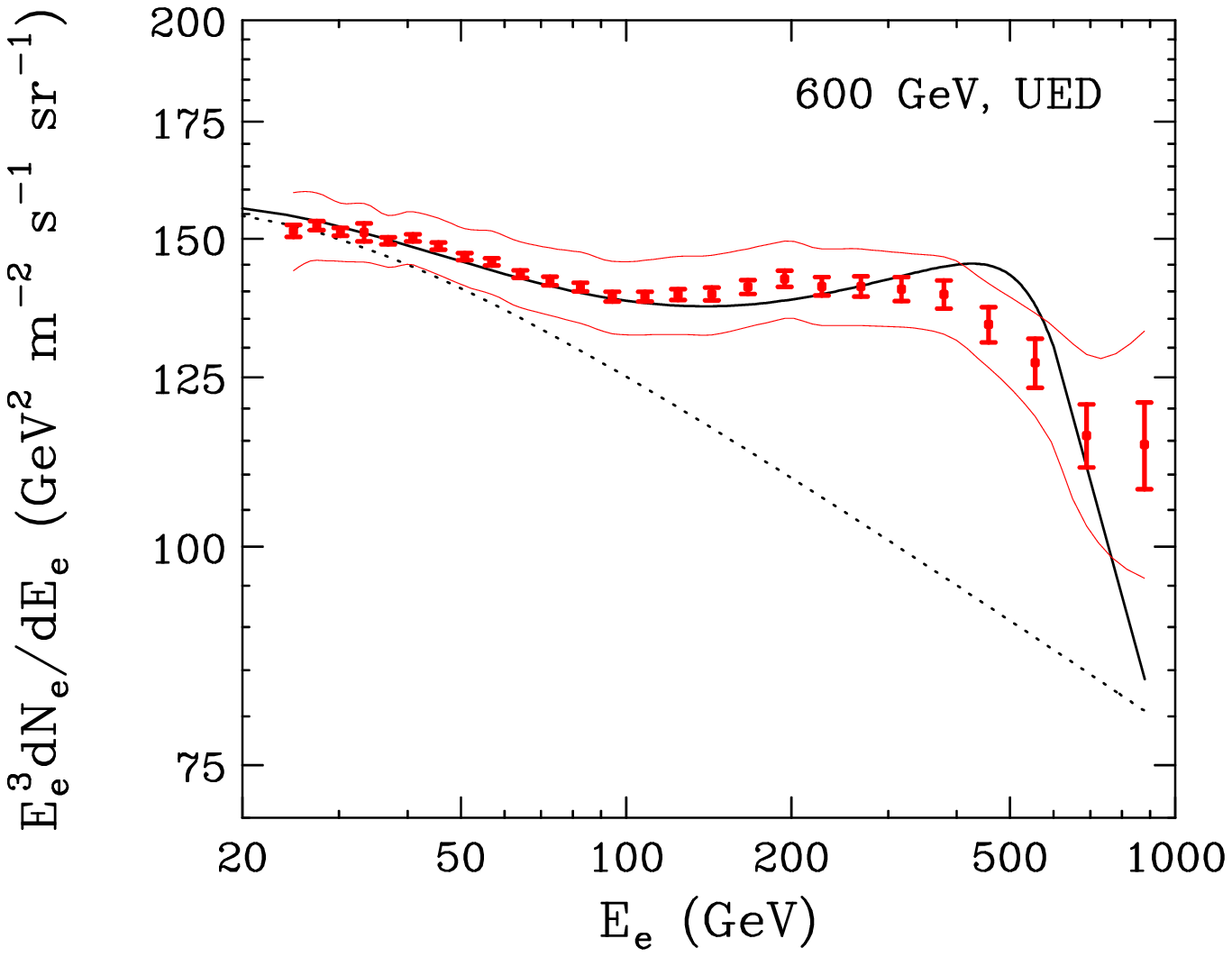}}
\hspace{0.2cm}
{\includegraphics[angle=0,width=0.45\linewidth]{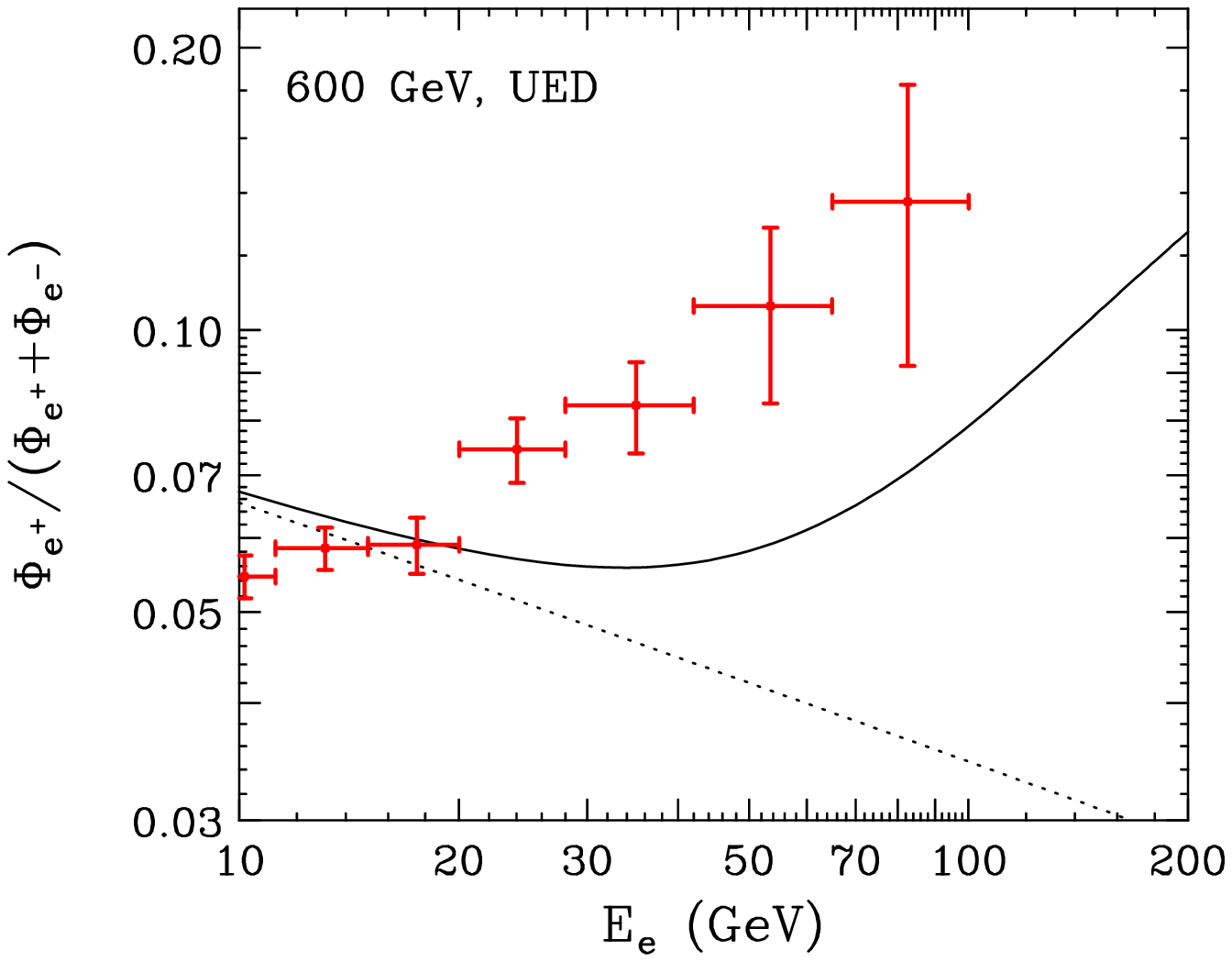}}\\
\caption{The best fits found to the PAMELA data above 10 GeV, and the (total) FGST data, for Kaluza-Klein dark matter along with a power-law spectrum of cosmic ray electrons from astrophysical sources. The dotted lines denote the astrophysical background used, without the contribution from dark matter. We find fairly poor fits to the data in this model. Fundamentally, this is the case because of the softer electron/positron spectrum produced by Kaluza-Klein dark matter relative to dark matter particles that annihilate uniquely to charged leptons. In each case shown, we have used propagation model B, and from top-to-bottom, have used $\phi_F=800$ and 750 MeV. See text for more details.}
\label{allued}
\end{center}
\end{figure}

Kaluza-Klein dark matter particles annihilate largely to charged leptons (approximately 20-23\% to each to $e^+e^-$, $\mu^+ \mu^-$ and $\tau^+\tau^-$), but also annihilate to quark-antiquark pairs a sizeable fraction (25-35\%) of the time. The hadronic final states tend to produce a flux of cosmic ray antiprotons exceeding observations unless the boundary conditions of the diffusion zone are particularly narrow (such as those in our propagation model B)~\cite{kkdmpamela}. In light of the FGST electron spectrum, we find that Kaluza-Klein dark matter faces other problems as well, as are demonstrated in Fig.~\ref{allued}. In particular, the contribution to the electron-positron spectrum from hadronic final states is less hard from than from charged leptons. To compensate for this, the astrophysical background of cosmic ray electrons must drop quite rapidly between approximately 10 and 100 GeV in order to accommodate the observed positron fraction. In Ref.~\cite{kkdmpamela}, for example, Kaluza-Klein dark matter along with a cosmic ray electron background of $dN_e/dE_e \propto E_e^{-3.40}$ was shown to provide a good fit to the PAMELA positron fraction data. This very soft background spectrum, however, does not agree with the FGST electron spectrum measurement. Even if we only require the model to fit the FGST data in the PAMELA range (below 100 GeV), we find that the best fits consistently underproduce the PAMELA positron fraction measurements. Furthermore, additional local contributions to the cosmic ray electron spectrum above 100 GeV from local sources do not significantly improve the overall fits in this model.

More generally speaking, we can conclude from this example that any 100-1000 GeV dark matter particle which annihilates to hadronic channels more than $\sim$~$10\%$ of the time will be unable to generate the positron fraction observed by PAMELA while also providing a consistent fit to the FGST electron spectrum, unless additional primary sources of cosmic ray positrons are also included. 

For an alternative Kaluza-Klein dark matter scenario in which this problem can be evaded, see Ref.~\cite{nojiri}.

\subsection{Annihilation to Light Intermediate States}

Next, we consider the case in which dark matter particles annihilate to light intermediate states that subsequently decay to charged lepton pairs. This phenomenology can be found within a number of recent models~\cite{theory,nmssm}, motivated in part by the fact that if the mass of the intermediate state is lighter than $\sim$0.1-1 GeV, this can explain why dark matter annihilations yield electrons and/or muons without a significant quantity of hadronic final states.

\begin{figure}[!]
\begin{center}
{\includegraphics[angle=0,width=0.45\linewidth]{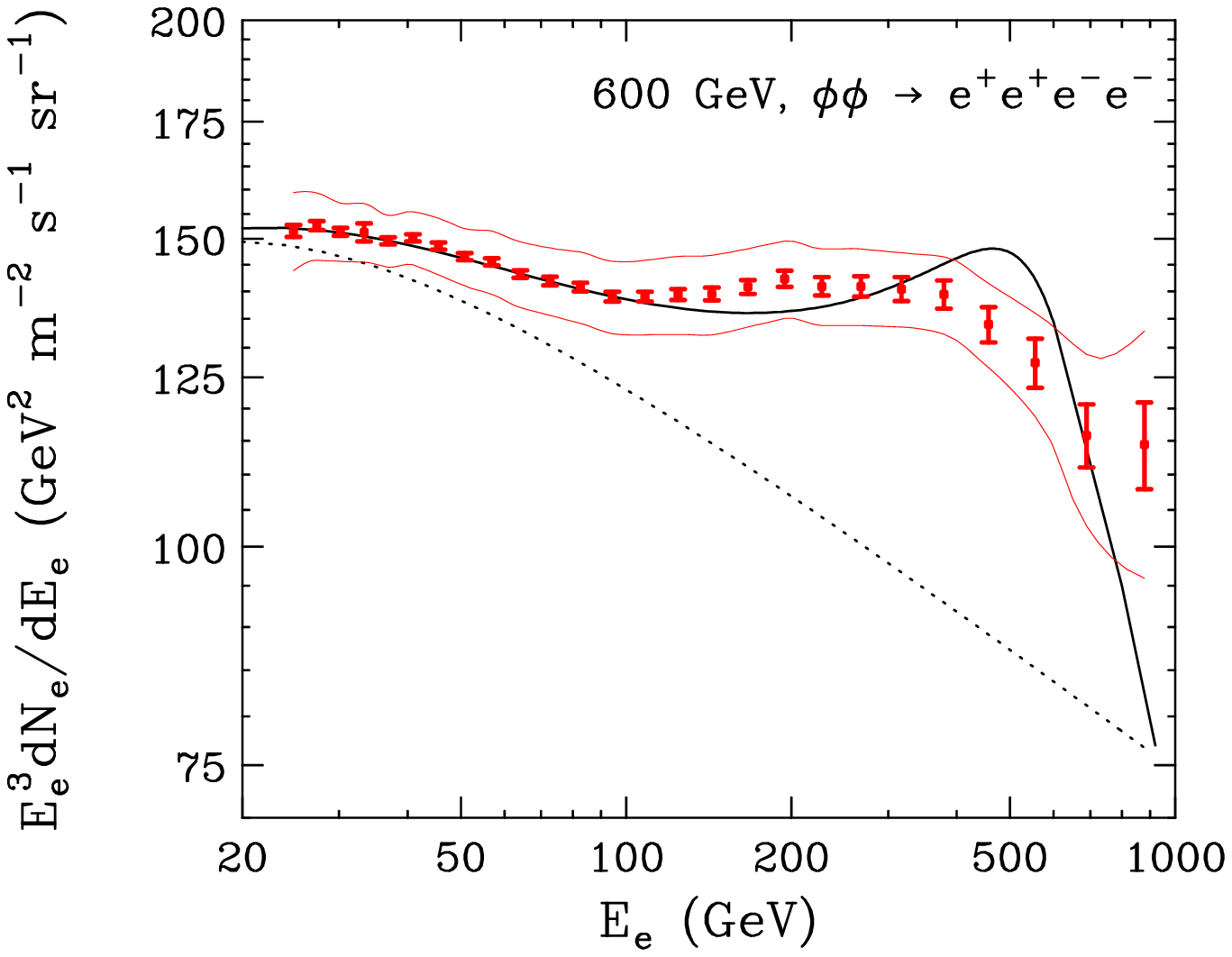}}
\hspace{0.2cm}
{\includegraphics[angle=0,width=0.45\linewidth]{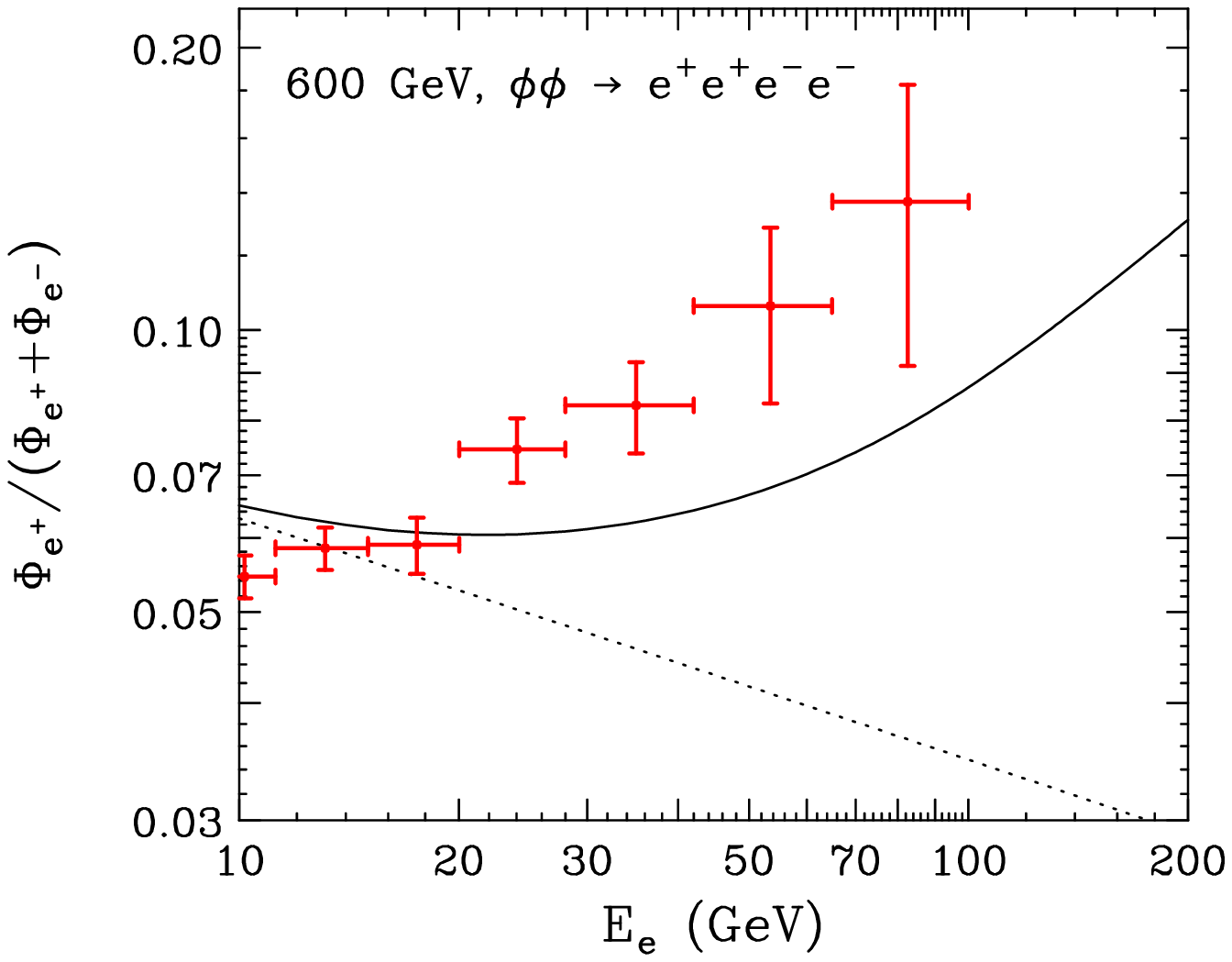}}\\
{\includegraphics[angle=0,width=0.45\linewidth]{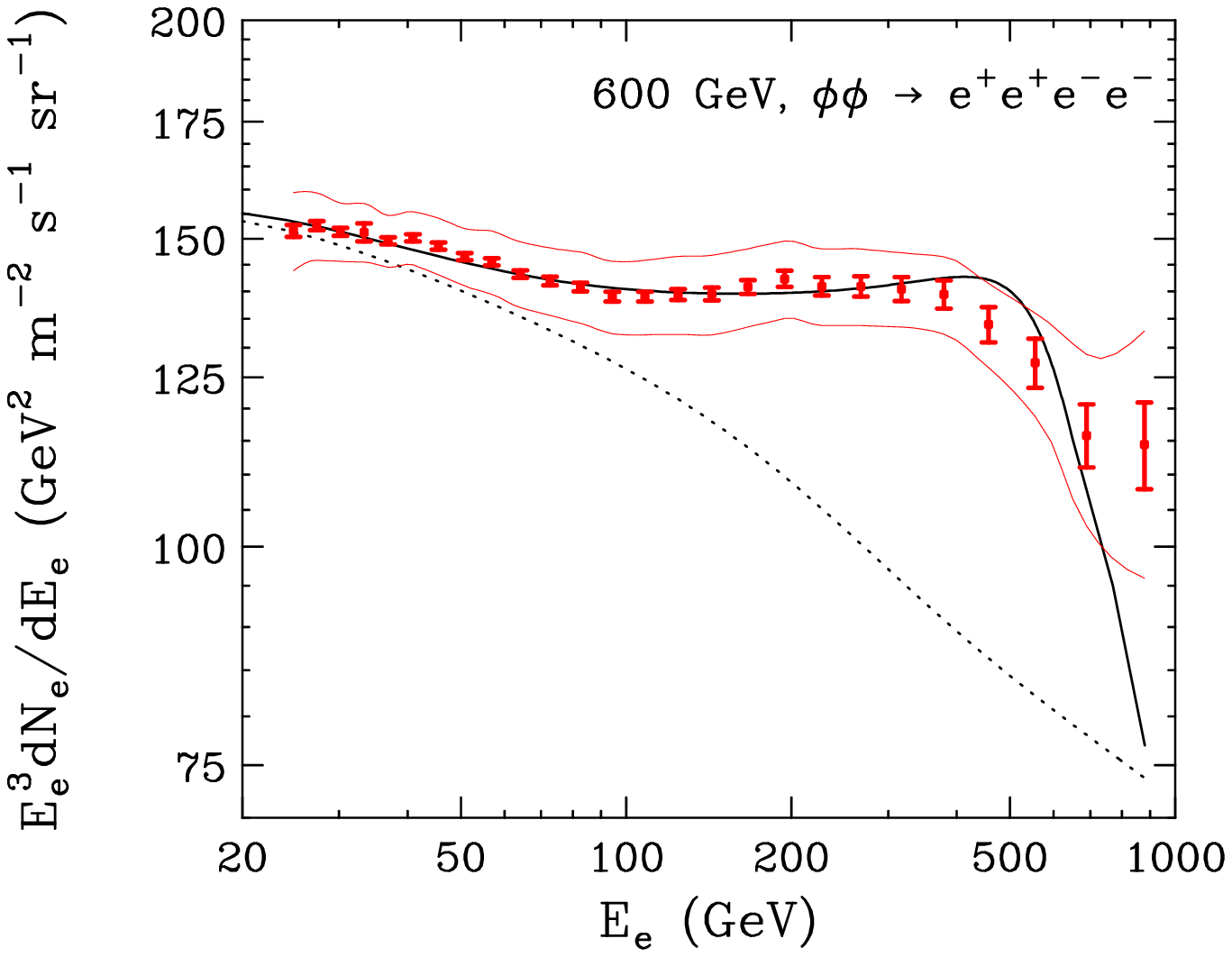}}
\hspace{0.2cm}
{\includegraphics[angle=0,width=0.45\linewidth]{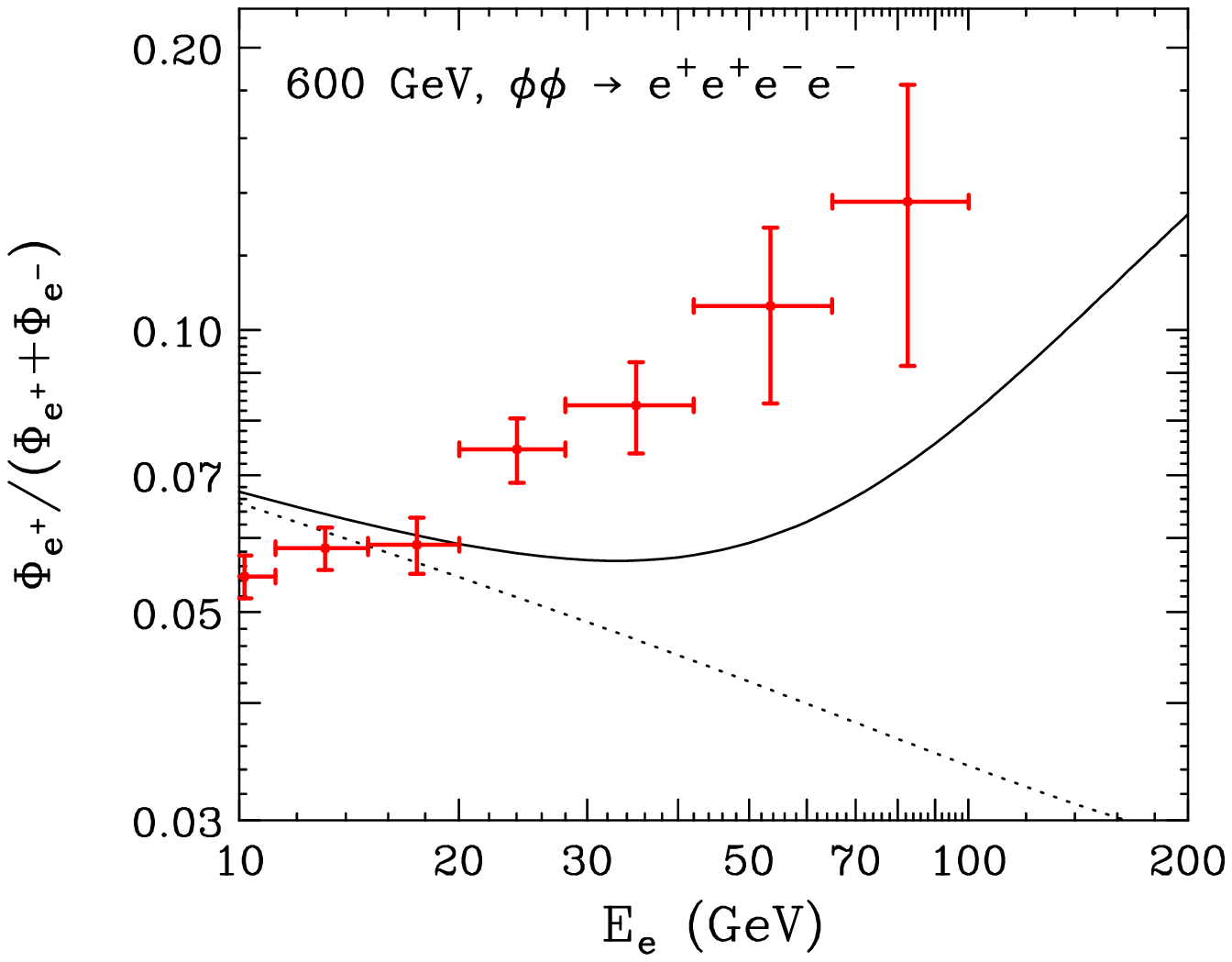}}\\
\caption{The best fits found to the PAMELA data above 10 GeV, and the (total) FGST data, for a 600 dark matter particle annihilating to a pair of intermediate light states ($\phi \phi$), which proceed to decay to electron-positron pairs, along with a power-law spectrum of cosmic ray electrons from astrophysical sources. The dotted lines denote the astrophysical background used, without the contribution from dark matter. The top (bottom) frames adopt an intermediate particle mass of 100 MeV (2 MeV). In the 100 MeV case, we find a very similar fit for the case of a 600 GeV dark matter particle which annihilate directly to $e+ + e^-$ (see Fig.~\ref{all}). This is because half of the resultant electrons and positrons have energies near the mass of the dark matter particle.  In the case of a 2 MeV intermediate particle, the distribution of electrons and positrons is distributed somewhat more evenly with energy, leading to a somewhat softer spectrum. See text for more details.}
\label{alleeee}
\end{center}
\end{figure}

In the limit that the intermediate state particle is much heavier than the mass of the final state leptons, half of the leptons produced will be highly boosted, leading to a spectrum that is nearly identical to that found in the case where the WIMP annihilate directly to two leptons. This is seen in the top frames of Fig.~\ref{alleeee}, where we consider the annihilation of dark matter to two 100 MeV intermediate states, which each decay to $e^+ e^-$. The result is similar to that shown in Fig.~\ref{all}. As we lower the mass of the intermediate state, more of the final state electrons have energies well below the mass of the dark matter particle. In the case shown in the lower frame of Fig.~\ref{alleeee}, this improves the fit to the FGST spectrum by suppressing the flux at $\sim 500-600$ GeV. This also softens the positron spectrum over the range of energies measured by PAMELA, making it more difficult to accommodate the observed positron fraction. We thus conclude that in the case of dark matter annihilations to $e^+ e^-$, four fermion final states provide only a marginally better fit to the combined data than annihilations to two fermion final states.



\section{Conclusions}

In this article, we have studied the possibility that sub-TeV dark matter particles annihilating in the Galactic Halo are responsible for producing the cosmic ray positron fraction observed by PAMELA, while also remaining consistent with the cosmic ray electron spectrum observed by the Fermi Gamma Ray Space Telescope (FGST). In our analysis, we include the possibility that stochastic variations in the local sources of cosmic ray electrons may significantly impact the cosmic ray electron spectrum above $\sim 10^2$ GeV, and also include the effects of FGST's finite energy resolution. With these considerations taken into account, we find acceptable fits to this combined set of data for WIMPs as light as $\sim 300 \mbox{ GeV}$. In particular, dark matter particles annihilating directly to $e^+ e^-$ can accommodate the data for masses between approximately 300 and 600 GeV, while WIMPs annihilating to $\mu^+ \mu^-$ or $\tau^+ \tau^-$ can provide good fits for masses between approximately 300 and 1000 GeV (and higher).

One challenge involved with fitting the PAMELA and FGST data with sub-TeV dark matter is in reconciling the relative hardness of the observed FGST spectrum with the rising positron fraction observed by PAMELA.  This combination of data implies that the positron signal must be very rapidly rising over the 20-100 GeV range (even more rapidly than previously implied by PAMELA alone). One consequence of this is that the fraction of dark matter annihilations that proceed to non-leptonic states must be very small ($\lsim 10\%$). Kaluza-Klein dark matter in a minimal model with one universal extra dimension, for example, fails in this regard to accommodate the combined PAMELA and FGST data. 



In summary, we conclude that dark matter with a wide range of masses below $\sim$1 TeV, annihilating to charged leptons, can accommodate the rising positron fraction observed by PAMELA, while remaining consistent with the cosmic ray electron spectrum reported by FGST. While multi-TeV dark matter particles can also provide good fits to this combined set of data~\cite{bergstrom,meade}, the results presented here enable us to continue to contemplate many theoretically attractive scenarios in which the PAMELA positron fraction results from particle physics associated with the electroweak scale.


\bigskip

We would like to thank the Aspen Center for Physics for their hospitality, where this work was initiated. This work has been supported by the US Department of Energy, including grant DE-FG02-95ER40896, and by NASA grant NAG5-10842.

\end{document}